\documentclass[preprint2]{aastex}

\usepackage{natbib}

\shorttitle{Ionized Gas in 30~Doradus}
\shortauthors{Indebetouw et al.}

\begin{document}

\newcommand{\arii}{[\ion{Ar}{2}]}
\newcommand{\ariil}{[\ion{Ar}{2}]$\lambda$7.0\um}
\newcommand{\ariii}{[\ion{Ar}{3}]}
\newcommand{\ariiil}{[\ion{Ar}{3}]$\lambda$9.0\um}
\newcommand{\neii}{[\ion{Ne}{2}]}
\newcommand{\neiil}{[\ion{Ne}{2}]$\lambda$12.8\um}
\newcommand{\neiii}{[\ion{Ne}{3}]}
\newcommand{\neiiil}{[\ion{Ne}{3}]$\lambda$15.5\um}
\newcommand{\siii}{[\ion{S}{3}]}
\newcommand{\siiil}{[\ion{S}{3}]$\lambda$18.7\um}
\newcommand{\siiill}{[\ion{S}{3}]$\lambda$33.4\um}
\newcommand{\siv}{[\ion{S}{4}]}
\newcommand{\sivl}{[\ion{S}{4}]$\lambda$10.5\um}
\newcommand{\feii}{[\ion{Fe}{2}]}
\newcommand{\feiil}{[\ion{Fe}{2}]$\lambda$26.0\um}
\newcommand{\hii}{{\sc H$\,$II}}
\newcommand{\niiil}{[\ion{N}{3}]$\lambda$57.3\um}
\newcommand{\oil}{[\ion{O}{1}]$\lambda$63.1\um}
\newcommand{\oiii}{[\ion{O}{3}]}
\newcommand{\oiiil}{[\ion{O}{3}]$\lambda$88.3\um}

\newcommand{\hua}{\mbox{Humphreys-$\alpha$}}
\newcommand{\hual}{\mbox{Humphreys-$\alpha$ $\lambda$12.37\um}}
\newcommand{\hug}{\mbox{Humphreys-$\gamma$}}
\newcommand{\hugl}{\mbox{Humphreys-$\gamma$ $\lambda$5.90\um}}
\newcommand{\molhtwo}{H$_2$ S(2)}
\newcommand{\molhtwol}{H$_2$ S(2) $\lambda$12.28\um}
\newcommand{\molhthree}{H$_2$ S(3)}
\newcommand{\molhthreel}{H$_2$ S(3) $\lambda$9.67\um}

\newcommand{\um}{$\mu$m}
\newcommand{\kms}{{km$\;$s$^{-1}$}}
\newcommand{\wms}{{W$\;$m$^{-2}\;$Sr$^{-1}$}}
\newcommand{\dor}{30$\;$Doradus}
\newcommand{\radec}[6]{{#1}$^h${#2}$^m${#3}$^s$ {#4}$^\circ${#5}\arcmin{#6}\arcsec}
\newcommand{\ee}[1]{\mbox{${} \times 10^{#1}$}}
\newcommand{\eten}[1]{\mbox{$10^{#1}$}}
\newcommand{\msun}{M$_\odot$}

\newcommand{\hr}{high resolution}
\newcommand{\lr}{low resolution}
\newcommand{\mc}{Magellanic Clouds}

\newcommand{\mpscl}{0.25} 
\newcommand{\mpsclo}{0.30} 
\newcommand{\mpvsp}{0.1in} 
\newcommand{\mphsp}{0.1in} 
\newcommand{\mpscla}{0.15} 
\newcommand{\ratioscl}{0.3} 
\newcommand{\TUscl}{0.25} 

\newcommand{\spit}{{\it Spitzer}}

\newcommand{\linepeak   }{``0''}
\newcommand{\bananasplit}{``A''} 
\newcommand{\parkerMI   }{``B''}    
\newcommand{\core       }{``C''}
\newcommand{\extinc     }{``D''}
\newcommand{\WN         }{``E''}
\newcommand{\trough     }{``F''}
\newcommand{\lowexcite  }{``G''}


\title{Physical Conditions in the Ionized Gas of 30 Doradus}

\newcounter{affil}
\setcounter{affil}{1}

\newcounter{uva}\setcounter{uva}{\value{affil}}\stepcounter{affil}
\altaffiltext{\arabic{uva}}{Department of Astronomy, University of Virginia, PO Box 3818, Charlottesville, VA 22903}

\newcounter{nrao}\setcounter{nrao}{\value{affil}}\stepcounter{affil}
\altaffiltext{\arabic{nrao}}{National Radio Astronomy Observatory, 520 Edgemont Rd, 
Charlottesville, VA 22903}

\newcounter{cea}\setcounter{cea}{\value{affil}}\stepcounter{affil}
\altaffiltext{\arabic{cea}}{Service d'Astrophysique CEA, Saclay, 91191 Gif Sur Yvette Cedex, France}

\newcounter{ua}\setcounter{ua}{\value{affil}}\stepcounter{affil}
\altaffiltext{\arabic{ua}}{Steward Observatory, University of Arizona, 933 North Cherry Ave., Tucson, AZ 85719}

\newcounter{toledo}\setcounter{toledo}{\value{affil}}\stepcounter{affil}
\altaffiltext{\arabic{toledo}}{Department of Physics and Astronomy, Mail Drop 111, University of Toledo, 2801 W. Bancroft St, Toledo, OH 43606}

\newcounter{stsci}\setcounter{stsci}{\value{affil}}\stepcounter{affil}
\altaffiltext{\arabic{stsci}}{Space Telescope Science Institute, 3700 San Martin Drive, Baltimore, MD 21218}

\newcounter{ucl}\setcounter{ucl}{\value{affil}}\stepcounter{affil}
\altaffiltext{\arabic{ucl}}{Department of Physics and Astronomy, University College London, Gower Street, London WC1E 6BT}

\newcounter{leiden}\setcounter{leiden}{\value{affil}}\stepcounter{affil}
\altaffiltext{\arabic{leiden}}{Leiden Observatory, Leiden University, PO Box 9513, 2300 RA Leiden, Netherlands}

\newcounter{ias}\setcounter{ias}{\value{affil}}\stepcounter{affil}
\altaffiltext{\arabic{ias}}{Astrophysique de Paris, Institute (IAP), CNRS UPR 341, 98bis, Boulevard Arago,  Paris, F-75014, France}

\newcounter{umd}\setcounter{umd}{\value{affil}}\stepcounter{affil}
\altaffiltext{\arabic{umd}}{Department of Astronomy, University of Maryland College Park, MD 20742}

\newcounter{cfa}\setcounter{cfa}{\value{affil}}\stepcounter{affil}
\altaffiltext{\arabic{cfa}}{Center for Astrophysics, 60 Garden St., MS 67 , Harvard University, Cambridge, MA 02138}

\newcounter{ames}\setcounter{ames}{\value{affil}}\stepcounter{affil}
\altaffiltext{\arabic{ames}}{NASA Ames Research Center, SOFIA Office,  MS 211-3, Moffet Field, CA 94035}

\newcounter{jpl}\setcounter{jpl}{\value{affil}}\stepcounter{affil}
\altaffiltext{\arabic{jpl}}{Jet Propulsion Lab, 4800 Oak Grove Dr., MS 264--767, Pasadena, CA 91109}


\author{
R. Indebetouw\altaffilmark{\arabic{uva},\arabic{nrao}},
G. E. de Messi\`{e}res\altaffilmark{\arabic{uva}},
S. Madden\altaffilmark{\arabic{cea}},
C. Engelbracht\altaffilmark{\arabic{ua}},
J. D. Smith\altaffilmark{\arabic{toledo}},
M. Meixner\altaffilmark{\arabic{stsci}},
B. Brandl\altaffilmark{\arabic{leiden}},
L. J. Smith\altaffilmark{\arabic{stsci},\arabic{ucl}},
F. Boulanger\altaffilmark{\arabic{ias}},
F. Galliano\altaffilmark{\arabic{umd}},
K. Gordon\altaffilmark{\arabic{ua}},
J. L. Hora\altaffilmark{\arabic{cfa}},
M. Sewilo\altaffilmark{\arabic{stsci}},
A. G. G. M. Tielens\altaffilmark{\arabic{ames}},
M. Werner\altaffilmark{\arabic{jpl}},
M. G. Wolfire\altaffilmark{\arabic{umd}}
}

\email{remy@virginia.edu, ged3j@virginia.edu}


\begin{abstract}

We present a mid-infrared spectroscopic data cube of the central part
of \dor, observed with {\it Spitzer}'s IRS and MIPS/SED mode.
Aromatic dust emission features and emission lines from molecular and
atomic hydrogen are detected but not particularly strong.  The
dominant spectral features are emission lines from moderately ionized
species of argon, neon, and sulphur, which are used to determine the
physical conditions in the ionized gas.  The ionized gas excitation
shows strong variations on parsec scales, some of which can plausibly
be associated with individual hot stars.  We fit the ionic line
strengths with photoionization and shock models, and find that
photoionization dominates in the region.  The ionization parameter $U$
traces the rim of the central bubble, as well as highlighting isolated
sources of ionization, and at least one quiescent clump.  The hardness
of the ionizing radiation field $T_{rad}$ reveals several ``hot
spots'' that are either the result of individual very hot stars or
trace the propagation of the diffuse ionizing field through the
surrounding neutral cloud.  Consistent with other measurements of
giant \ion{H}{2} regions, log($U$) ranges between -3 and -0.75, and
$T_{rad}$ between 30000 and 85000K.

\end{abstract}



\section{Introduction}

The \dor\ region of the Large Magellanic Cloud (LMC) is an ideal
laboratory in which to study the effect of massive star formation and
its feedback on the circumcluster interstellar medium (ISM).  Star
formation processes in the \mc\ are potentially a template for the
early universe, where small irregular galaxies were commonplace and
the overall metallicity lower.  In particular, the star formation rate
measured relative to molecular mass may be high in the LMC compared to
our Galaxy, but that measurement is complicated by
interpretation of CO data in molecular clouds that are known to be
more porous and deeply penetrated by ultraviolet radiation, and an
uncertain CO-H$_2$ conversion factor
\citep[e.g.][]{poglitsch,sageism}.

\begin{figure}
\plotone{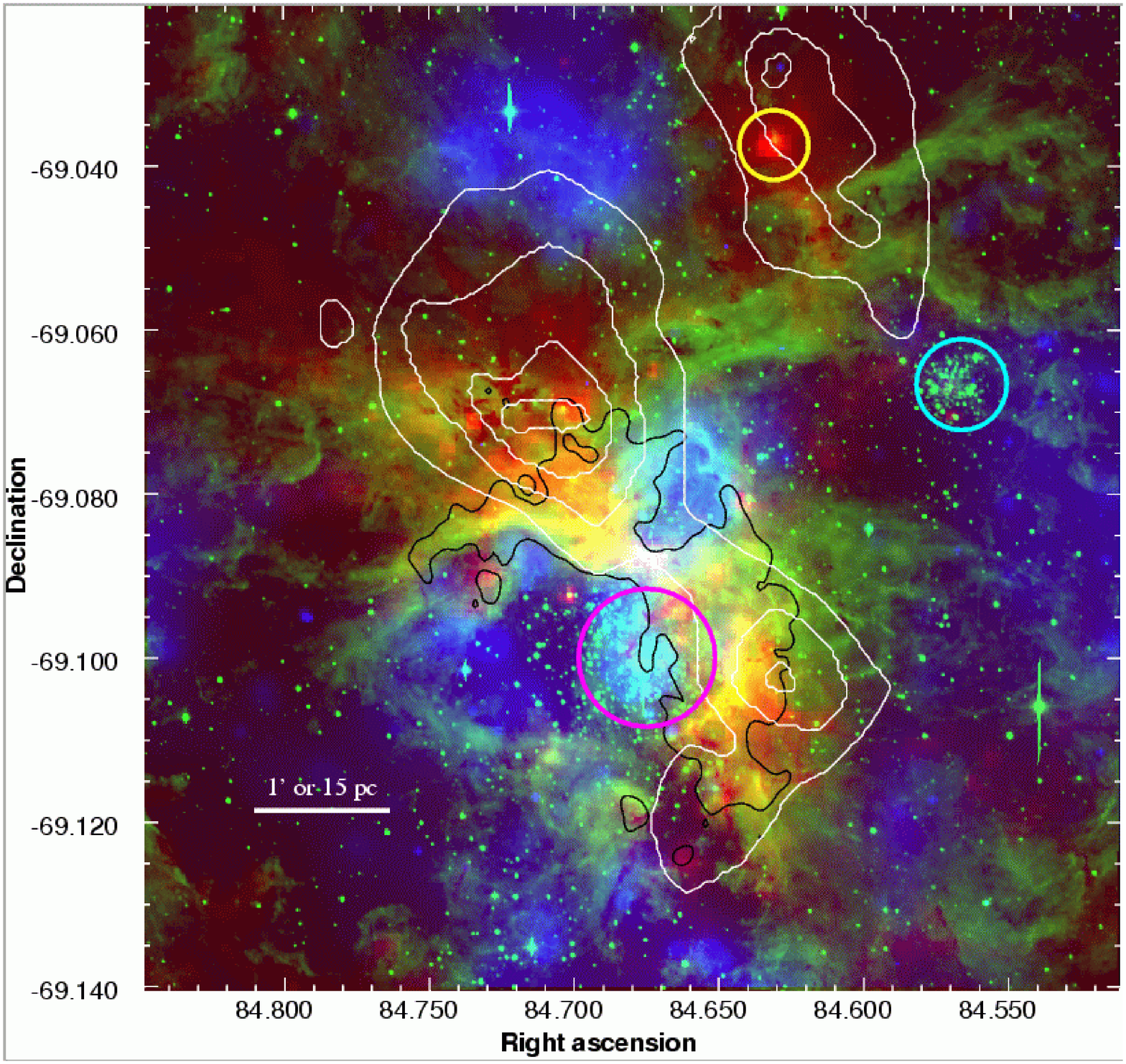}
\caption{\label{threecoloroverview}The \dor\ nebula.  Red: IRAC 8\um\ image 
  (SAGE).  Green: ESO B-band image.  Blue: broad-band soft X-ray image, 
  $0.5 - 2$ keV \citep[private correspondence and ][]{townsley06}.  All 
  images are on a linear scale.  White contours: $^{12}$CO(1-0) emission 
  \citep{johansson}.  Black contour: a single 
  level of 3cm radio emission, to guide the eye (see Figure \ref{bwoverview}).
  Magenta mark: the star cluster R136, core of NGC 2070, 
  at \radec{5}{38}{42}{-69}{06}{00}.  Cyan mark: the star cluster Hodge~301, 
  at \radec{5}{38}{16}{-69}{04}{00}.  Yellow mark: an IR point source 
  in the molecular cloud [JGB98] Dor-06 \citep{johansson}, at 
  \radec{5}{38}{31.63}{-69}{02}{14.6}.  This source is also marked in white 
  in Figure \ref{saturationmap} and discussed in \S\ref{artifacts}. }
\end{figure}

\begin{figure}
\plotone{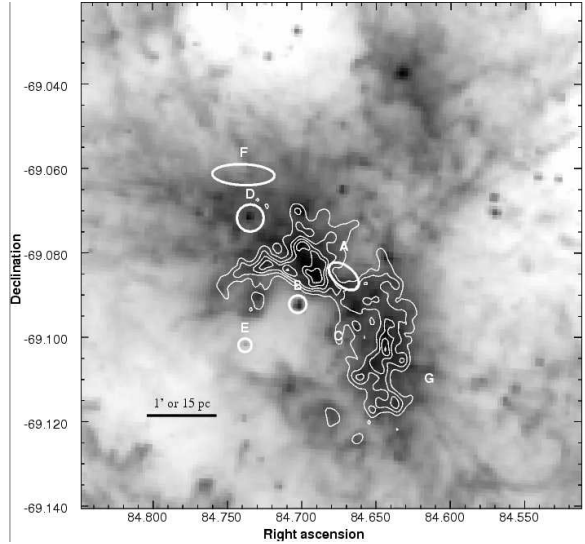}
\caption{\label{bwoverview}IRAC 8\um\ image (log scale).  White
  contours: high-resolution 3cm continuum from \citet{lazendic}.  Contour 
  levels: 6, 9, 12, 15, 18, and 21 \ee{-3} Jy/bm.
  Sources of interest marked here are listed in Table~\ref{sources_table},
  described in \S\ref{sources_text}, and their spectra shown in
  Figure~\ref{sources_spectra}.}
\end{figure}

\dor\ itself, the brightest star formation region in the LMC, contains
several \eten{5}\msun\ of molecular hydrogen traced by $^{12}$CO(1-0)
\citep{johansson}, much of which is in two elongated clouds that form
an arc or ridge in the center of the nebula (see
Figure~\ref{threecoloroverview} for a multiwavelength view and
Figure~\ref{bwoverview} for the distribution of CO).  That molecular
material is probably only the remnant of the cloud that formed the
central cluster and thousands of OB stars.  Observations of higher
energy CO transitions suggest that this remnant ridge of molecular gas
is quite warm and dense \citep{kim07}, and near-infrared observations
have revealed that it is actively forming new stars
\citep{hyland92,rubio98,maercker05}

The dominant star-forming cluster of the \dor\ \hii\ region is NGC 2070,
whose dense center, R136, is usually considered the nearest super star
cluster (SSC).  R136 contains 39 O3 stars, a total stellar mass of
$\simeq$5\ee{4}\msun\ within 2.5 pc, and stellar densities exceeding
5\ee{4}\msun{}pc$^{-3}$ \citep{hunter,walborn}.
Extragalactic SSCs are the hosts of one of the most extreme modes of
star formation in the universe, may develop into globular clusters,
and play a key role in galaxy formation \citep{johnson}.  At a
distance of 50$\pm$2.5 kpc \citep[see discussion of the LMC distance
  and uncertainty in][]{schaefer}, \dor\ is close enough to study at
the parsec-scale resolution that is required to understand the
formation and feedback effects of individual stars, and a prime target
for detailed study of the same mechanisms which operate in more
distant and massive SSCs.

Observations at mid-infrared (MIR) wavelengths offer several
advantages for studying star formation and its interaction with
circumcluster dust and gas.
MIR (here roughly defined as 5-50\um) continuum emission is generally
dominated by radiation from very small dust grains (VSGs); the broad
shape of the continuum is sensitive to the VSG size and temperature
distribution, and hence indirectly to the radiation intensity in
photon-dominated regions, and to the destruction of smaller grains as
would be expected above 2000K and in \hii\ regions.  
Broad ($\sim$1\um) dust emission features are also present between 3
and 19\um.  These previously named ``unidentified infrared bands''
result from distortion (bending, stretching) modes of aromatic
molecules containing tens to hundreds of carbon atoms.  (Most features
have been attributed to polycyclic aromatic hydrocarbons, PAHs, but
unique astrophysical identification is a work in progress).  Analysis
of relative strengths of PAH features reveals their size and
ionization state which is expected to change in intense radiation
fields.
Extinction by dust in the infrared is low compared to other wavelength
regimes.  Observations in the MIR can pierce cold molecular clouds and
reveal the star-forming regions that they shroud.  There are two major
bands of absorption by silicate dust at 9.7 and 18\um, whose shape and
strength provide further diagnostics of the dust in the region.
Mid-infrared (MIR) spectroscopy of the entire \dor\ region with {\it
  ISO-SWS} \citep{sturm00} revealed a continuum-dominated spectrum
with very weak silicate absorption and also only modestly strong
aromatic emission features.  The 4 spectra taken in \dor\ with ISOPHOT
\citep{vermeij02} similarly show low ratios of PAH strength to IR
continuum.

The most recent and sensitive MIR continuum observations of the \mc\ 
were obtained as part of the \spit\ \citep{werner04} Legacy program
``Surveying the Agents of a Galaxy's Evolution'' (SAGE), using IRAC
\citep[3--8\um;][]{fazio_irac} and MIPS
\citep[24-160\um;][]{rieke_mips}.
The goals of SAGE are to conduct a detailed study of the dust
processes in the ISM, to complete a census of newly formed stars in
order to find the star formation rate in the LMC, and to study evolved
stars with high mass loss rates in order to determine the rate of mass
injection into the ISM \citep{meixner}.
Figure~\ref{threecoloroverview} places the MIR emission from \dor\ in
its multiwavelength context.  8\um\ emission traces the same arc-shaped
ridge seen in the optical, delineating the edge of a bubble probably
blown by R136 and filled with hot X-ray emitting plasma.  The
remaining molecular material in the region is also located in that
ridge.

This paper presents a new spatially filled spectral cube of \dor\ with
the low-resolution modules of \spit/IRS \citep{houck_irs} and with the SED mode of \spit/MIPS \citep{rieke_mips,lu_sed}.  The
sensitivity of this dataset greatly exceeds the previous ISOCAM-CVF
data \citep{madden06}, which only provided maps of the few strongest
ionic lines, but tentatively showed a gradient of decreasing
excitation in the \neiii/\neii\ ratio as a function of distance from
R136, and signs of PAH destruction in the central regions of the
nebula.

The data are described in sections \S\ref{observations}.  In
particular, \S\ref{datared} and Appendix~\ref{appendix1} discusses
reduction and artifacts, \S\ref{fitting} the line and feature fitting
procedure, and Appendix~\ref{appendix2} quality assurance tests
including a comparison with high-resolution IRS data from
\citet{lebouteiller08}.  
In \S\ref{maps} we describe the general results -- the spatial
distribution of emission lines and their ratios.  \S\ref{density}
describes the derived distribution of matter -- electron density and
dust, and \S\ref{excitation} the distribution of radiation evident in
the excitation of the gas.  We compare the data with photoionization
and shock models, and summarize and discuss implications in
\S\ref{conclusions}.


\section{Observations}\label{observations}

The {\it Spitzer}'s Infrared Spectrograph has $\gtrsim$100 times greater
spectroscopic sensitivity than the previous premier infrared
observatory, {\it ISO-SWS}, and $\gtrsim$10 times higher spatial
resolution than the SWS aperture size.
We used the four \lr\ modules of the IRS.  The spectral resolution
ranges from 60 to 120 \citep{som}\footnote{Available at:
  http://ssc.spitzer.caltech.edu/documents/SOM/}, with reliable
wavelength coverage as noted in 
section \ref{finaladjustments}.
The four modules are the short
wavelength / \lr\ second order (SL2), short wavelength / \lr\ first
order (SL1), long wavelength / \lr\ second order (LL2), and long
wavelength / \lr\ first order (LL1).  Our data is divided into eighteen
Astronomical Observing Requests (AORs), each containing many BCD
(basic calibrated data) frames, which are pipeline-processed images of
the IRS chip.

\begin{figure}
\plotone{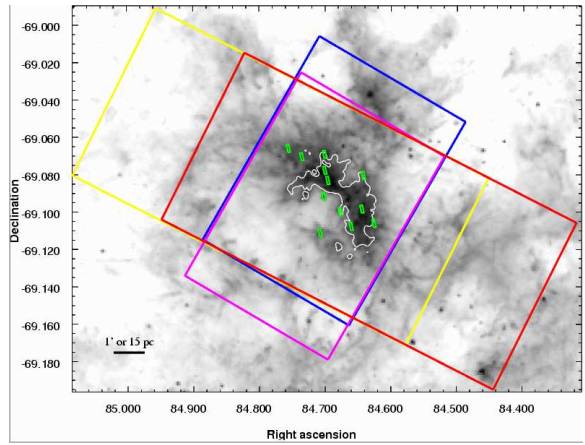}
\caption{\label{irscoverage}IRAC 8\um\ image with the scope of coverage
  in each of the IRS modules shown.  Red: LL1.  Yellow: LL2.  Blue:
  SL1.  Magenta: SL2.  Green: \hr\ GTO apertures (see
  Appendix~\ref{qasec}).}
\end{figure}

We observed \dor\ over five days in September of 2006, using {\it
Spitzer's} spectral mapping mode in order to obtain detailed spatial
information.  The total amount of time on target was 74 hours.  We
used 3440 slit pointings, covering 40.5 square arcminutes in the
short-wavelength modules and 69.1 square arcminutes in the
long-wavelength modules (see Figure \ref{irscoverage}).  Each slit
overlaps half of the preceding one, and each row of slits overlaps
half of the preceding row, so every point on the map was observed at
four separate slit pointings.  There were three repetitions per
pointing for the SL observations, at 14 seconds each (for a total time
of 168 s for each point on the SL maps), and four repetitions per
pointing for the LL observations, at 6 seconds each (for a total time
of 96 s for each point on the LL maps).  The result is four large and
dense spectral cubes, one for each module.

We also made separate observations of a nearby region of comparatively
blank sky, bracketing our target observations in time, to comprise
background observations and to better characterize rogue pixel
response.  The total time on background was 44 minutes.

\subsection{Reduction of IRS data cube}\label{datared}

We used the basic calibrated data (BCD) from the Spitzer Science
Center pipeline version S14.4.0.  The main tool for the assembly and
reduction of our data cube was CUBISM \citep{smith07cubism}, which is
designed for spatially dense IRS maps.  The functions of CUBISM include
tools to reduce the data (subtracting background, applying a slit loss
correction function, trimming the slit in the spatial direction,
masking pixels that are flagged by the pipeline, and applying
algorithms to identify rogue pixels) and to extract spectra and maps
from the data cube.

Appendix~\ref{appendix1} describes in detail our procedure for
calibrating and adjusting the flux levels such that the different
spectrograph modules could be combined into smooth spectra.  In brief,
we used contemporaneous off-source background observations to subtract
the thermal background. We used measurements of the slit-loss
correction function provided by the {\it Spitzer} Science Center to
undo that part of their point-source-based calibration (since our
observations are more similar to a uniform diffuse source than a point
source).  Finally, we adjusted spectra from different modules to match
each other.

Appendix~\ref{appendix2} describes our comparisons of high and
low-resolution spectra in \dor\ to test the fidelity of our fitting
process.  We find that the strengths of even somewhat blended lines
are recovered well in the low-resolution spectra.


\subsection{Artifacts in IRS data cube}
\label{artifacts}

There are two major artifacts which appear in a 2-D map of our data
cube at any given wavelength.  In all four modules, there are numerous
faint stripes, one pixel wide, which cross the map in the direction
that the slit scans.  In the two SL modules, there is a broad
intensity discontinuity and associated bright region in one area of
the map.  The affected region varies with wavelength.
The faint stripes appear to be mainly caused by hot pixels on the IRS
chip.  As the slit was stepped across \dor, each hot pixel was
``dragged'' across the map, creating a bright stripe.  They can be
eliminated via two cleaning methods: wavsamp trimming and rogue pixel
masking.  The intensity discontinuity is caused by saturation in the
peak-up image, and must be corrected by fitting a correction factor to
the regions of the chip between the SL modules.

The Spitzer Science Center has defined a polygon, called the wavsamp,
defining the active area to extract from the IRS chip for each module.
In the case of \dor, the default wavsamp tends to be too generous in
the spatial direction, including a few pixels where the spectral
response is reduced.  By trimming it, we were able to eliminate some
of the faint stripe artifacts.

It is possible to determine which pixels on the IRS chip contribute to
a given point in the data cube, using a CUBISM tool called
backtracking.  Backtracking from the artifacts in the maps
demonstrated that most, if not all, of the remaining faint stripes are
caused by individual rogue pixels, whether hot or cold.  In addition
to masking the pixels automatically flagged by the Spitzer Science
Center pipeline, we employed CUBISM's automatic rogue pixel masking
algorithms at both the global and record levels.  At the global level,
we masked any pixel which deviated by at least 2.5 sigma from the
median pixel level in at least 35\% of the records.  At the record
level, it was necessary to be much more conservative to avoid masking
out real spectral features.  We masked any pixel which deviated by at
least 7.5 sigma in 70\% of its occurrences in the cube.  The procedure
missed some pixels which were clearly hot or cold, so we also manually
marked a set of global and record level rogue pixels.  In the
resulting cubes, the faint stripes were greatly reduced.

\begin{figure}
\plotone{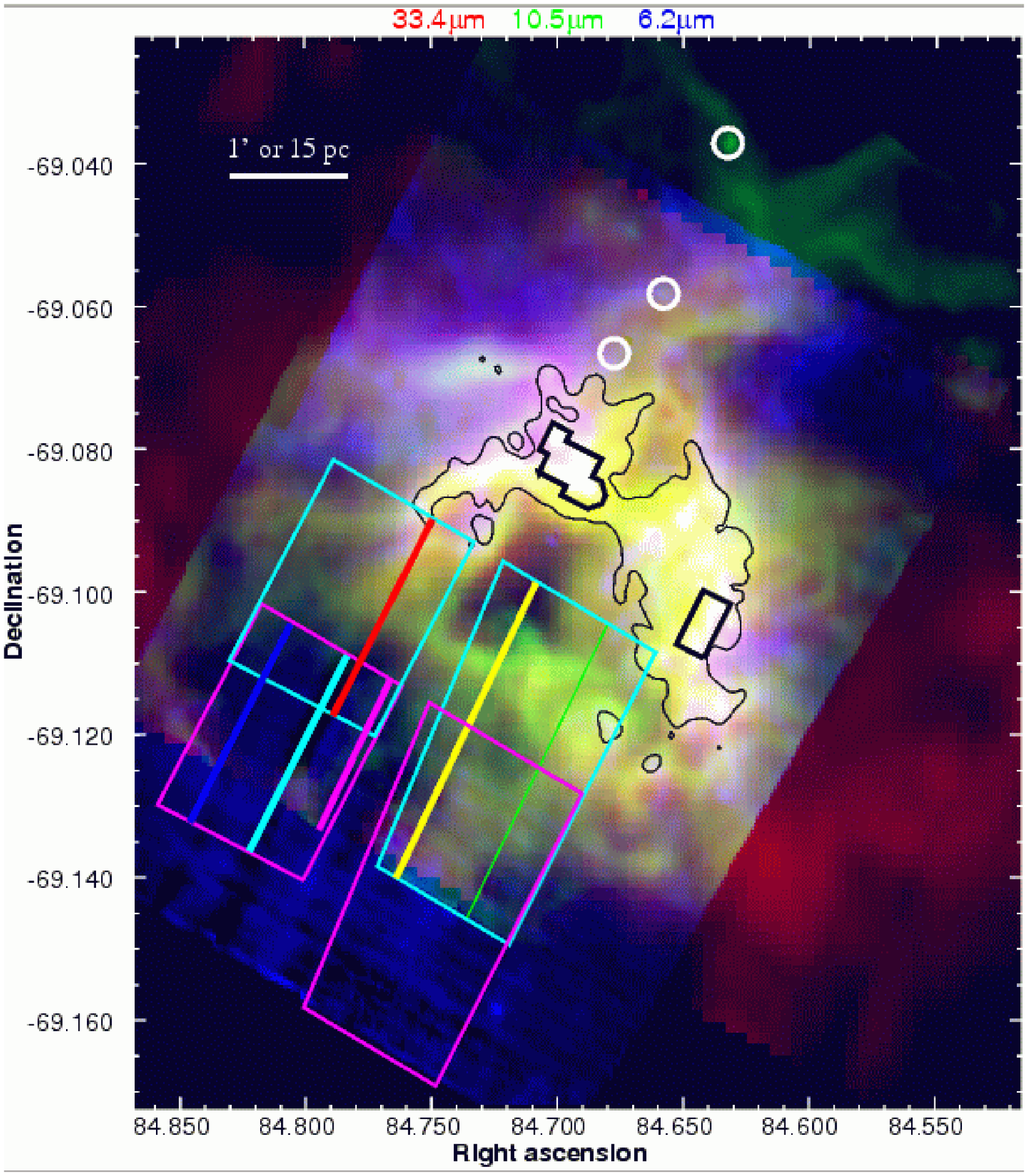}
\caption{\label{saturationmap}Three-color log-scale image generated
  from the IRS cube (see \S\ref{datared}) with regions of interest
  overlaid.  Black contour: a single radio contour to guide the eye
  (see Figure \ref{bwoverview}).  The emission line maps are 
  continuum-subtracted, and each is plotted at a scale that brings 
  out detail.  Red: \siiill, in the
  LL1 module.  Green: \sivl, in the SL1 module.  Blue: PAH at 6.2\um,
  in the SL2 module.  The areas outlined in bold black are vulnerable to
  falling response (saturation) and strong fringing in the LL1 module
  (see Appendix~\ref{qasec}).  Continuum fits in these areas should be
  regarded with caution, as should the fitted line strengths of
  long-wave emission lines.  The sources marked in white
  saturated the peak-up camera and caused dark stripes and other
  artifacts in the SL modules elsewhere in the map. The cyan regions
  represent the total affected area for the SL1 module, while the
  magenta regions represent the total affected area for the SL2
  module.  However, the main effect of the saturation is a stripe
  across the map that changes with wavelength.  There are generally no 
  severe effects in the rest of the map for a given wavelength.  The 
  approximate location of this stripe is marked by a series of bold 
  lines, for several wavelengths corresponding to major emission lines 
  in our spectra.  Blue: \ariil.  Cyan: PAH at 6.2\um.  Magenta: \hugl.  
  Green: \ariiil.  Yellow: \sivl.  Red: \neiil.  See \S\ref{artifacts} 
  for more detailed discussion.}
\end{figure}

The other major artifact, the intensity discontinuity, is caused by
saturated sources on the peak-up image (PUI), primarily by bright
point sources in the northern part of the region which are marked in
white in Figure \ref{saturationmap}.  The northernmost of these is
the source marked in yellow in Figure \ref{threecoloroverview}.  By 
examining the records
associated with the affected region on the map, we can see that where
a source saturates several neighboring pixels in the peak-up image,
there is a bleed-over effect that reduces the response of the rest of
the IRS chip (BCD image) in those rows, and may increase the response
in neighboring rows.  The result is broad, uneven dark or bright
stripes in a 2D map extracted from the cube.  The location varies with
wavelength because as the PUI is scanned across the bright source in
space, the artifact changes position on the BCD frame, i.e. changes
wavelength.  Simply masking all affected pixels at the record level
results in holes in the maps at many wavelengths.  Instead, we chose
to mend the response.  We selected all rows that intersected a group
of saturated pixels on the peak-up image, with a margin of two rows on
either side.  These are the affected rows.  We found the continuum
level in adjacent rows, excluding major emission lines, and linearly
interpolated across the affected rows.  In the case of affected rows
that corresponded to continuum emission, we replaced the values with
the interpolated values.  In the case of affected rows that
corresponded to atomic or PAH emission, we attempted to restore the
base continuum level below the emission line by adding the
interpolated value, and subtracting out the median value of the
affected rows in the region between the SL1 and SL2 chips, which is
the bias level of the affected rows.  The artifact is reduced, though
not eliminated.  The result is a complete cube, but with localized regions 
that have larger uncertainty in the SL1 and SL2 modules (see the cyan and 
magenta regions marked in Figure \ref{saturationmap}).  For the purposes of 
presentation, the areas where the artifact's effect is still distinct have 
been cropped out of each map.

At some wavelengths in the SL cubes, a large region of somewhat
elevated emission is still visible after our corrections.  This
artifact appears to be another result of the saturation in the peak-up
images.  After a saturation event, the response of the chip is
elevated for the remainder of the AOR.  The resulting artifact, which can
be seen in maps at a variety of wavelengths (for example, 5.2, 6.3 and 
10.7\um), is localized in space.  This low-level artifact is also generally 
reduced in continuum-subtracted images, because the effect is broad on the 
BCD frame, and thus broad in wavelength as well.

\subsection{Spectral Features and Line Fitting}\label{fitting}

Hot dust emission is responsible for the overall shape of the
continuum in \dor\ and its strong rise toward the red.  Using the
fitting package PAHFIT \citep{smith07pahfit}, we fit the continuum
with an assortment of four thermal functions from 40 to 300K and a
stellar blackbody of 5000K (the fit is insensitive to exact
temperature as long as it peaks blueward of our shortest wavelength
5\um).  The integrated spectrum of the entire region is shown below in
Figure~\ref{entirespec}, and astrophysically interesting spectra are
described below in \S\ref{sources_text}.

The broad PAH bands and unresolved atomic lines are blended in some
parts of the spectrum.  In order to decompose them, we fit the whole
suite of emission lines at once.  Detailed analysis of the dust
features will be addressed in a future work.
The feature of greatest potential concern is blending of \neiil\ with
the 12.7\um\ PAH feature.  The PAH contribution is much lower than the
atomic line over much of \dor.
In particular, \citet{sturm00} found that at the higher spectral
resolution of {\it ISO-SWS}, the 12.7\um\ PAH feature was undetectable
relative to \neiil\ within the two {\it ISO-SWS} apertures (located on
the ridge) -- the lowest ratio of PAH to \neii\ of any \hii\ region
they studied.  We used comparisons with high-resolution spectra (see
Appendix~\ref{qasec}) and tests in parts of \dor\ that should be
completely PAH-free, like the core of R136, to determine that the
joint fitting of those two features is robust in \dor\ and
uncertainties are properly accounted for.


\subsection{MIPS SED cube}

\begin{figure}
\epsscale{1.05}
\plotone{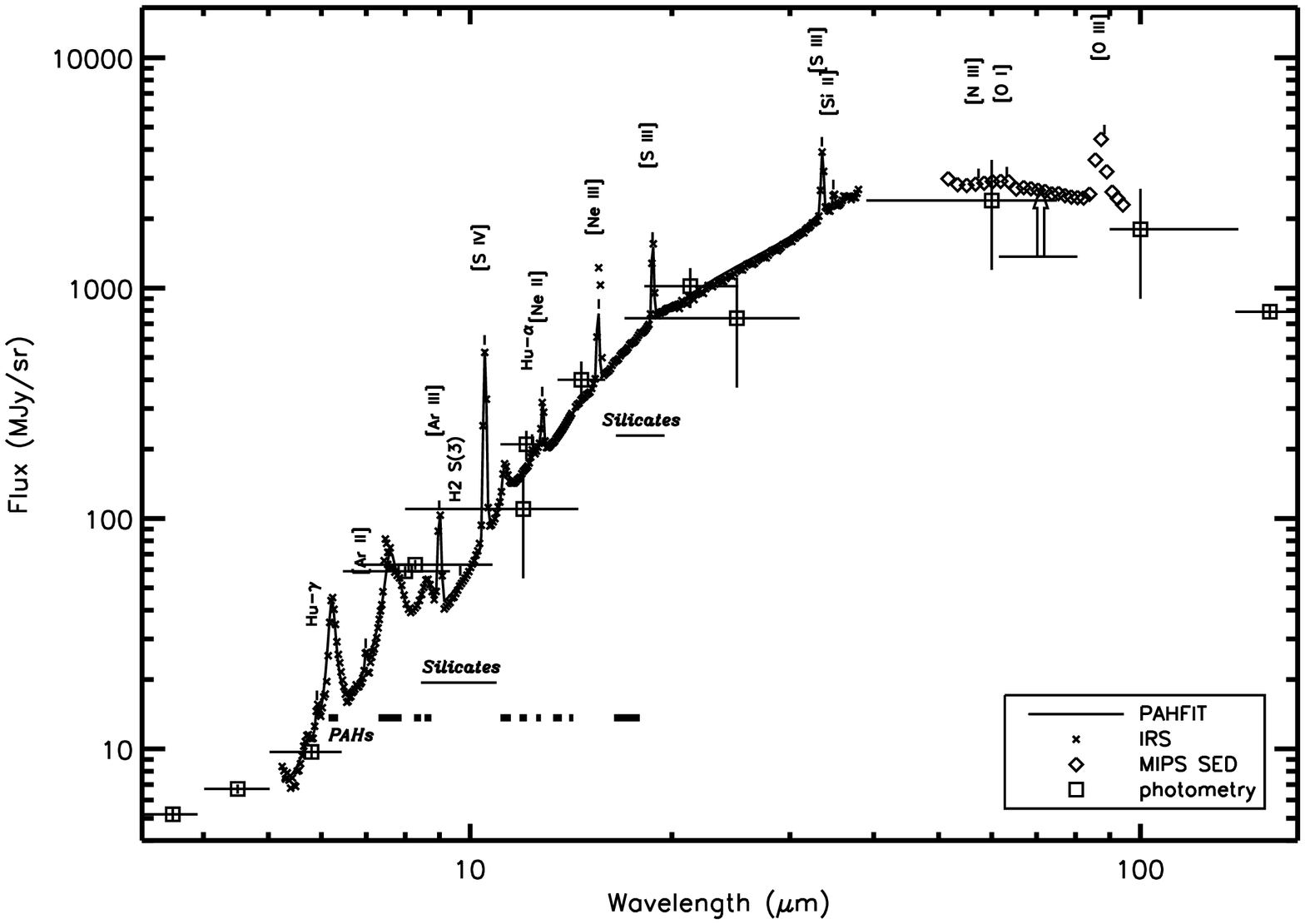}
\caption{\label{entirespec} The spectrum of the entire \dor\ nebula (a
  square about 4' or 60 pc on a side centered on
  \radec{05}{38}{44}{-69}{05}{28}).  The observed spectra from IRS and
  MIPS/SED, and infrared photometric points, are plotted with symbols,
  while the fit we made to the IRS data is displayed with a solid
  line.}
\end{figure}

We also made contemporaneous observations\footnote{AOR keys 18633728
  and 18634240} of \dor\ using the SED mode of the MIPS
\citep{rieke_mips,lu_sed}.  The raw data are responsivity-corrected,
dark-subtracted, and illumination-corrected using the MIPS germanium
pipeline \citep{gordon_ge}, with the difference that the illumination
correction is derived from Galactic cirrus rather than zodiacal light.
Wavelength correction is derived from observations of bright planetary
nebula.  Since \dor\ is so bright, we use the chopped ``background''
area simply as additional on-source observation, and assume that the
thermal background is much less intense than the emission from the
nebula itself.
As with the IRS data, mapping the area with offsets equal to half of
the slit width, perpendicular to the slit direction, yields a fully
sampled spectral cube, with resolution
$\lambda/\delta\lambda\simeq$25.  Even with such poor resolution, we
detect [\ion{N}{3}]$\lambda$57.3\um\ with high significance over most
of the map, and tentatively [\ion{O}{1}]$\lambda$63.1\um\ in the very
center (\S\ref{emission}).  The shape of the dust continuum will be
discussed in a future publication on the dust content of \dor, but we
note that the SED peaks shortward of 75\um, consistent with very high
average dust temperatures (Figure~\ref{entirespec}).

\section{Results}
\label{results}

Figure~\ref{entirespec} shows the integrated spectrum of \dor, with
dust and atomic features labeled.  The overall continuum shape is
typical of hot small dust grains in \hii\ regions.  The MIPS/SED data
clearly show that the spectral energy distribution of the entire
nebula peaks shortward of 70$\mu$m, indicating a quite warm average
dust temperature.  A full discussion of dust properties will be the
subject of a subsequent publication.  We note that the aromatic
features are well-detected but not extremely strong - comparable to
Galactic compact \hii\ regions or compact blue galaxies, and weaker
than the integrated spectra of entire spiral galaxies or the diffuse
emission of the Milky Way \citep{galliano,peeters}.  The equivalent
width of the 6.2\um\ feature, averaged over this large aperture, is
0.61\um.  In particular, the weak 17$\mu$m feature may indicate a dearth
of large neutral PAHs \citep{smith07pahfit} in this intensely ionized
environment.
The strongest spectral features are the ionic emission lines which are
sensitive to physical conditions in the ionized gas.  

\begin{deluxetable}{lrrrr}
\tabletypesize{\scriptsize}
\tablecolumns{5}
\tablewidth{0pc}
\tablecaption{Ionization Potentials.\label{ip}}
\startdata
\tableline
    & I & II & III & IV \\
\tableline
O  & 13.62 & 35.12 & 54.94 & 77.41 \\
Ne & 21.56 & 40.96  & 63.45  &  97.11 \\
S  & 10.36 & 23.33  & 34.83  & 47.30  \\
Ar & 15.76 & 27.63  & 40.74  & 59.81 \\
\enddata
\tablecomments{Units are eV.}
\end{deluxetable}

In the following sections we describe the spatial distribution of
emission lines and their ratios, and show spectra of notable compact
regions.


\subsection{Spatial distribution of emission}
\label{emission}

\begin{figure}
\epsscale{0.8}
\plottwo{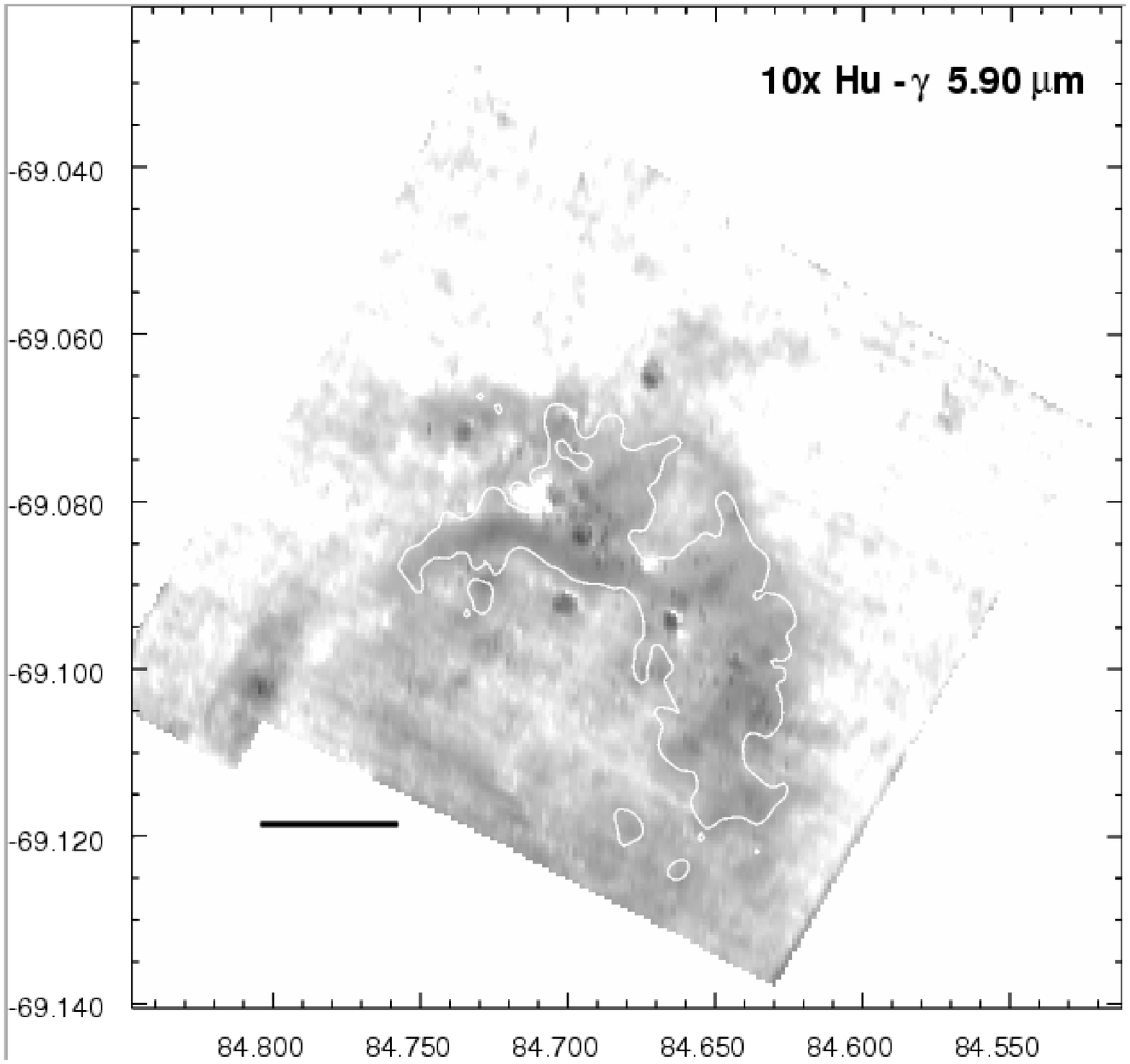}{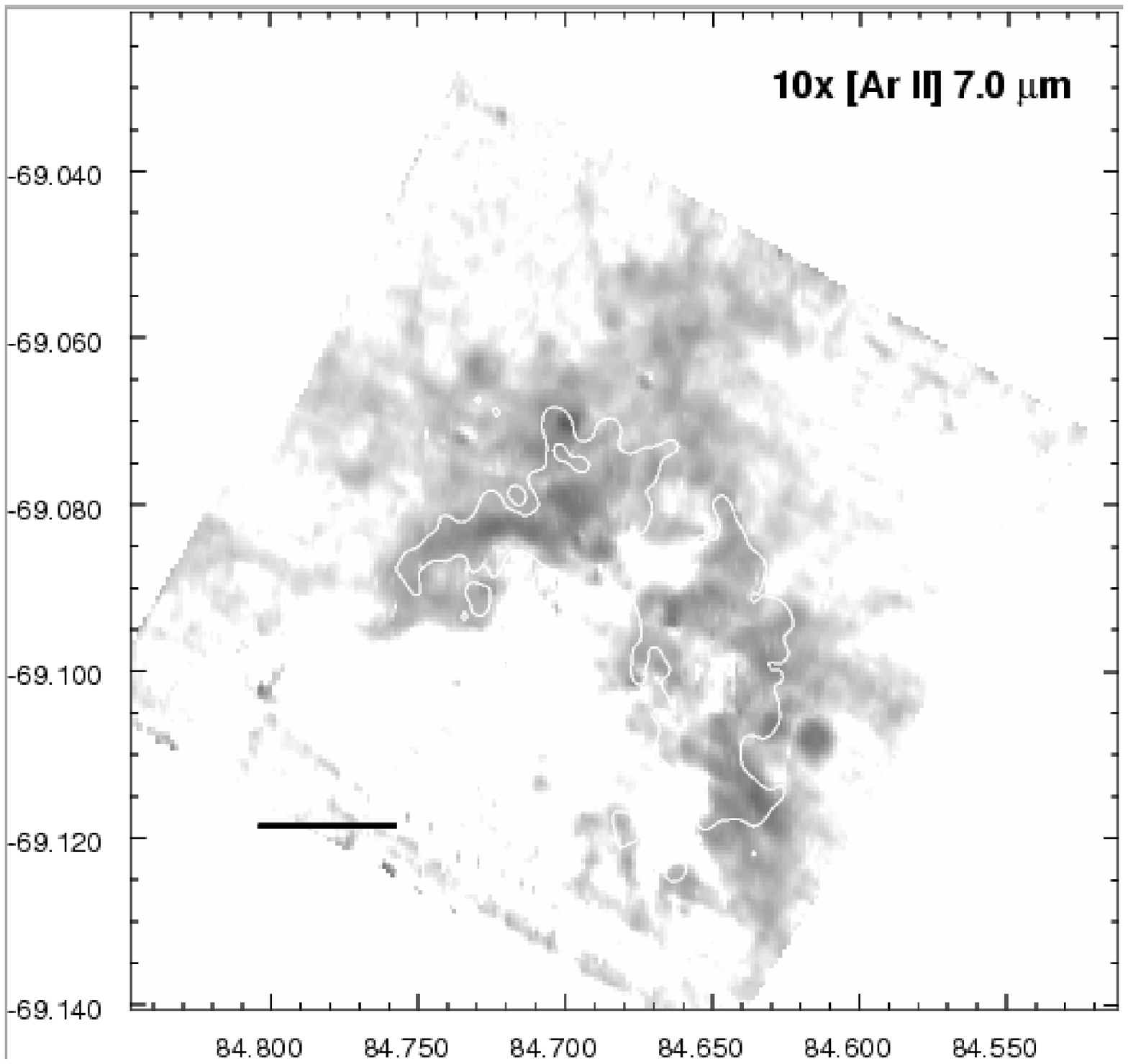}
\caption{\label{atlinemaps1} through Figure~\ref{atlinemaps6}: 
Fitted atomic line maps.  All plots are on the 
same log scale, ranging from $10^{-7}$ to $10^{-5}$ $W m^{-2} sr^{-1}$. 
Note that four of the fainter maps have been multiplied by 10, and the 
brighter \oiii\ map has been divided by 100.  The length of the black bar 
is 1'' or 15 pc.  White contours: a single level of 3cm radio 
emission, to guide the eye (see Figure \ref{bwoverview}).}
\end{figure}

\begin{figure}
\plottwo{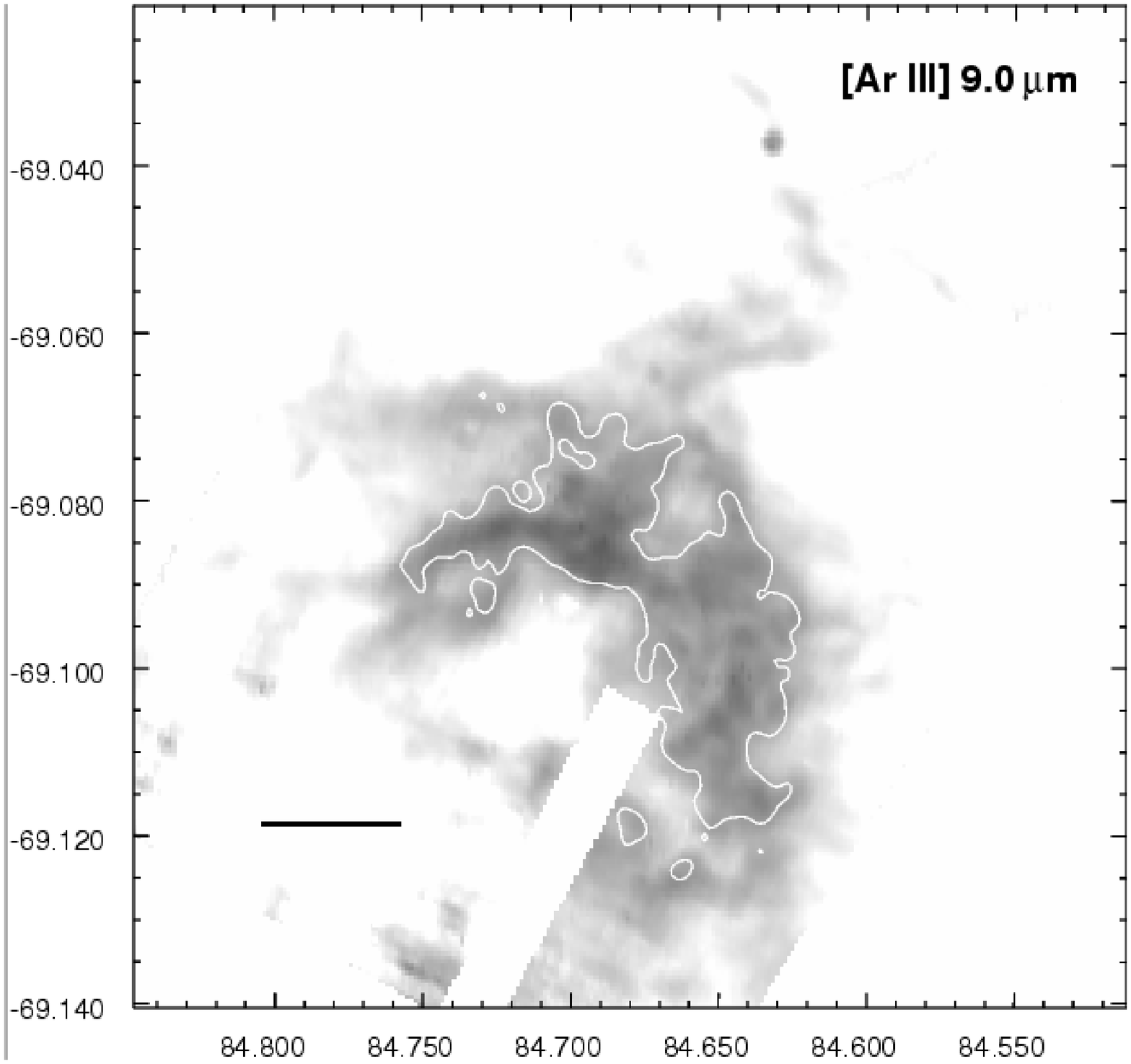}{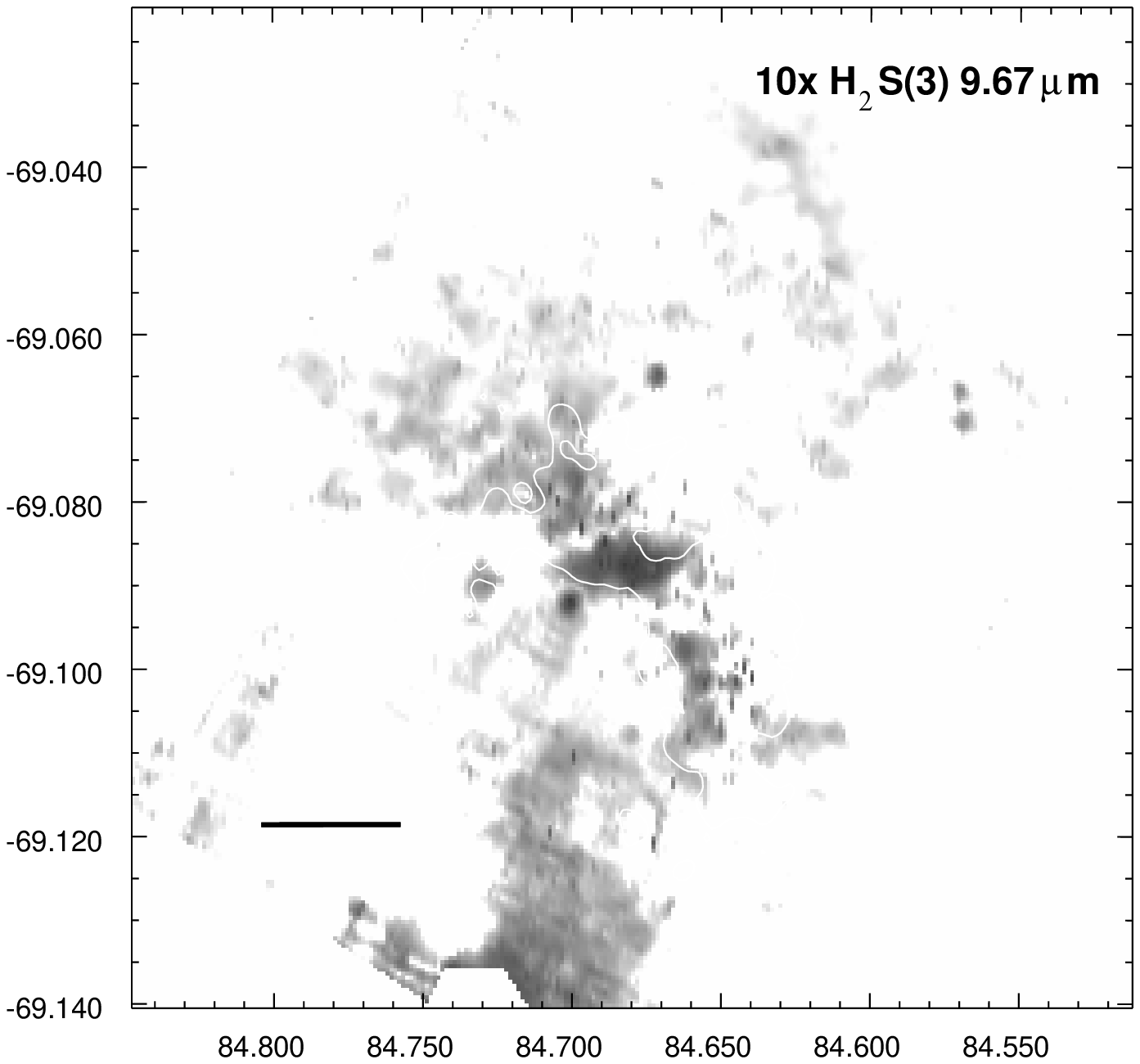}
\caption{\label{atlinemaps2}Fitted line maps (see caption, Figure 
\ref{atlinemaps1}).}
\end{figure}

\begin{figure}
\plottwo{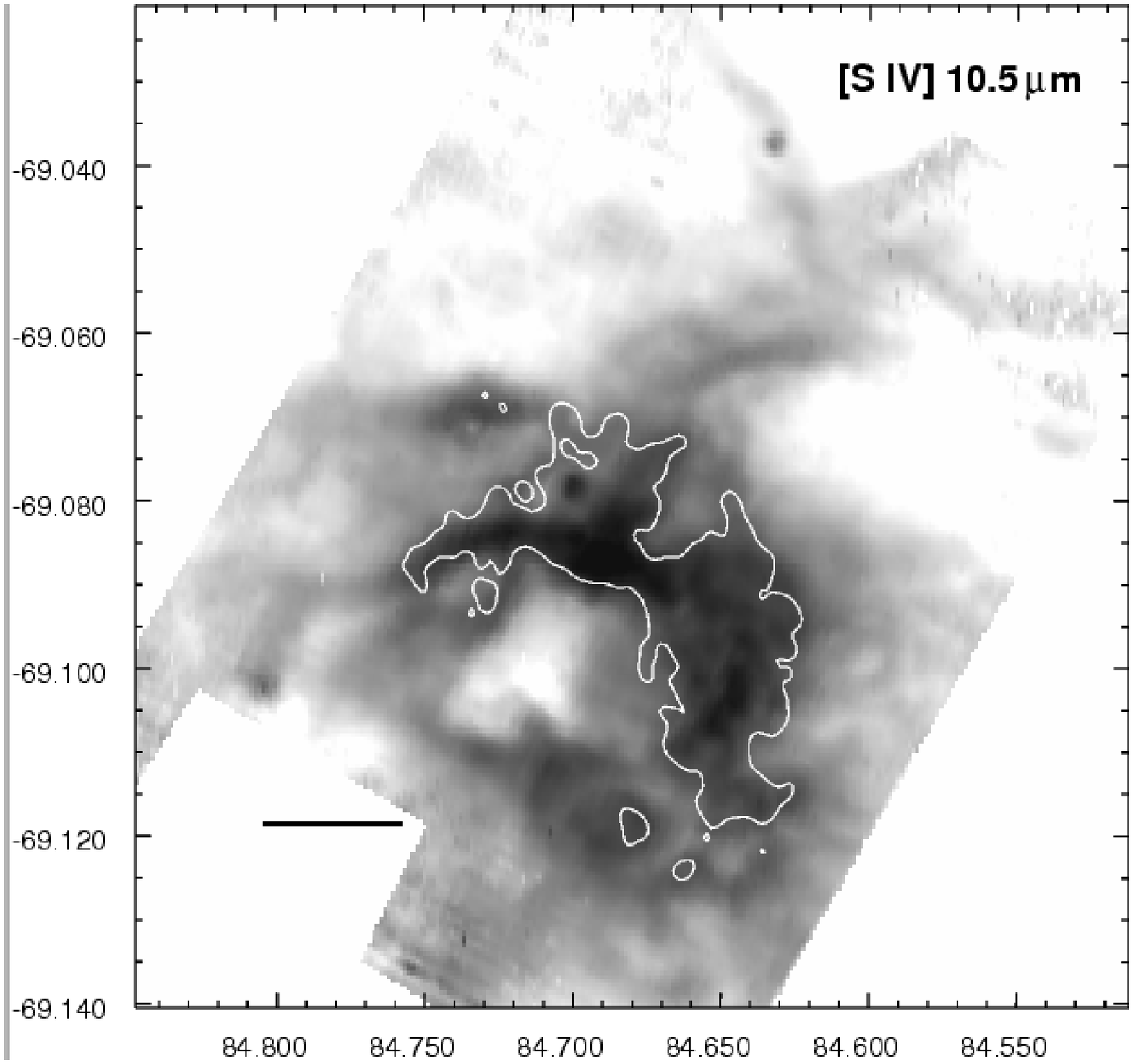}{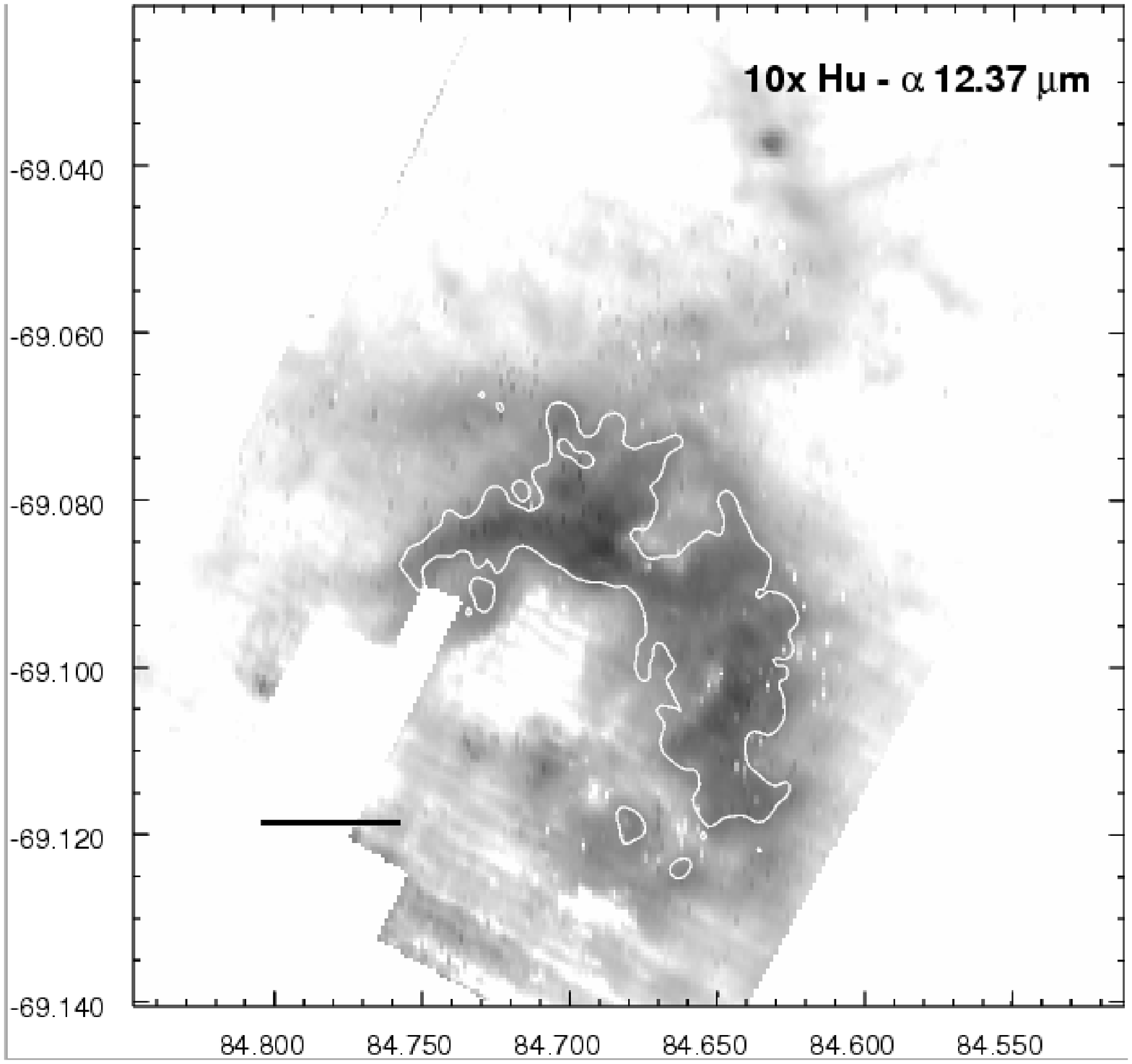}
\caption{\label{atlinemaps3}Fitted atomic line maps (see caption, Figure 
\ref{atlinemaps1}).}
\end{figure}

\begin{figure}
\plottwo{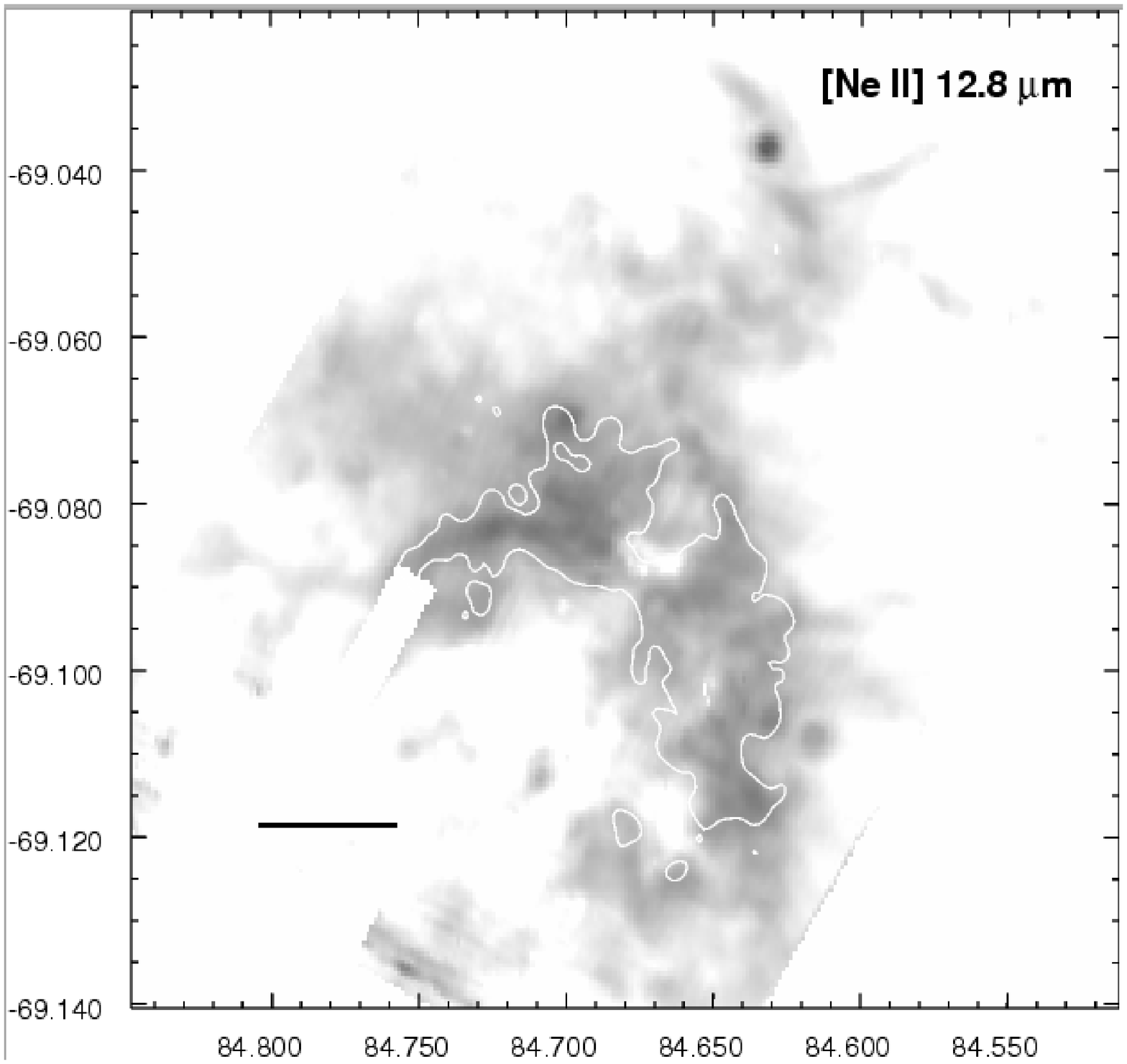}{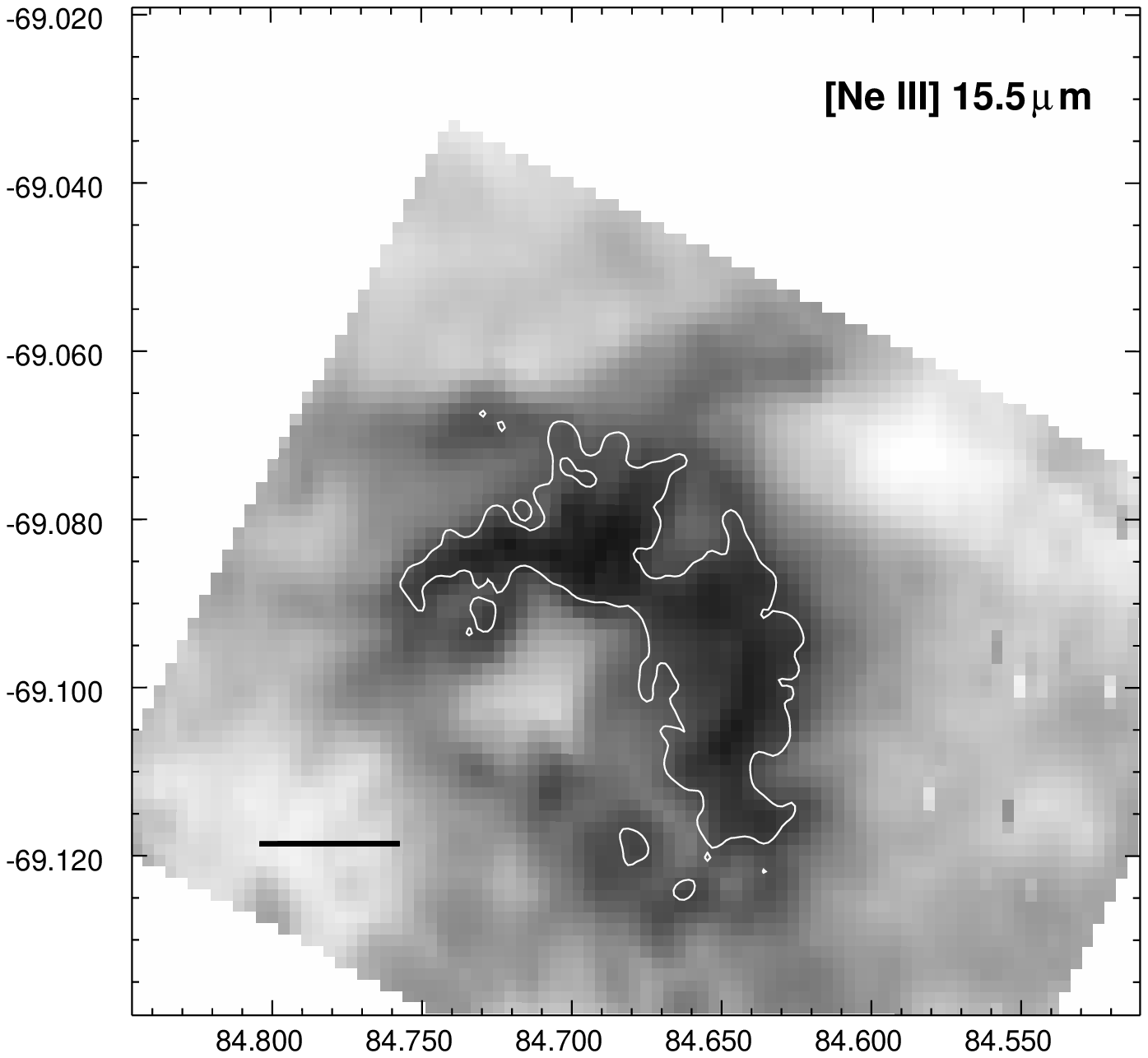}
\caption{\label{atlinemaps4}Fitted atomic line maps (see caption, Figure 
\ref{atlinemaps1}).}
\end{figure}

\begin{figure}
\plottwo{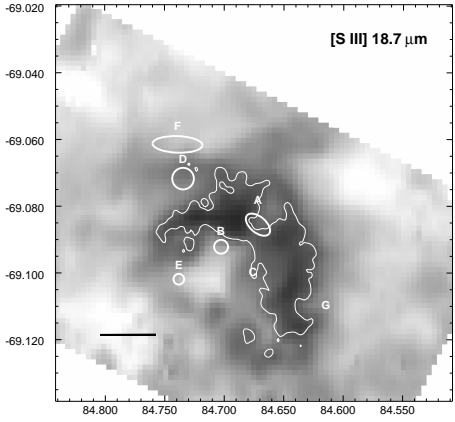}{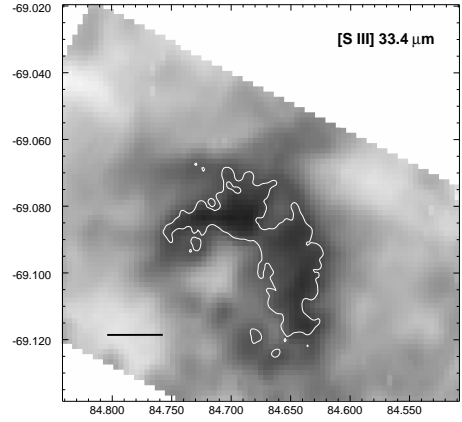}
\caption{\label{atlinemaps5}Fitted atomic line maps (see caption, Figure 
\ref{atlinemaps1}).  The \siiil\ map is marked with regions of interest listed 
in Table~\ref{sources_table}.}
\end{figure}

\begin{figure}
\plottwo{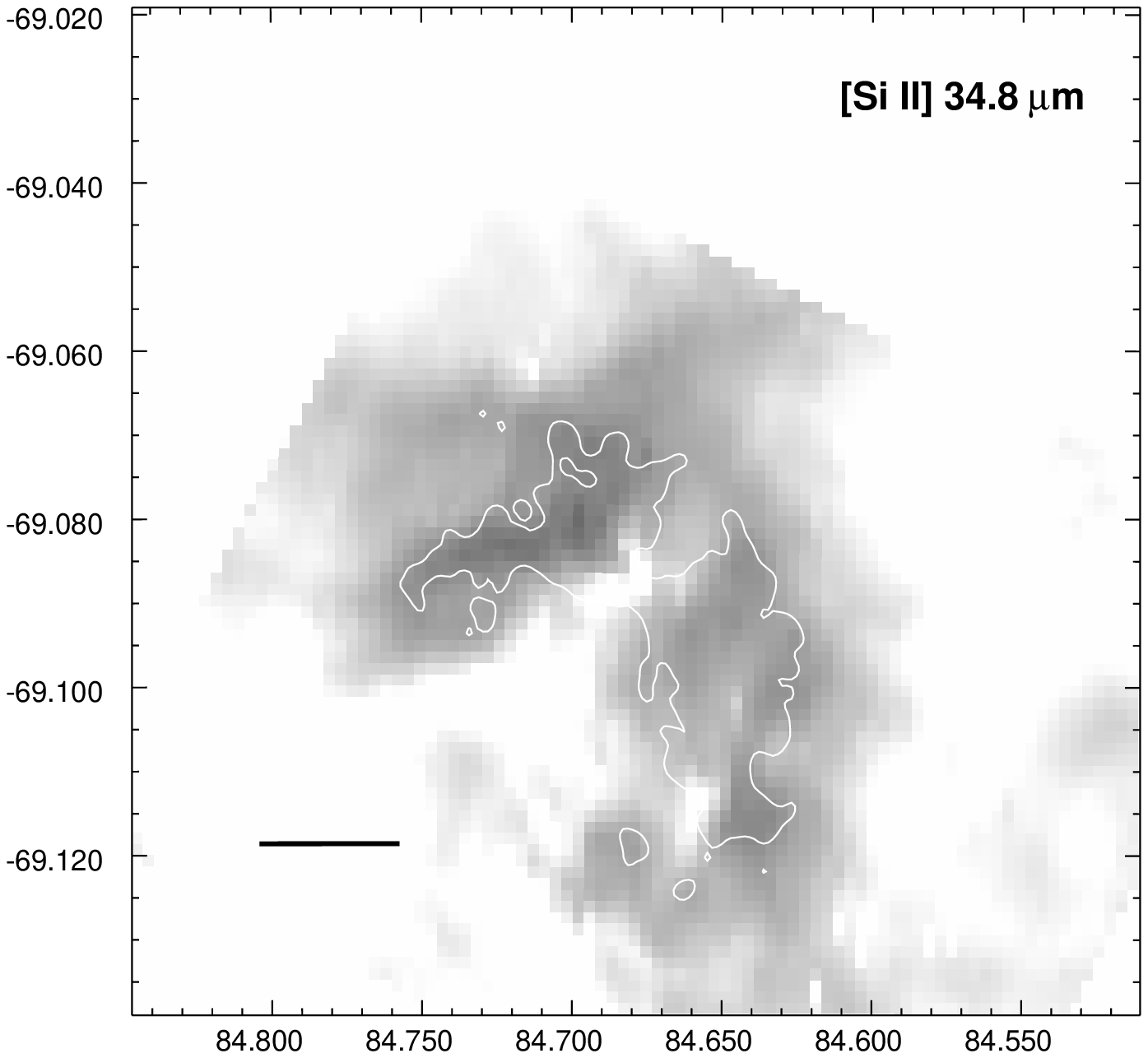}{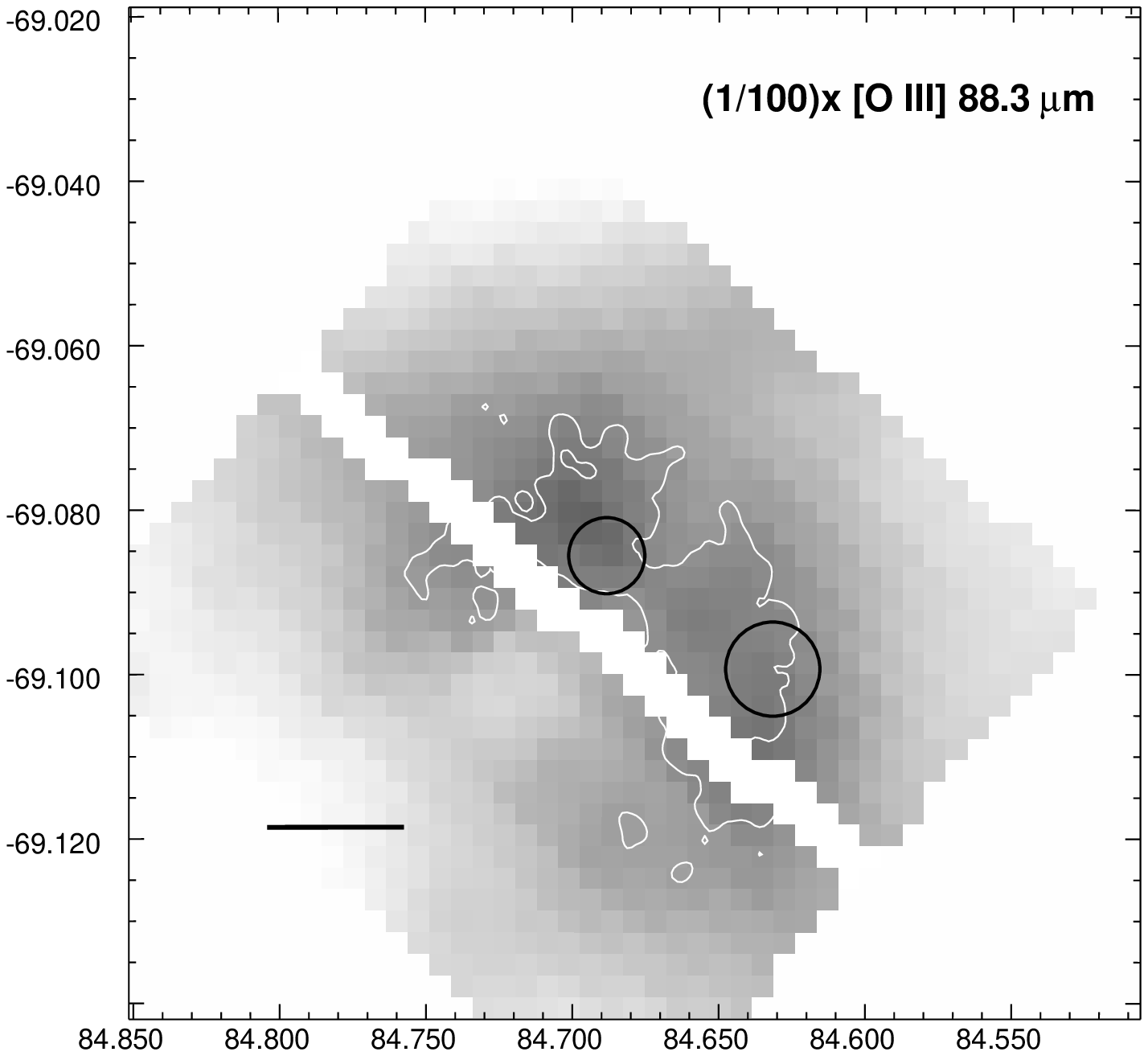}
\caption{\label{atlinemaps6}Fitted atomic line maps (see caption, Figure 
\ref{atlinemaps1}).  The black circles on the \oiiil\ map
mark the detected peaks of \oil\ and \niiil.}
\end{figure}

\begin{figure}
\plottwo{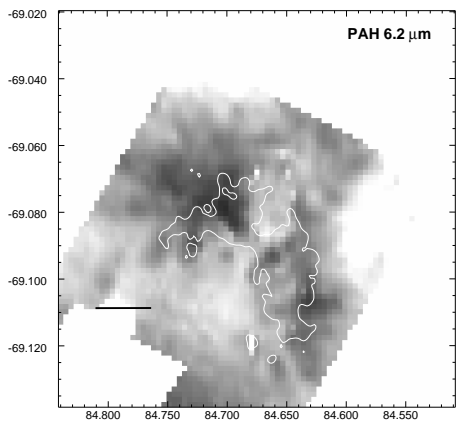}{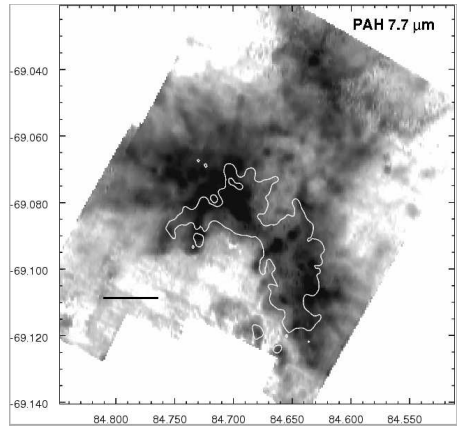}
\caption{\label{pahlinemaps1} through Figure~\ref{pahlinemaps3}: Fitted PAH feature maps.  For details, see caption, Figure \ref{atlinemaps1}.}
\end{figure}

\begin{figure}
\plottwo{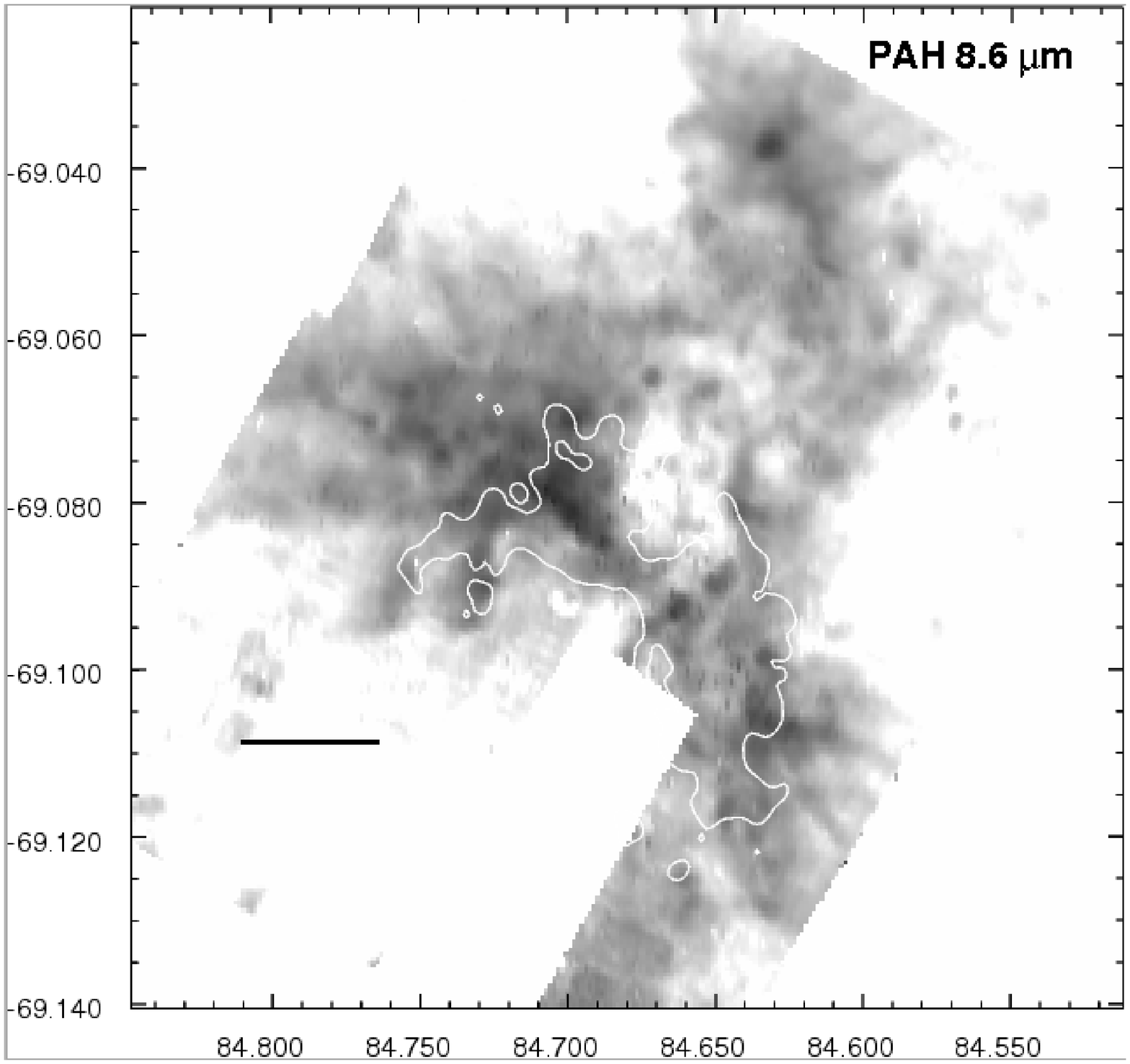}{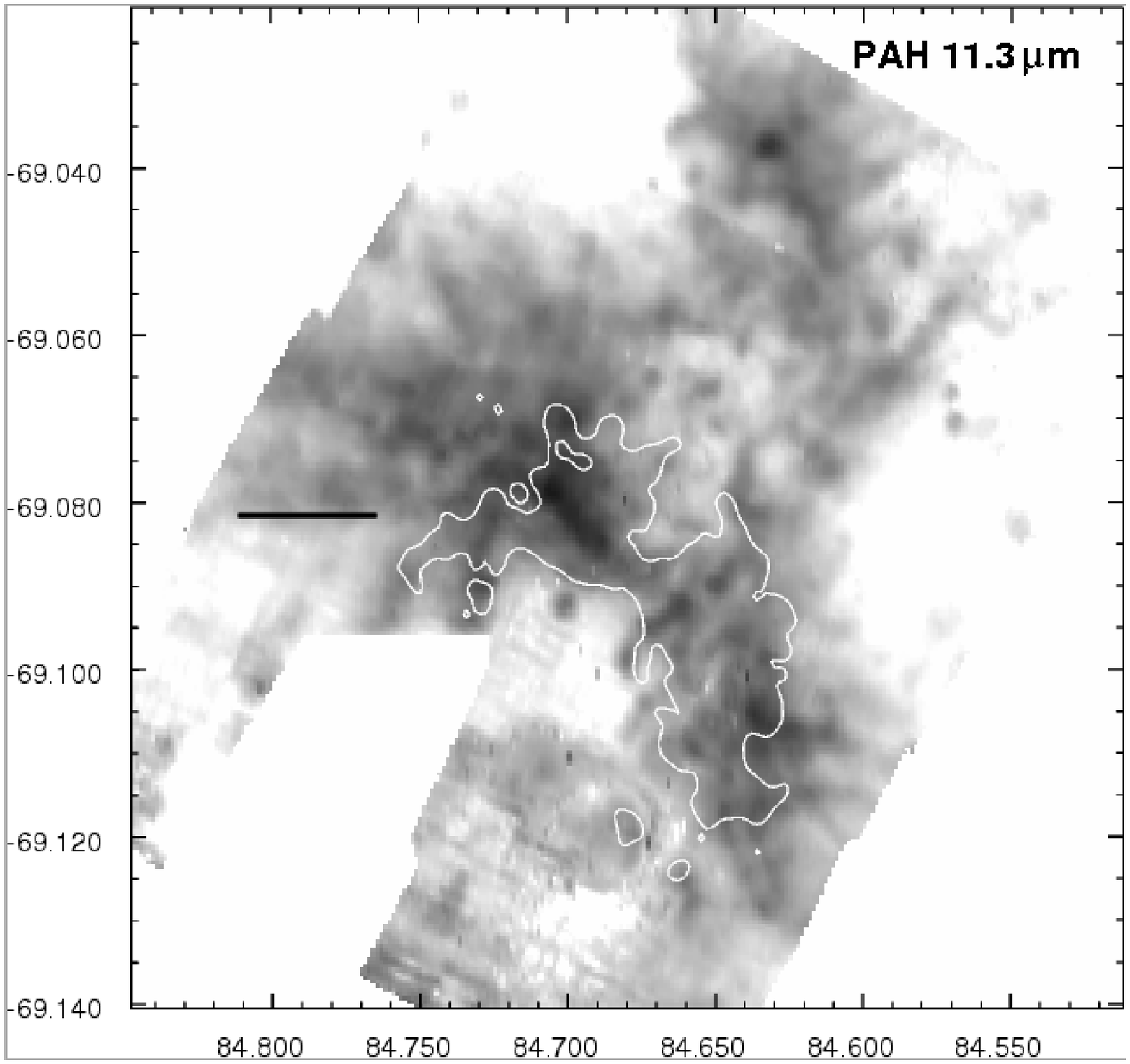}
\caption{\label{pahlinemaps2}Fitted PAH feature maps (see caption, Figure \ref{atlinemaps1}).}
\end{figure}

\begin{figure}
\plottwo{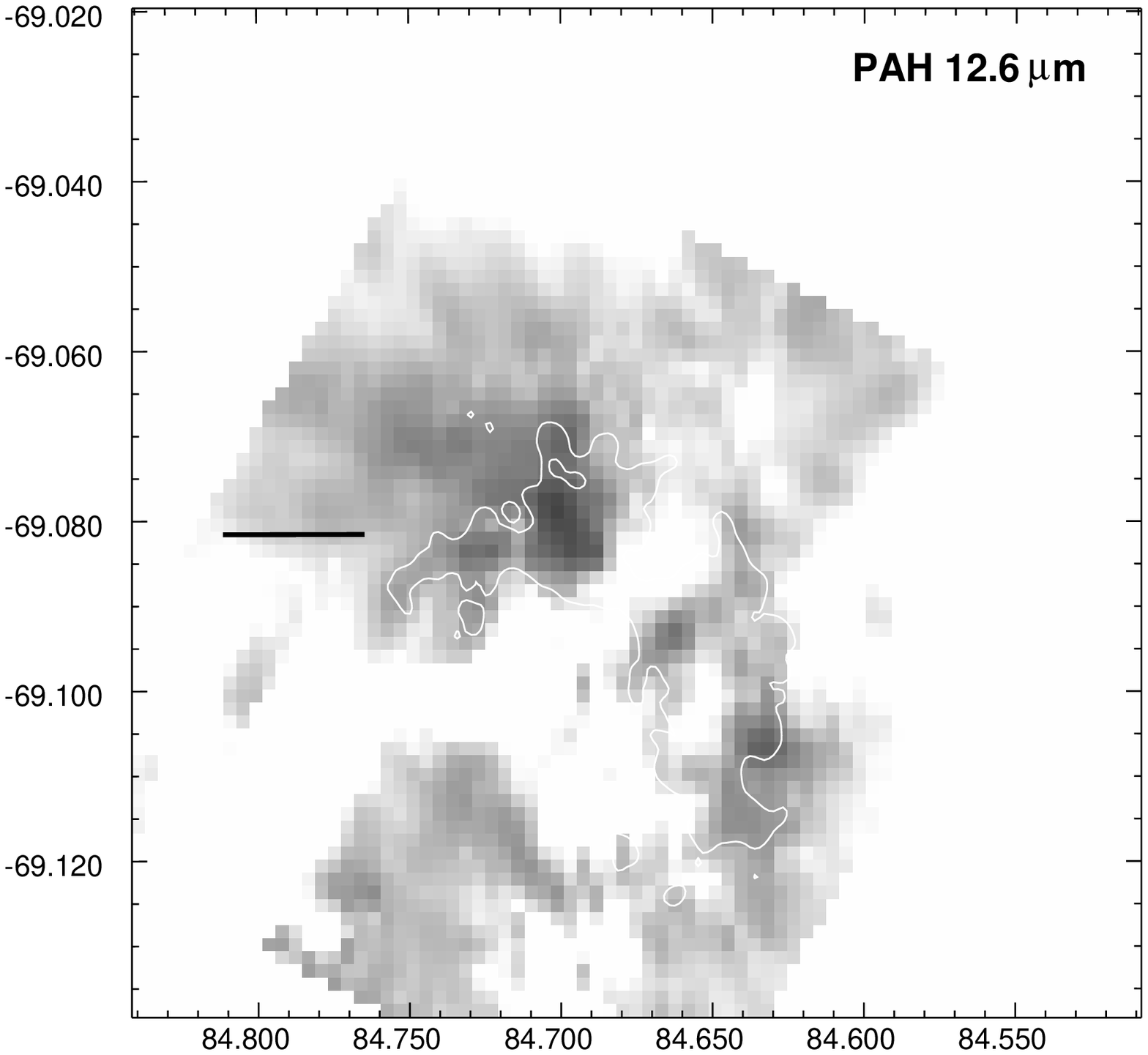}{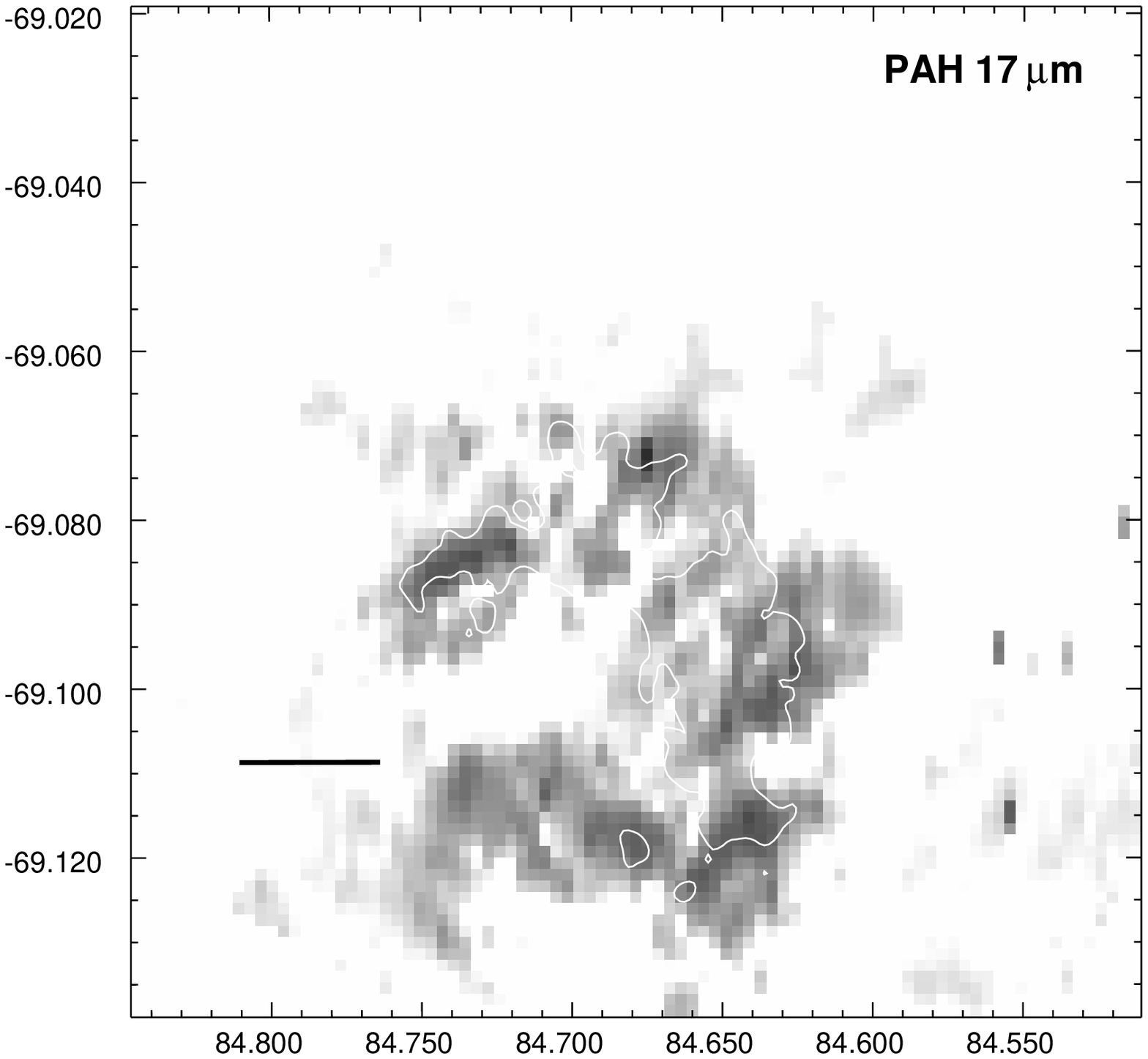}
\caption{\label{pahlinemaps3}Fitted PAH feature maps (see caption, Figure \ref{atlinemaps1}).}
\end{figure}

Figures~\ref{atlinemaps1}-\ref{pahlinemaps3} show maps of ionic
line emission in \dor.  To first order, line emission follows the
diffuse emission pattern seen at other wavelengths, of a broad arc or
``ridge'' surrounding an evacuated hole.  The overall excitation of
the region is immediately clear from the distribution of \arii\ and
\ariii\ -- there is no detected \arii\ in the evacuated hole.  With an
ionization potential of 15.7eV, the photoionization cross section and
charge transfer with hydrogen ensures very little Ar$^+$ within an
\hii\ region \citep[e.g.][]{sofia98}.  (Ionization potentials are
listed in Table~\ref{ip} for reference.)  We note that the Ar$^{++}$
recombination rate is highly uncertain, so absolute calibration of
\ariii/\arii\ has large systematic uncertainty
\citep{morisset04,stasinska}.
The similarity of overall morphology amongst the Ar, Ne, and S lines
suggests that any variation of excitation across the nebula will be
correlated with the structure seen in continuum from radio to
infrared, although as we will explore thoroughly below, the shape of
the hole varies somewhat, and there are some distinct locations of
high excitation.  The [\ion{Si}{2}] distribution appears somewhat
displaced away from R136 and the cavity, as would be expected since
the line is strong in PDRs as well as ionized gas.

Lower ionization species have tentative detections in a few places,
but never very strongly or widely distributed. Of particular interest
is a peak near \radec{5}{38}{44.9}{-69}{05}{13} (source \linepeak,
\S\ref{sources_text}), at which we also detect [\ion{N}{3}]$\lambda$57.3\um\  
and [\ion{O}{1}]$\lambda$63.1\um\ in MIPS SED observations (see the 
northeastern region marked in Figure \ref{atlinemaps6}).  The spectra in 
this part of the map are affected by fringing in the LL1 module ($>$20\um), 
so a \feiil\ detection in the same region is considered tentative.
The ratios of \neiii/\neii\ and \siv/\siii\ (and the MIR continuum) are
high in this general area, but this region in the center of the ridge
is complex, and the higher ionization species do not peak in exactly
the same location as these lower ionization species.  In fact, the
peak of low-ionization species falls between two peaks of the
\neiii/\neii\ map.

Molecular hydrogen emission is detected in the central molecular ridge
and marginally detected in several other locations.  Only the S(3)
9.67$\mu$m line is reliably detected over much of the map.  Tentative
detections in our low-resolution spectra of the S(0) 28\um\ line can
be ruled out by examining the handful of high-resolution spectra in
\dor.  Detection of the S(2) 12.28\um\ line independently of the
neighboring \hual\ line is not very reliable with low-resolution
spectra as described in more detail below, and from the
high-resolution spectra we find that the molecular line is weaker than
the atomic line everywhere, and less than 15\% over most of the
central part of the map.

Higher excitation species [\ion{O}{4}] and [\ion{Ne}{5}], which have
strong lines in the IRS bandpass, are not detected in our maps.  As
discussed in \S\ref{shock}, this favors photoionization over
shocks as the dominant physical process in the region.

\onecolumn
\label{maps}

\begin{figure}
\includegraphics[scale=\ratioscl]{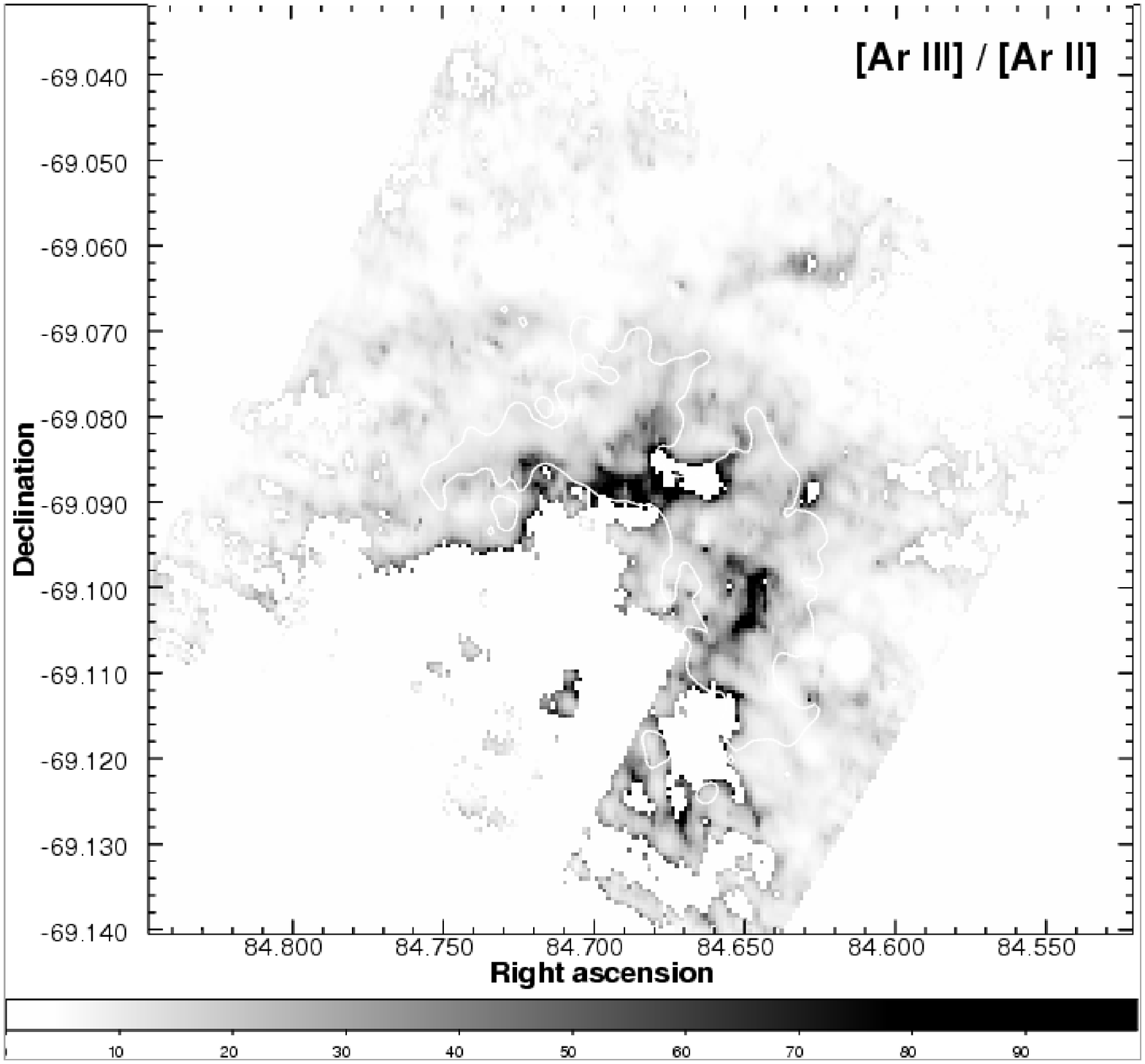}
\includegraphics[scale=\ratioscl]{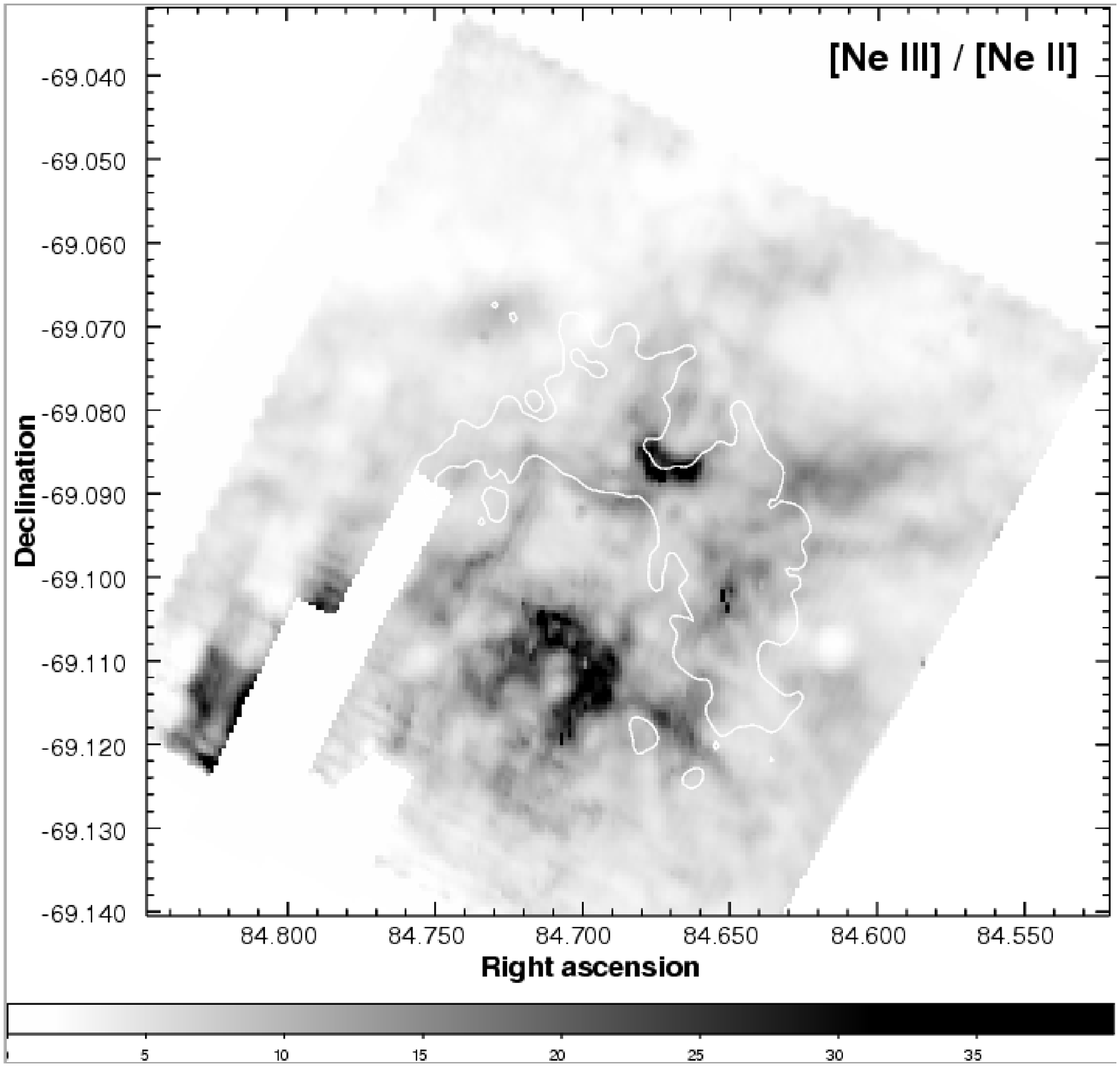}
\vspace{\mpvsp} \\
\includegraphics[scale=\ratioscl]{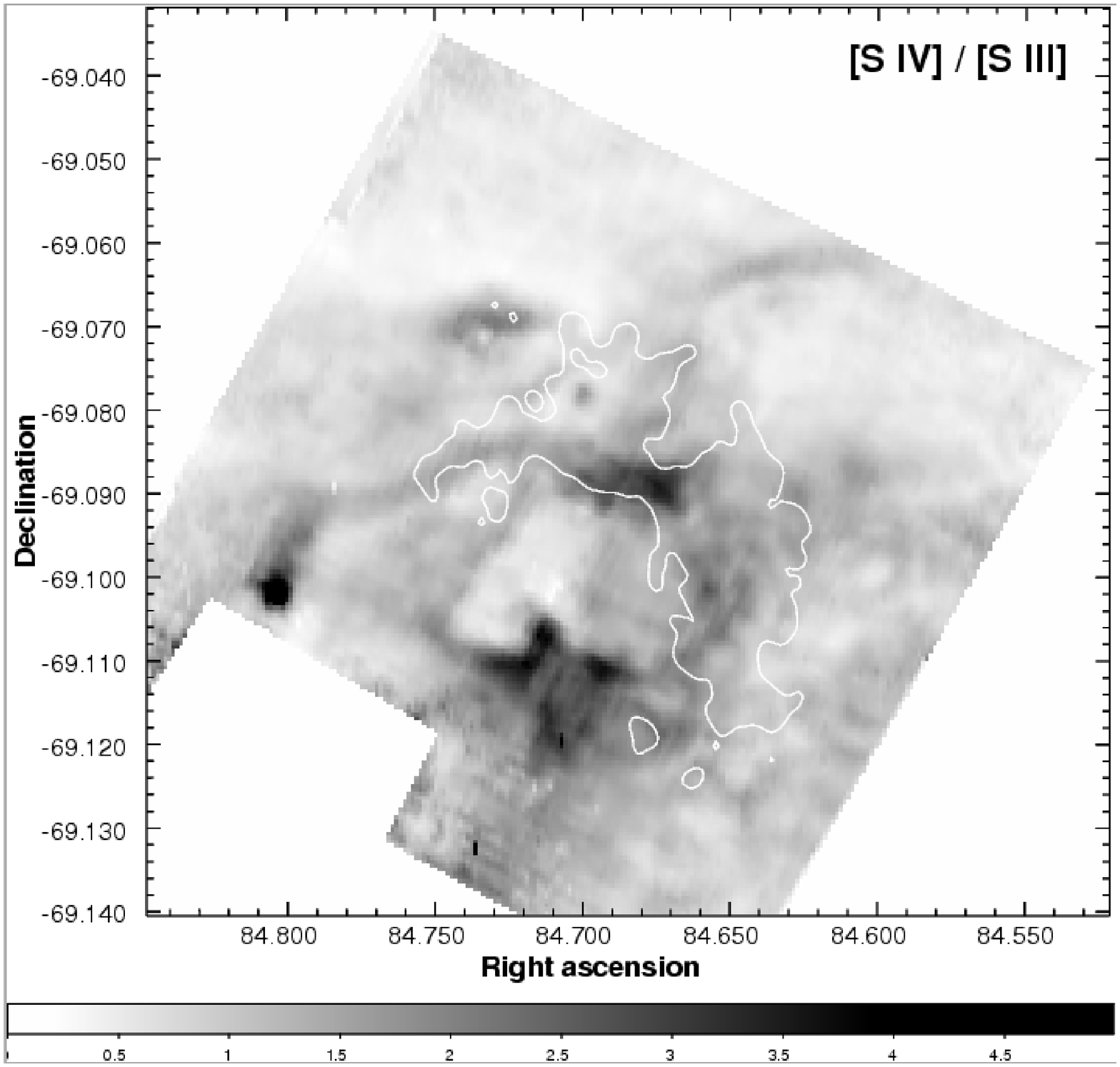}
\caption{\label{ratiofig} Line ratios on a linear scale.  White contours: 
  a single level of 3cm radio emission, to guide the eye (see Figure 
  \ref{bwoverview}). The white region in the middle of the argon ratio plot
  indicates nondetection of \ariil\ and a very high value of the ratio.
  Small masked regions in the lower left of the plots are affected by 
  artifacts (see Fig.~\ref{saturationmap}).}
\end{figure}

\twocolumn

\subsection{Line Ratios}
\label{ratios}

Figure~\ref{ratiofig} shows the three line ratio maps \ariiil/ \ariil,
\neiiil/ \neiil, and \sivl/ \siiill.  These species all have
ionization potentials above 13.6eV, and thus are sensitive to the
shape or hardness of the ionizing spectrum (we will call this
$T_{rad}$, because it is often parameterized by the temperature of 
the best-fitting
black-body over $\lambda<$912\AA) and the intensity of ionizing
radiation.  The latter is usually quantified as the dimensionless
ionization parameter
\[ U = {1\over{n}}\int_{13.6eV}^{\infty}{{F_\nu}\over{ch\nu}}d\nu, \]
or ratio of ionizing photon density to atom density.  Any single line
ratio cannot distinguish between elevated $U$ or $T_{rad}$, but
because the spacing of the ionization potentials differs with atom,
different ratios have different $U$ and $T_{rad}$ dependencies, and
measuring two ratios can break the degeneracy. \citep[This has been
discussed by many authors, see especially the discussions in][,and
\S\ref{cloudy} and Fig~\ref{confidence}
below.]{morisset04,martinh02,dopita06}.

All ratios increase with $U$, but locations where the ratios are not
well correlated can indicate changes in the hardness of the ionizing
field.  In \dor\ the ratios are very well correlated (correlation
coefficient of 0.7 between the neon ratio and argon ratio map, 0.8
between the neon and sulphur ratios, and 0.7 between the argon and
sulphur ratios).  The Ne and S ratios are telling us similar things
over much of the region.  For example, there is high excitation in a
``hot spot'' between the two lobes of the ridge (source \bananasplit,
\S\ref{sources_text}), and in the region to the south of the bubble.
There is a low excitation ridge to the north (source \trough,
\S\ref{sources_text}).  The location noted above for low argon
emission (source \lowexcite, \S\ref{sources_text}) is significantly
low in all three line ratios.

Interestingly, the Ne and S line ratios show quite different behavior
on the eastern edge of the bubble, where \neiii/\neii\ is high but
\siv/\siii\ is low (the red region in Figure~\ref{fitmap} below,
centered on source \WN, \S\ref{sources_text}).  The S ratio is more
affected by extinction since \sivl\ is in the silicate dust absorption
feature, so that could be a region of very high extinction or a region
of particularly hard ionizing radiation.  In this case, the sulphur
ratio does not vary dramatically throughout the bubble, and it is the
neon ratio which is higher at the western end, which argues for harder
ionizing radiation.  An extinction effect would require simultaneous
increase in the ionization parameter or strength of the ionizing field
{\it and} increased extinction.  The specific extinction difference
(A(\sivl)-A(\siiil))/A(2.1\um) equals 0.42$\pm$0.03 \citep[uncertainty
  reflects differences between published extinction curves; we used
  that in PAHFIT, ][]{smith07pahfit}, whereas the effect of extinction
on the neon ratio is small (A(\neiiil)-A(\neiil))/A(2.1\um) =
0.0$\pm$0.07.  Reproducing the observed line ratios without changing
the {\it hardness} of the ionizing field would require
A(2\um)$\simeq$0.75, or A$_V\simeq$6.  Such significant extinction
seems unlikely given the overall relatively low extinction in the
region, and is not detected in our extinction maps
(\S\ref{extinction}).

\begin{figure}
\plotone{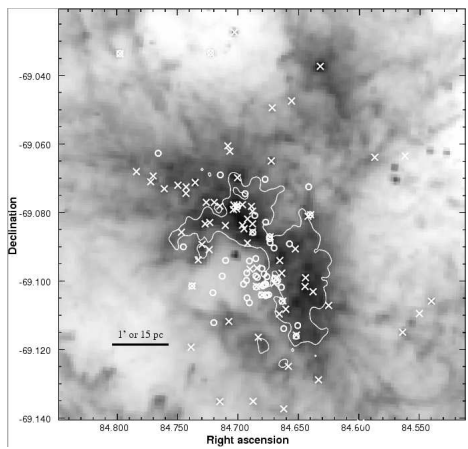}
\caption{\label{maerckerfig} IRAC 8\um\ image (log scale) with two types
  of relevant point sources: Circles show the most energetic stars in the
  region - Wolf-Rayet stars and early O stars, from
  \citet{breysacher,parker}, and X's show protostellar candidates
  determined from 3\um\ excess emission by \citet{maercker05}.  
  White contours: a single level of 3cm radio emission, to guide the eye 
  (see Figure \ref{bwoverview}).
}
\end{figure}

One question that is possible to investigate with this dataset is
whether the optically known hot stars in \dor\ can completely account
for the ionization structure, or whether embedded star formation
affects gas excitation measured in the infrared that otherwise might
be concealed by at other wavelengths.
On Figure~\ref{maerckerfig} we plot show the location of Wolf-Rayet
stars and early O stars (there are discrepancies between spectral
types determined by different authors, and our intent is to show the
most massive stars, not provide the most precise spectral typing
possible).  There is a cluster of WR stars between the two lobes of
the ridge, coincident with very high excitation gas.  There is not a
dramatic increase of excitation centered on R136, although the
excitation is generally high in the area.
Figure~\ref{maerckerfig} also shows the 3.5\um-excess sources of
\citet{maercker05}, representing a crude selection of possible
protostellar candidates. Other kinds of sources, including some of the
previously identified WR stars, can also display infrared excess.
Nevertheless, this selection of sources does trace what is known from
more precise studies with incomplete spatial coverage
\citep[e.g.][]{rubio98,brandner01}, that the young embedded sources in
this region are concentrated along the IR-bright ridge.  The
protostellar candidates shown here do not show any particularly
striking correlation with regions of high excitation -- in fact the
southern part of the region has high excitation and few protostellar
candidates.  We conclude that the ionization structure in \dor\ is
primarily determined by the (optically) known hot stars.

\subsection{Notable Regions}
\label{sources_text}

\begin{figure}
\includegraphics[scale=0.36]{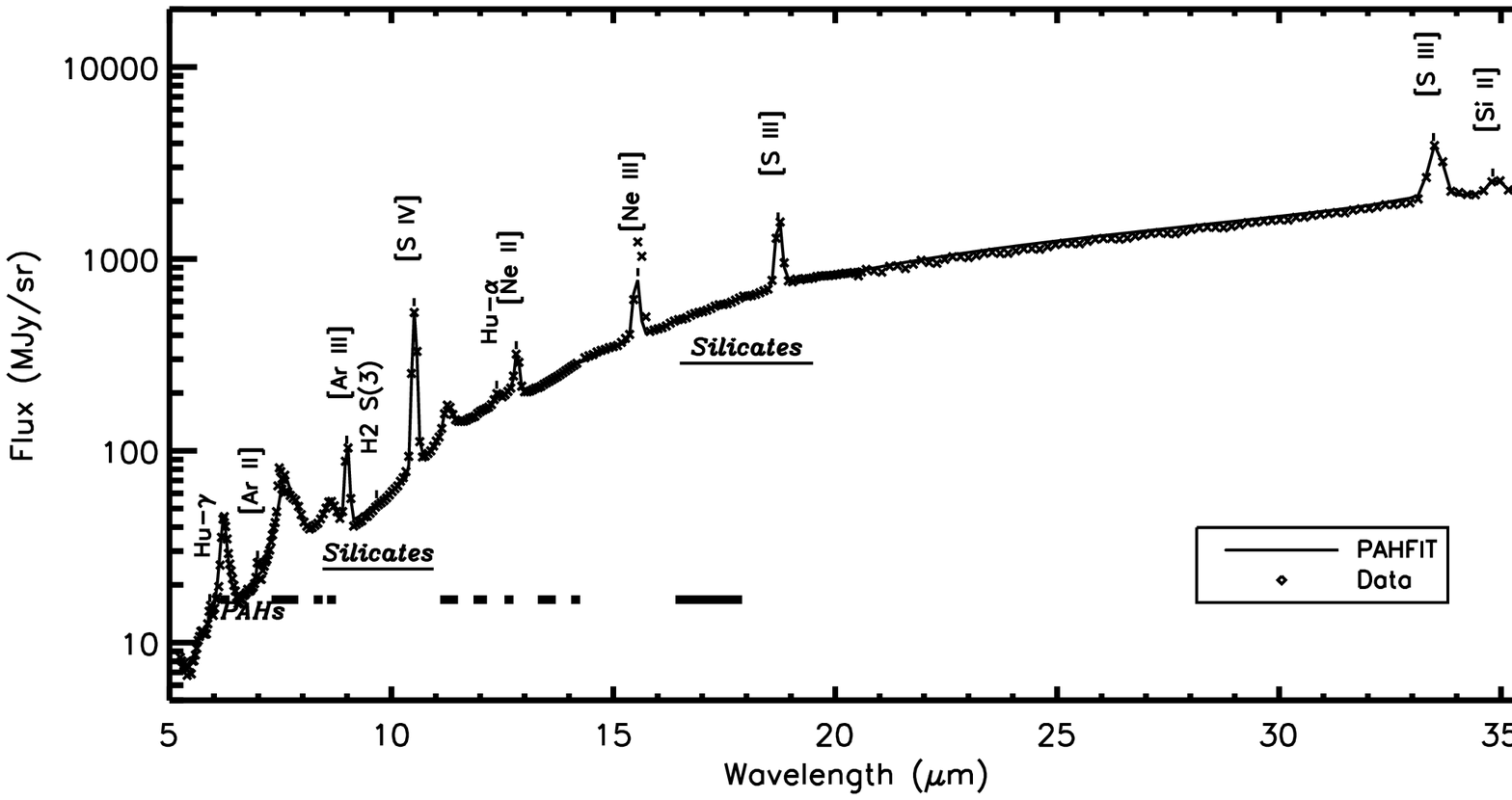}\\
\includegraphics[scale=0.5]{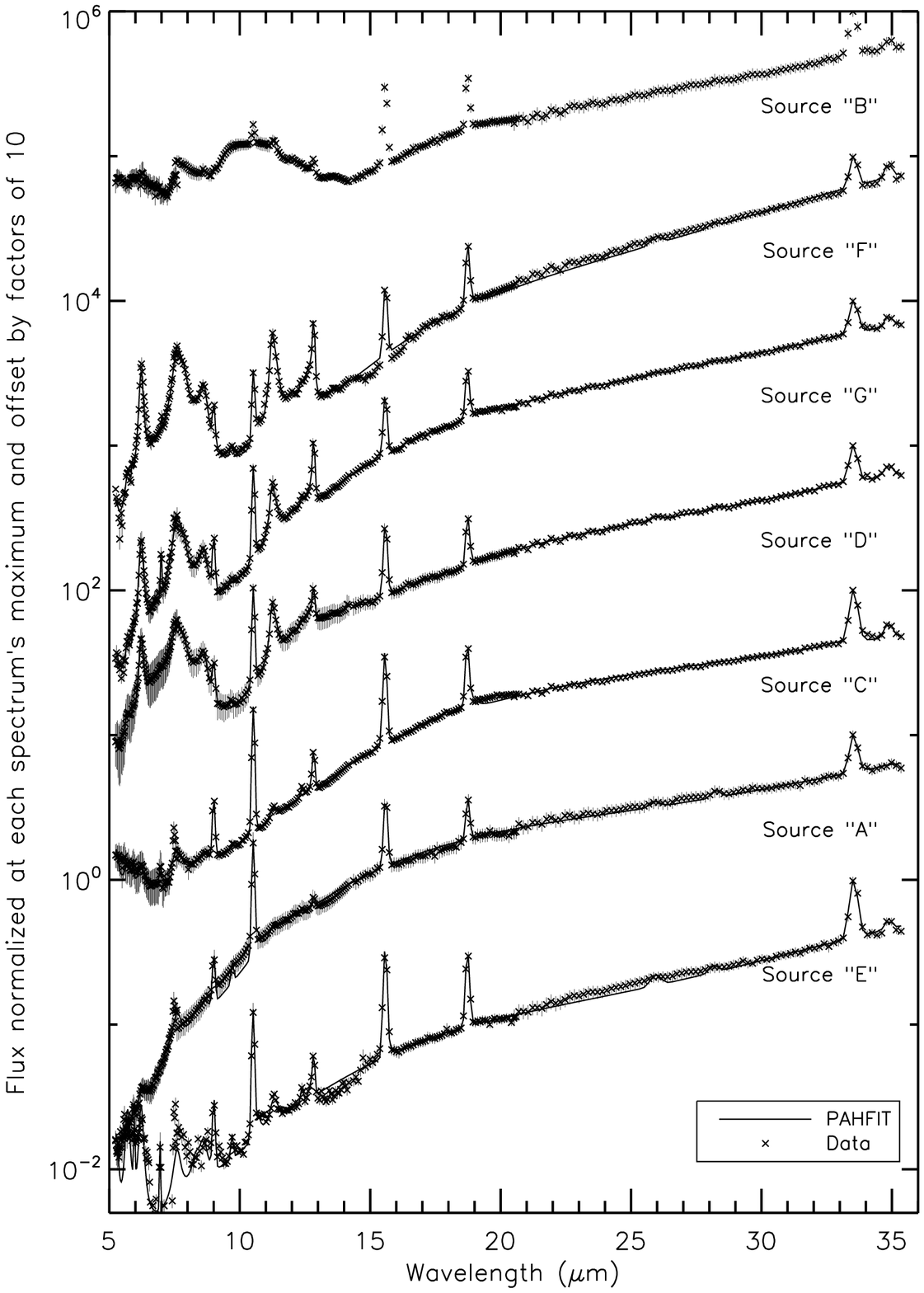}
\caption{\label{sources_spectra} Spectra of the seven regions described 
  in \S\ref{sources_text}, with errorbars.  For comparison, the spectrum 
  of the entire region (Fig~\ref{entirespec}) is repeated at the top, with 
  feature labels.  }
\end{figure}

Figure~\ref{sources_spectra} shows the spectra of small regions which stand
out in the continuum (Fig~\ref{threecoloroverview}), in the feature
maps (Figures~\ref{atlinemaps1} through \ref{pahlinemaps3}), or in the line 
ratio maps (Fig~\ref{ratiofig}).  They have been chosen to illustrate the 
range of excitation, extinction, and continuum shape in the \dor\ nebula.  
These regions of interest were marked on Figure~\ref{bwoverview}, and are 
listed for reference in Table~\ref{sources_table}.  

\begin{deluxetable}{lccl}
\tabletypesize{\scriptsize}
\tablecolumns{4}
\tablewidth{0pc}
\tablecaption{Notable Sources\label{sources_table}}
\tablehead{
\colhead{label} & \colhead{RA} & \colhead{Dec} & \colhead{description}
}
\startdata
\linepeak    & 5h38m45s & -69d05m13s & brightest point on ridge \\
\bananasplit & 5h38m42s & -69d05m05s & hot spots between molecular clouds \\
\parkerMI    & 5h38m49s & -69d05m32s & \citet{parker} \#1445 MI star \\
\core        & 5h38m42s & -69d06m00s & R136 \\
\extinc      & 5h38m56s & -69d04m18s & example of high extinction \\
\WN          & 5h38m57s & -69d06m07s & isolated WN6 star R145 \\
\trough      & 5h38m57s & -69d03m42s & low-excitation trough \\
\lowexcite   & 5h38m28s & -69d06m30s & SW low-excitation region \\
\enddata
\end{deluxetable}

Source \linepeak\ is the brightest point on the ridge in most tracers,
including infrared continuum, many lines, and centimeter continuum.
We detect [\ion{N}{3}]$\lambda$57.3\um\ and
[\ion{O}{1}]$\lambda$63.1\um\ in MIPS SED observations at this
location, and tentatively \feiil, although this small region of the
map is affected by residual fringing and saturation in the LL1 module,
raising the uncertainty of the \feiil\ detection.

Source \bananasplit\ is a diffuse region between the two lobes of the
molecular ridge.  We describe it as a ``hot spot'' because it has
locally elevated excitation seen in the line ratios
(Fig~\ref{ratiofig}), and when analyzed with photoionization models,
is best fit with a hotter or harder radiation field (Fig~\ref{fitmap},
\S\ref{cloudy}). There are known hot stars in the vicinity, but it is
not completely clear if they are responsible for the locally higher
excitation in the \ion{H}{2} region.

Source \parkerMI\ is a prominent point source from optical to MIR
wavelengths, and has been spectroscopically classified as an M
supergiant \citet[][; see also refs therein]{parker}.  Their quoted
absolute V magnitude appears to place the star at twice the distance
to the LMC, but other authors note that it is likely in the
foreground.  It seems unlikely in any case that it is located in the
bubble or affecting the ionization structure.  Because this source
exhibits silicate emission at 9.8 and 18 \um\ rather than absorption,
PAHFIT cannot be applied to it using the parameters used on the rest 
of the spectral cube.  It is the only source in 
Figure~\ref{sources_spectra} without the fitted spectrum overlaid on 
the data points and errorbars.

Source \core\ is the R136 cluster core. 

Source \extinc\ is a prominent mid-infrared point source outside the
main cluster and bubble.  The surrounding ionized gas has quite high
atomic line ratios and thus appears quite highly excited, but we were
unable to find a hot star cataloged in the literature which might be
responsible for this local excitation.
This region also shows relatively high extinction compared to the rest
of \dor.  Extinction can raise the sulphur line ratio but has
relatively small effect on the neon line ratio, so in our opinion the
extinction is not responsible for the apparently high excitation.

Source \WN\ is a known WN6 star R145, fairly isolated from other
catalogue hot stars at the eastern end of the bubble.  The high
fitted radiation temperatures in the neighborhood of this source
(\S\ref{cloudy}) are consistent with excitation by the harder expected
radiation field from a Wolf-Rayet star.

Source \trough\ is an east-west extended region that appears to be a
trough or low point in the excitation and line ratios.  The electron
density is not remarkable, neither does there appear to be any
molecular gas traced by CO emission in the vicinity.  It is possible
that the feature is in the foreground of this three-dimensional
nebula.

Source \lowexcite\ is a region associated with two peaks in the CO
emission [JGB98] 30 Dor-12 and 13 \citep{johansson}, just to the west
of the main ridge.  There is a peak in both argon line maps at this
location,  and the region is a low point in all
three excitation ratios \ariii/\arii, \neiii/\neii\ and \siv/\siii.
There are in fact several discrete parsec-sized regions of low
excitation in the vicinity
%
There is little 3cm continuum at those locations, which lie almost
exactly on the opposite side of a molecular cloud from R136.  One spot
coincides with a knot in the infrared diffuse continuum (\#29) noted
by \citet{hyland92}. There is a 3.5\um-excess protostellar candidate
\citep[][and Fig.~\ref{maerckerfig}]{maercker05} nearby, but not
coincident with the low-excitation spots.  High resolution and
sensitivity molecular observations may reveal dense, starless
molecular clumps in this part of \dor, (self)-shielded from the
intense radiation.


\section{Distribution of Matter: Extinction and n$_e$}
\label{density}


\subsection{Extinction}\label{extinction}

Previous estimates of the extinction in \dor\ found
A$_V$=1.1$\pm$0.1$mag$ \citep[][and refs therein]{dickel94}, with two
possible locations of higher extinction at \radec{05}{38}{39}{-69}{07}{30} and
\radec{05}{38}{32}{-69}{06}{22}.  
%
%
\citet{rosa} found extinction corrections at H$\beta$ of 0.5-1.0 
(A$_V$=0.4-0.8) at ten locations in the outer parts of the region 
we are studying.  
%
We make three estimates of extinction in the region, two directly from
our dataset.  All methods suffer from systematic uncertainty and
modest signal-to-noise, but we can be confident of higher
extinction in the regions that all three methods agree.

\begin{figure}
\epsscale{1.05}
\plotone{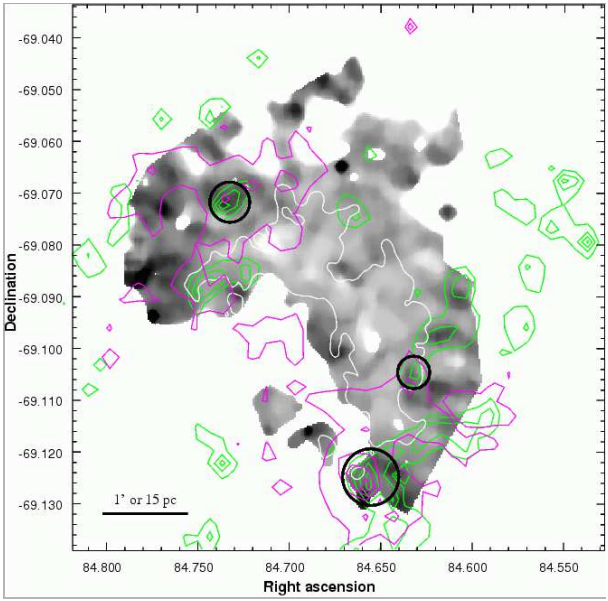}
\caption{\label{threecolorextinction} Map of extinction in 30 Doradus.  
  Grayscale linear image: the IRS \hug/\hua\ ratio, which is proportional 
  to extinction.  Dark areas of the map indicate high extinction.  The 
  Humphreys lines are weak, and this map has been severely cropped to remove 
  areas of the worst signal-to-noise. Green contours: the optical depth of 
  silicate absorption from the IRS spectral map, $\tau = 0.10, 0.27, 0.43, 
  0.60$.  Areas of low signal-to-noise have been masked out of this map as well.
  Magenta contours: The ratio of the 3cm continuum \citep{lazendic} 
  to H$\alpha$ emission (MCELS, {http://www.ctio.noao.edu/$\sim$mcels/}), 
  also proportional to attenuation.  The northernmost magenta
  source corresponds with the molecular cloud marked in yellow in Figure~
  \ref{threecoloroverview}.
  White contours: a single level of 3cm radio emission, to guide the eye 
  (see Figure \ref{bwoverview}).
  The areas where all three maps tend to agree on high extinction have been 
  marked in black.  The northeasternmost corresponds with source \extinc\ in 
  Figure~\ref{bwoverview} and \S\ref{sources_text}.}
\end{figure}

First we use the ratio of centimeter continuum to H$\alpha$ emission
to derive a map of extinction in \dor, following the procedure used in
\citet{lazendic} and using their centimeter data that they kindly
provided (magenta contours, in Fig.~\ref{threecolorextinction}).  We 
assume that
all of the centimeter continuum emission in the \dor\ region is
thermal.  Single dish measurements estimated that the nonthermal
component contributes less than 2\% at 6cm \citep[][and refs
  therein]{shaver83}. \citet{lazendic} identified two possible
supernova remnants from comparison of cm synthesis images and optical
recombination lines, but their own analysis and subsequent followup
with optical and Xray imaging and optical spectroscopy suggests that
these are merely extinguished \ion{H}{2} regions \citep{chu04snr}.
The relationship between thermal bremsstrahlung centimeter continuum
and hydrogen recombination line emission depends weakly on the
electron temperature, but \citet{rosa} and \citet{peck} both found
that T$_e$ variations are small in 30~Doradus ($\lesssim\pm$300K), so
it is unlikely that the calculated extinction variations are actually
misinterpreted T$_e$ variations.

We derived two more maps of extinction directly from the IRS spectral
cube.  Amorphous silicate dust is responsible for two broad bands of
absorption at 9.7 and 18\um.  The PAHFIT package reports the fitted
optical depth at 9.7 \um.  The resulting map for 30 Doradus is shown
in Figure \ref{threecolorextinction}, in green contours.  The fit to the
absorption feature is sensitive to noise in the spectrum.  We cropped
away southern parts of the map that were clearly artifacts, and any
part of the map where the signal-to-noise was less than 2.  Finally,
we median-smoothed the resulting map with a window of 3 pixels.  The
result is a sparse map showing the regions of comparatively reliable
elevated silicate absorption.

The ratio of Hydrogen recombination line strengths can also be used to
calculate extinction, most commonly by assuming an intrinsic
(unextincted) ratio from Case~B recombination.  Our dataset includes
\hua\ (7-6) and \hug\ (9-6).  Due to modest signal-to-noise and to 
a systematic tendency of PAHFIT to slightly underestimate the continuum 
level near these wavelengths, 
the absolute value of extinction calculated from this ratio has a
large systematic uncertainty.
It is also difficult to resolve the weak \hua\ line from H$_2$
S(2) 12.28$\mu$m in low-resolution spectra.  We quantified the amount
of potential contamination to the \hua\ line strength in several ways:
We fit our low-resolution spectra with and without the molecular
hydrogen line (the central wavelengths of both lines are very tightly
constrained by PAHFIT), and found that the 
\hua\ line strength was only decreased by 15\% (that flux was attributed to H$_2$) when both lines were
included.  We also fit the high-resolution GTO spectra, in which the
lines are easily separated and reliably measured, and found that over
most of our area, H$_2$ was less than 20\% of \hua.  We examined fits
to both the high and low-resolution spectra in the few high-resolution
apertures where the H$_2$ and \hua\ strengths are comparable (H$_2$
never exceeds \hua\ in strength) and found that the low-resolution
\hua\ strength used in our analysis was at most overestimated by 40\%
by the presence of the H$_2$ line on the wing of the \hua\ line.
While some level of contamination of \hua\ by H$_2$ S(2) may be
present in our maps, we do not expect that to change any of our
conclusions including the regions of high extinction identified in
Figure~\ref{threecolorextinction}.

Further systematic uncertainty arises in the choice of extinction
curve, since the lines lie on the wings of the silicate absorption
features, which vary in amplitude and width in different studies.
However, all extinction curves that we considered show greater
relative extinction at 
Humphreys-$\alpha$ $\lambda$12.37\um\ than at 
Humphreys-$\gamma$ $\lambda$5.90\um.  We examined the
extinction curves of \citet{chiar06} for the Galactic Center and the
ISM, and the average LMC extinction curve\footnote{Available at:
  www.astro.princeton.edu/\~{}draine/dust/dustmix.html} based on the
carbonaceous-silicate grain model of \citet{weingartner01}.  All of
the extinction curves roughly agree in the vicinity of the \hua\ 
emission line.  Near the \hug\ line, the Chiar \& Tielens curves agree
well with the near-infrared extinction for the ISM found by
\citet{remyextinct}.
Thus, extinction correlates with the flux ratio of \hugl\ to \hual, and
we can still find the relative level of extinction across \dor\ from
the \hua/\hug\ ratio, even though the absolute normalization is
uncertain.

We calculated the signal-to-noise of the ratio based on the RMS
variance of the individual maps of emission line strength, and removed
all regions of the map where the signal-to-noise was less than 2.  We
cropped away the parts of the map where an additive correction was
made to the SL spectra (see \S\ref{finaladjustments}), because this
subset of the map encloses some spectra which have an unphysical
plunge at the red edge of the SL2 module.  The fit to those spectra
generally underestimates the continuum near the \hug\ line, and thus
overestimates the ratio of \hug\ to \hua.  We also cropped away the parts 
of the map where the reduced $\chi^2$ of the fit was greater than 5.  
Finally, even after applying these filters, it was necessary to trim away 
some regions from the low-signal edge of the map where the noise levels 
were still untrustworthy.  The ratio map shown in grayscale in Figure
\ref{threecolorextinction} is now mainly limited to those areas of 30
Doradus where we obtained both good detections of these two faint
lines and a good fit.

Despite the level of noise in the maps of extinction, particularly the 
map derived from the \hua\ and \hug\ lines, there are three areas where all 
three methods agree on especially high extinction.  These areas have been 
marked on Figure~\ref{threecolorextinction}.  Generally, there is not much
extinction by dust in \dor.  As discussed above, extinction has a 
small effect on the Neon excitation
ratio, but raises the \siv/\siii\ (due to the shape of the extinction
curve), which in turn will get interpreted as higher T$_{rad}$ in our
photoionization models.  The regions of interestingly high T$_{rad}$
do not turn out to correspond to regions of clearly high extinction,
but the effect should be kept in mind in interpreting the data.

\subsection{Electron Density}

\begin{figure}
\plotone{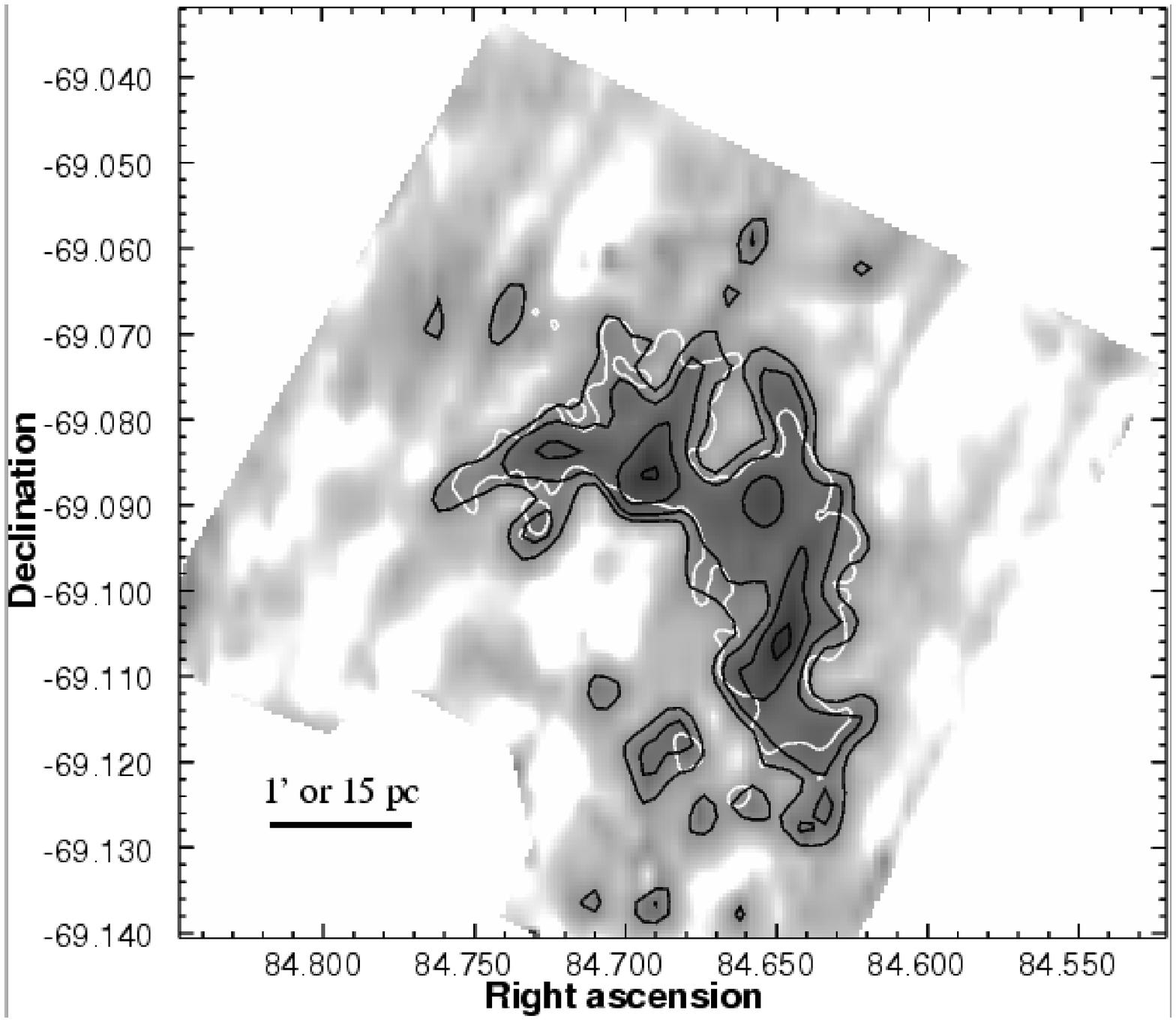}
\caption{\label{densityfig} n$_e$ calculated from \siiil/\siiill.  Black
  contours: log n$_e$ = 2.4, 2.5, 2.6, 2.7.  White contour: a single level 
  of 3cm radio emission (see Figure \ref{bwoverview}).  The
  ridge is prominent, as well as a region of increased density to the
  south.  }
\end{figure}

Figure~\ref{densityfig} shows the electron density calculated from
\siiil/\siiill.  The \siiil\ map was convolved to the lower resolution
of the \siiill\ map, and the line ratio converted to electron density
using the conversion in \citet{dudik} at $T_e$=10$^{4}$K (their
Figure~9 and section 6: Those authors calculated the line ratio as a
function of density and temperature for a five level atom using the
collision strengths from \citet{tayal} and radiative transition
probabilities from \citet{mendoza}).  The calculated ratio is not a
strong function of temperature.  The ridge is prominent in the \siii\ 
ratio or n$_e$ map, and in fact the n$_e$ map is quite similar to the
3cm continuum in morphology.  The density is elevated to the south of
the bubble and R136, in a region of relatively high excitation.

\subsection{Abundances}

Variations in elemental abundances can in principle change the
diagnostic ratios that we are using to measure physical conditions in
the \dor\ nebula.  Previous studies have found very small internal
abundance variations in \dor\ and other giant \hii\ regions.
\citet{peck} found no significant variations on 15\arcsec\ scales in
the He abundance $Y^+ = He^+/H^+$ = 0.13$\pm$0.02 measured from radio
recombination lines.  \citet{rosa} found little variation between
several positions measured with optical spectroscopy.  Of more direct
relevance to this work, \citet{lebouteiller08} found less than 0.01
dex dispersion of Ne/H, S/H, and Ar/H, using the high resolution
GTO Spitzer spectra mentioned above.

Small scale abundance variations in \hii\ regions including \dor\ 
\citep{tsamis05} have been proposed to explain discrepancies between
optical and infrared abundance determinations.  We might hope to
detect sub-parsec scale abundance variations in this dataset by using
three infrared line ratios to solve simultaneously for $U$, $T_{rad}$,
and $Z$.  In practice, we find that the excitation variations can be
adequately explained without abundance variations, which would be seen
as residuals in our fitting of $U$ and $T_{rad}$.  The signal-to-noise
especially in our argon line ratio is insufficient to detect abundance
{\it variations} at the 0.1 dex level predicted by \citet{tsamis05}.

Overall, abundances in \dor\ are not particularly low: \citet{rosa}
found Ne/O, S/O, and Ar/O ratios close to solar.
\citet{lebouteiller08} found Ne/H and Ar/H of
12+log(X/H)=7.76$\pm$0.02 and 6.32$\pm$0.06, respectively, within 0.25
dex of the range of solar values
\citep{lodders07,lodders03,asplund05}.
They found S/H of 6.77$\pm$0.03, only about 0.3 dex sub-solar.  We
also find that on average, half-solar abundances 
result in modestly better
agreement between the three infrared line ratios in our
photoionization models than solar or 0.1-solar models.  In practice, systematic effects such as the
argon recombination rate have a larger effect on this agreement than
the abundances \citep{morisset04,stasinska}.

%



\section{Distribution of Radiation and Gas: Excitation}
\label{excitation}


\subsection{Photoionization models}
\label{cloudy}

As mentioned above, if photoionization is assumed to be the dominant
physical process, the ionic line ratios depend on $U$, $T_{rad}$, and
to a lesser degree metallicity.  If one assumes constant abundances
and hardness or $T_{rad}$, then the line ratio maps in
Figure~\ref{ratiofig} are maps of the ionization parameter $U$,
varying from log $U\simeq$-3 to log $U\simeq$-1.5.


A somewhat more sophisticated analysis is to solve for $T_{rad}$ and
$U$ simultaneously.  We prepared a grid of photoionization models
using Cloudy \citep{ferland} as a 0-dimensional tool to solve for the
ionization structure and line emissivities given a specified radiation
field and ionization parameter (we used the output of the first zone
in each simulation).  We explored different input spectra, including
ATLAS \citep{atlasref}, Tlusty \citep{tlusty}, CoStar OB and Wolf-Rayet atmospheres
\citep{costarref,lindasmith}, and black-bodies.  
In the end we used a grid of Tlusty atmospheres
calculated at half-solar metallicity, extrapolated to hotter effective
temperatures using black-body atmospheres to set the functional
dependence of line ratios on effective temperature, and normalizing
the ratios to those of the hottest Tlusty models. (We found that adopting
different stellar atmospheres primarily changes the line ratios by a
constant multiplicative factor, and has very little effect on the
functional dependence of the line ratios on effective temperature and
ionization parameter.)
We were not able to find freely available grids of more modern
atmospheres (e.g. WMBASIC, CoStar) at subsolar metallicity, but we
performed careful comparisons of our modeling at solar metallicity to
understand the systematic effects.
If nebular abundances are set at solar levels in the photoionization
models, and stellar atmospheres calculated at solar abundance are used
for self-consistency, the derived ionization parameter $U$ decreases
systematically by about 0.1 dex, and the derived radiation temperature
increases systematically by 10\%.
At solar metallicity, WMBASIC atmospheres result in about a factor of
two lower \neiii/\neii ratio and a 50\% lower \siv/\siii ratio.  That
would increase the derived ionization parameter in 30~Doradus
systematically by $\sim$0.15 dex, and lower the derived radiation
temperature by 5--10\%.  These effects have been explained in detail
by other authors, especially \citet{morisset04}.
None of the changes in stellar atmospheres or metallicity that we
explored would result in qualitative changes in our conclusions,
merely small systematic shifts in derived parameters.
We also varied the
dust prescription in terms of abundance and grain size.  Neither had a
strong effect on the line ratios as a function of $T_{rad}$ and $U$,
provided that $U$ was calculated locally, i.e. from integrating the
diffuse ionizing field in the simulation above 1~Ry, rather than
assuming some geometry-dependent expression such as $N_\star/c4\pi
r^2n_e$.

\begin{figure}
\plotone{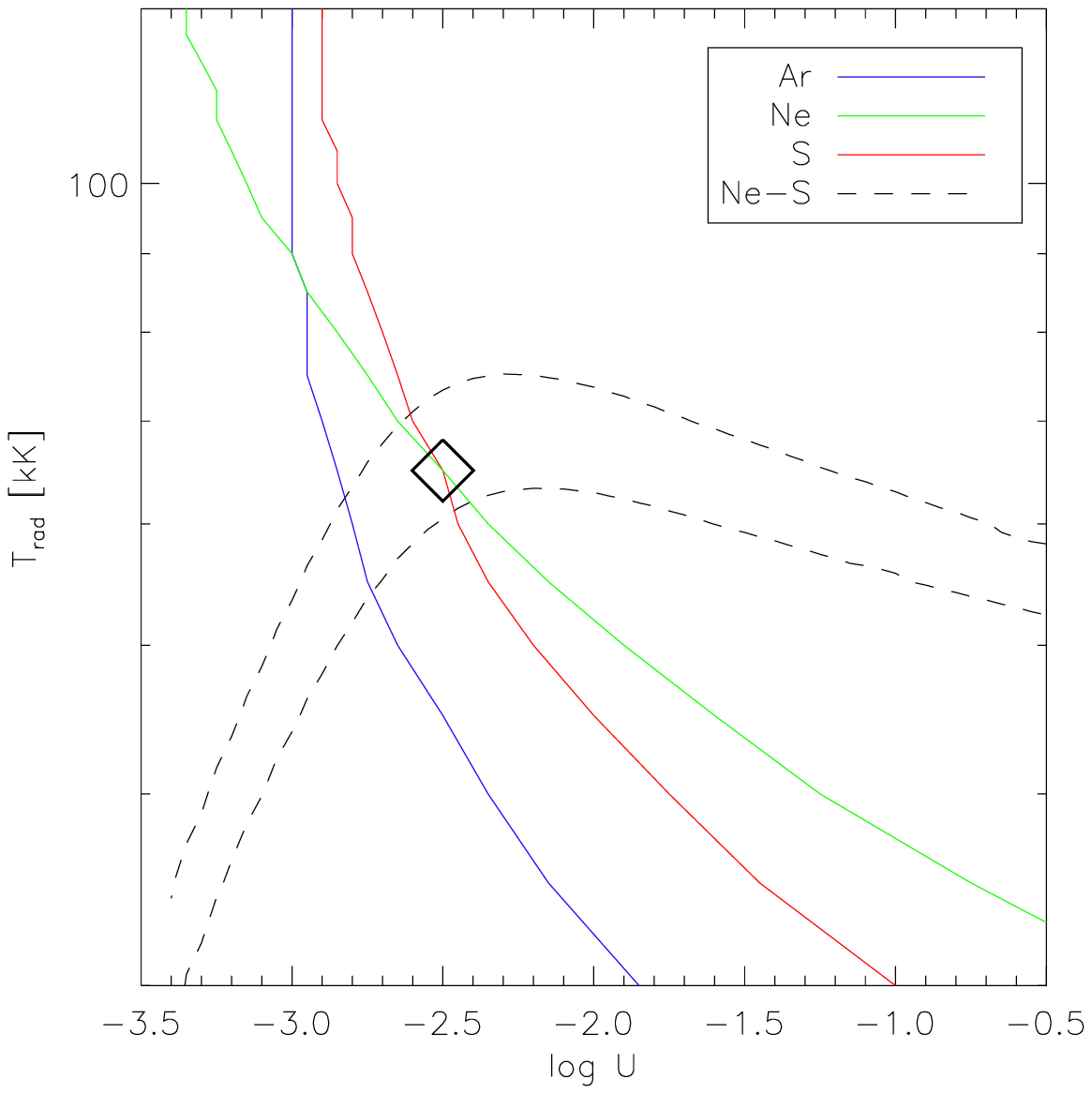}
\caption{\label{confidence} Fits to $U$ and $T_{rad}$ at a
  representative location \radec{05}{38}{37}{-69}{06}{12}.  Loci of
  consistency with the data are shown for fitting Ar, Ne, and S ratios
  alone (lines from upper left to lower right), the difference between
  the S and Ne ratios (log(\siv/\siii)-log(\neiii/\neii), dashed
  lines, 1-sigma confidence interval shown), and the combination of
  all ratios (bold diamond).  See text \S\ref{cloudy} for discussion.}
\end{figure}

Figure~\ref{confidence} shows typical behavior - each measured ratio
defines a curve in $U-T_{rad}$ space, but the lines have different
slopes because the ionization potentials are spaced differently for
the different atoms.  This is equivalent to the nearly parallel arrows for
$U$ and $T_{rad}$ in \citet{morisset04} \citep[see
  also][]{martinh02,dopita06}.  Two measured ratios can better
constrain $U$ and $T_{rad}$.  We will focus on the Ne and S ratios
since \arii\ is not detected over large parts of the \hii\ region.\

\begin{figure}
\includegraphics[scale=\TUscl]{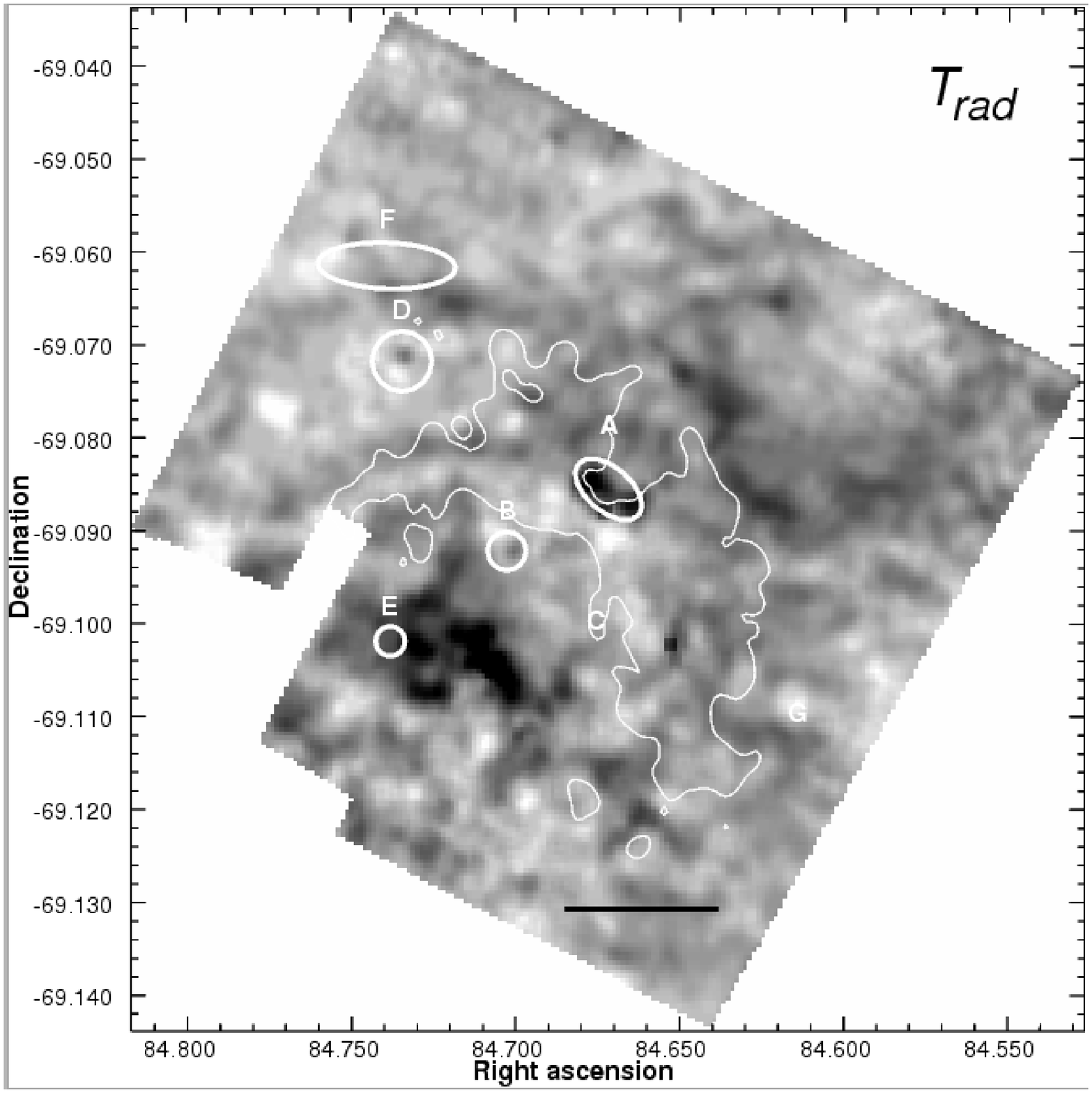}
\hspace{\mphsp}
\includegraphics[scale=\TUscl]{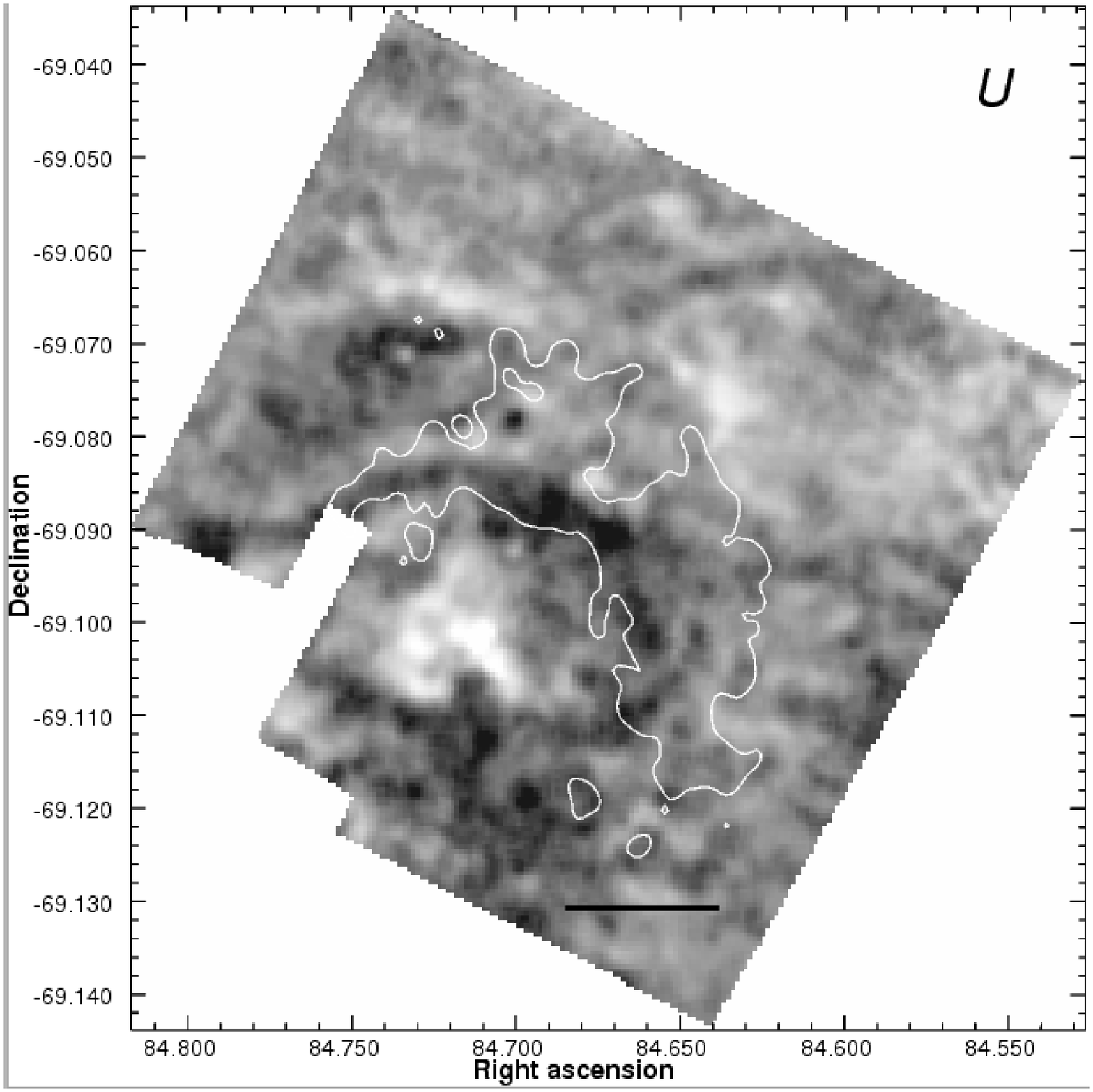}
\vspace{\mpvsp} \\
\includegraphics[scale=0.21]{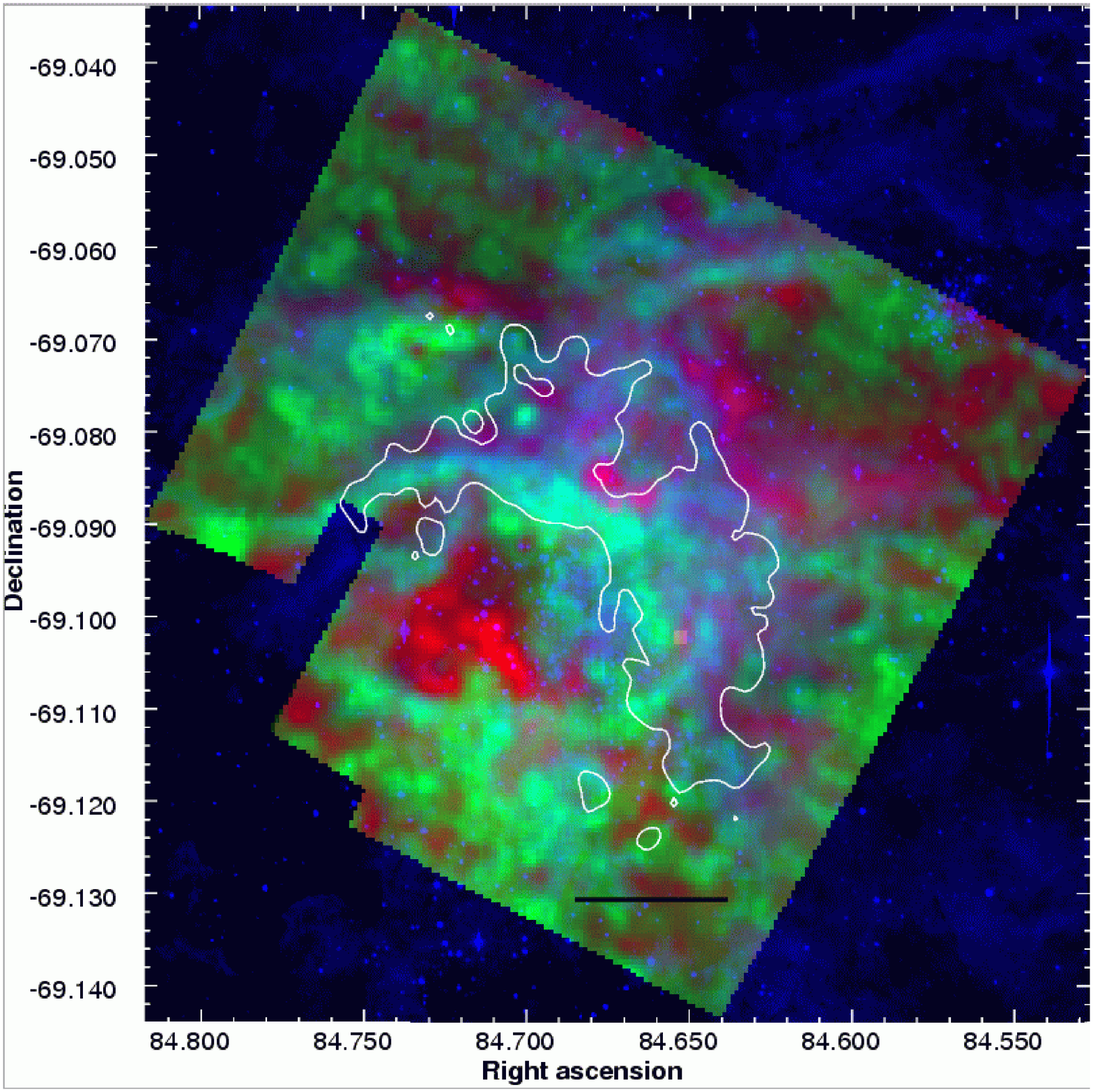}

\caption{\label{fitmap} Top: fitted $T_{rad}$ based on the S and Ne
  line ratios.  Middle: fitted $U$.  Higher values are dark in both 
  grayscale images.  Bottom: fitted $T_{rad}$ (red) and $U$ (green), 
  plus optical B band (blue).  All images are in log space except for the 
  B band map.  White contours: a single level of 3cm radio emission (see 
  Figure \ref{bwoverview}).  
  The black lines indicate a scale of 1' or 15 pc.  Sources of interest
  (see Table~\ref{sources_table}) are labeled on the $T_{rad}$ map.
  The derived parameters, $T_{rad}$ and $U$, should be compared
  with the ratio maps in Figure~\ref{ratiofig}.}
\end{figure}

Figure~\ref{fitmap} shows the fitted $U$ and $T_{rad}$ across \dor.
The ridge is a region of high ionization parameter, as previously
noted from simple examination of the line ratios.  The region around
the relatively isolated source at \radec{5}{38}{56.5}{-69}{04}{17} to
the north of the ridge is also highly excited, probably due to the
local effects of that star (source \extinc, Figure~\ref{bwoverview},
\S\ref{sources_text}).  We note two regions of high $U$ in particular.
The most prominent is between the two parts of the ridge (just south of 
source \bananasplit, \S\ref{sources_text}), where young stars may be 
locally
ionizing the gas and beginning to lower the density and disperse the
ridge (see Figure~\ref{maerckerfig} for the location of protostars and
the most energetic optically identified stars).  Alternately, the
density is simply lower there and ionizing radiation can more easily
leak out from the bubble region around R136. The northern side of this
``hot spot'' shows evidence for hardening of the radiation field,
which could result of the radiation originates in the bubble or on its
rim, and is propagating northward, and being absorbed by gas and dust
(both of which will harden the field).  The high degree of porosity
and mixing between molecular and ionized material in \dor\ is well
known \citep[see e.g.][]{poglitsch}.
On the south side of the bubble is another region of high ionization
parameter - this region also shows high electron density in the \siii\ 
ratio, so the field must be locally strengthened, perhaps by the WR
stars known in that region (Fig~\ref{maerckerfig}).
Particularly interesting is the region on the eastern side of the
bubble, which shows up in the fitted parameter maps as high $T_{rad}$,
but not particularly high $U$.  There is a single catalogued
Wolf-Rayet star (R145, WN6 type, source \WN, \S\ref{sources_text}) in
the center of that area, which may be energizing the eastern end of
the bubble.

\begin{figure}
\resizebox{!}{3in}{\includegraphics{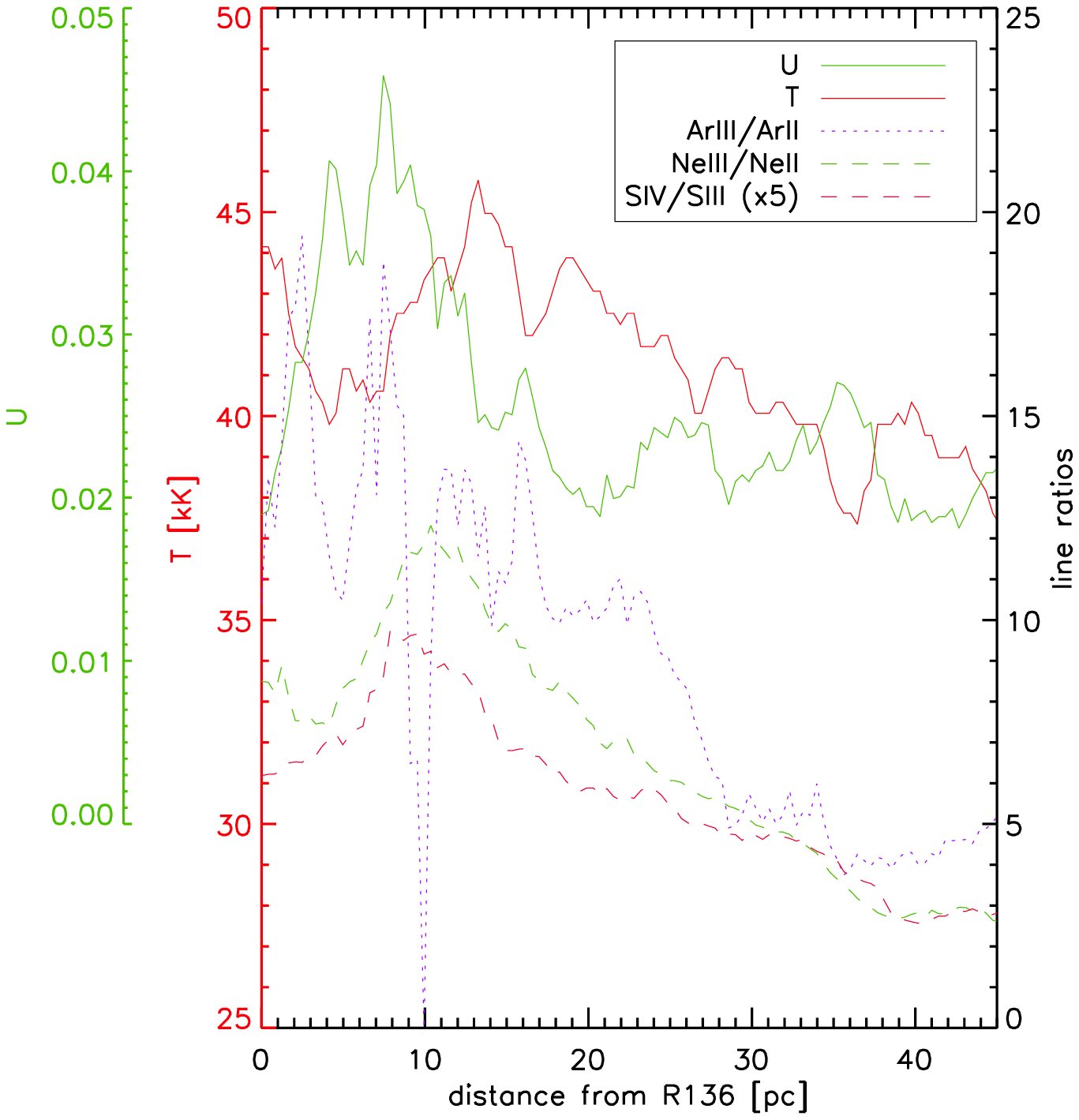}}\\
\resizebox{!}{3in}{\includegraphics{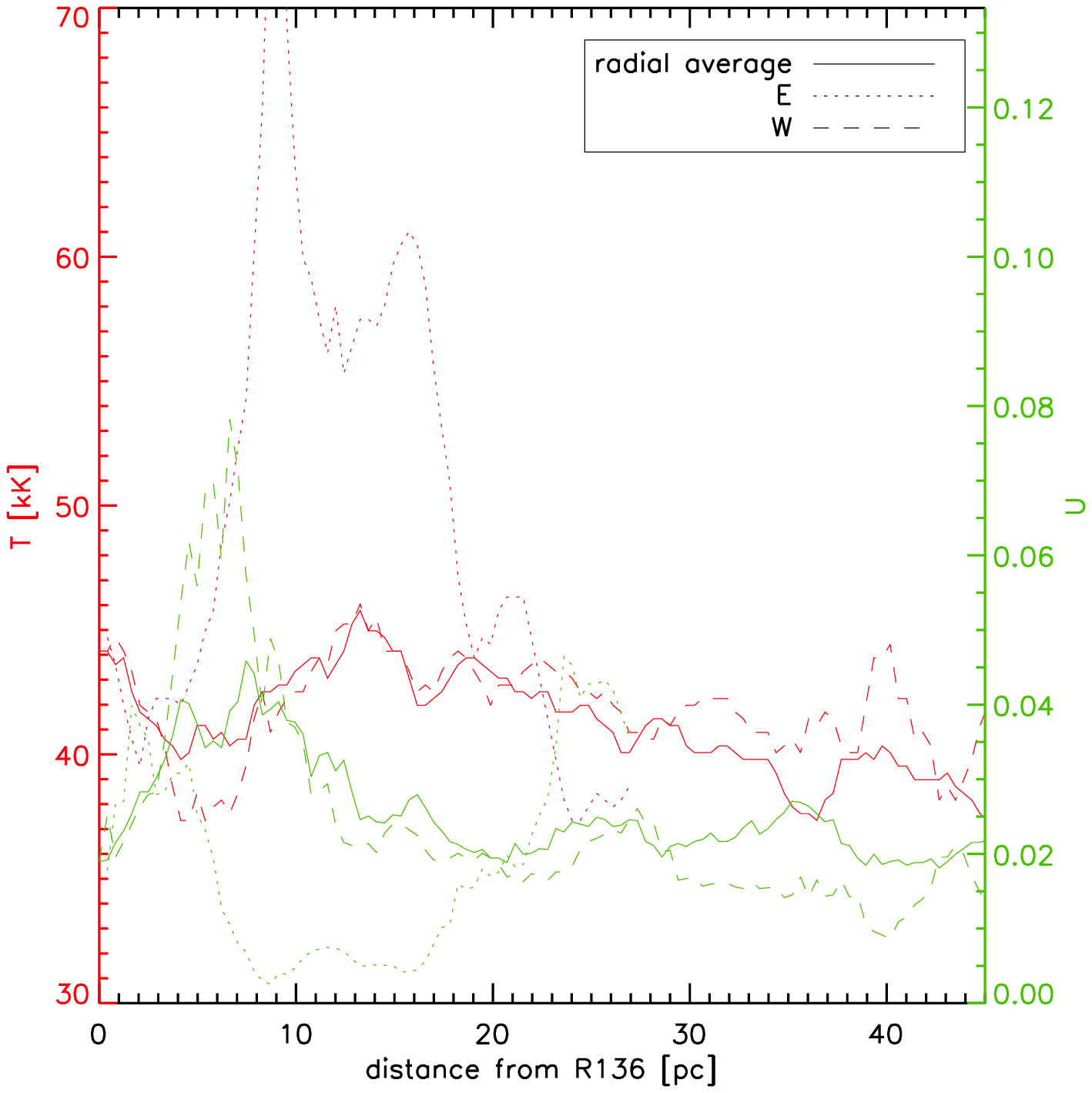}}
\caption{\label{radial} (top) Radial dependence of conditions in the
  ionized gas as a function of radial distance from R136.  The
  different lines show azimuthal median values of the ionic line
  ratios and of the fitted $T_{rad}$ and $U$ from photoionization
  models. 
(bottom) Radial parameters in the eastern and western directions.
  R136 is an off-center blister \ion{H}{2} region, much closer to the
  bubble wall in the eastern than western direction.}
\end{figure}

The radial variation of physical conditions in \dor, as a function of
distance from R136, is of particular interest, to determine whether
feedback in the nebula is dominated by that cluster core, or whether
individual hot stars scattered throughout the region are equally
important.  We have already seen that the latter is true, for example
the hot eastern end of the bubble apparently excited by the WN6 star
source \WN.  Within several tens of parsecs from R136, however, there
does appear to be a global effect.  The top panel of
Figure~\ref{radial} shows the radial dependence of the ionic line
ratios and fitted $T_{rad}$ and $U$.  The bubble at $\sim$10pc radius
is clearly evident in lowered $T_{rad}$ and raised $U$. Outside the
bubble there is a modest gradient in the hardness of the fitted
radiation field, but not much gradient in the ionization parameter
$U$.  This is sensible in a region which has the hottest stars in the
center, but some ionizing sources distributed more widely.  As R136 is
apparently driving an off-center ``blister'' type \ion{H}{2} region,
it is interesting to separate the radial dependencies of the ionized
gas parameters in the eastern and western direction.  The second panel
of Figure~\ref{radial} shows this comparison, showing clearly the
bubble wall at $\sim$5pc distance in the east, and the hot bubble
between R136 and the $\sim$25pc distant bubble wall in the west.
Unfortunately the effects of saturation make the measurements less
reliable beyond the bubble wall in the west (see \S\ref{artifacts}).

\subsection{Shock Models}
\label{shock}

Protostellar shocks can also produce MIR fine-structure emission, but
only if the excitation (shock velocity) is high enough.  \neii\ 
requires J-shocks with v$\gtrsim$60\kms\ \citep{hollenbach89} and
\neiii\ velocities in excess of 100\kms\ \citep{molinari02}.
\citet{lefloch03} detect \neiil\ and \neiiil\ in HH~2, but only from the
highest excitation working surface.  Similar results are being found
with {\it Spitzer}, detecting \neiil\ but not \neiiil, and \siiill\ but
not \sivl\ in HH46/47 \citep{alberto04}, Cep~E \citep{alberto04b}, and
HH7-11 and 54 \citep{neufeld06}.  Even the bow shock near a runaway
O9.5 star \citep{france07} and the stronger shocks in SNR
\citep{neufeld07} show similar relatively low-excitation MIR
emission.

The (few) MIR observations of regions containing both shocks and
photoionization tend to show photoionization dominant.  In \neii\ and
\siv\ maps of the massive star formation region W51~IRS2, high spectral
resolution mapping with TEXES \citep{lacy07} allows the high velocity
emission to be separated from that at the velocity of the molecular
cloud.  The authors do not note dramatically different line ratios in
the high velocity emission from that at the systemic velocity, but
rather confirm earlier ground-based observations of MIR fine structure
lines consistent with photoionization by late O-type stars
\citep{okamoto01}.  \citet{lacy07} interpret their observations as a
neutral jet emerging from the molecular cloud and being subsequently
photoionized.

More recent work with {\it Spitzer} in the Galactic center
\citep{simpson07} cannot produce the observed O$^{+3}$ abundance with any
reasonable photoionization model, even including hot supergiant
atmospheres and 10$^{5-6}$K blackbodies representing diffuse X-ray
emission.  The proposed alternative of $\sim$100\kms\ shocks can match
the observed [\ion{O}{4}] line emission.  The authors do not {\it require}
shock excitation to explain their Ne and S line ratios, but do note that the
highest excitation gas is found between the two stellar clusters,
which is indicative of shock excitation.

\begin{figure}
\centerline{\resizebox{!}{2.8in}{\includegraphics{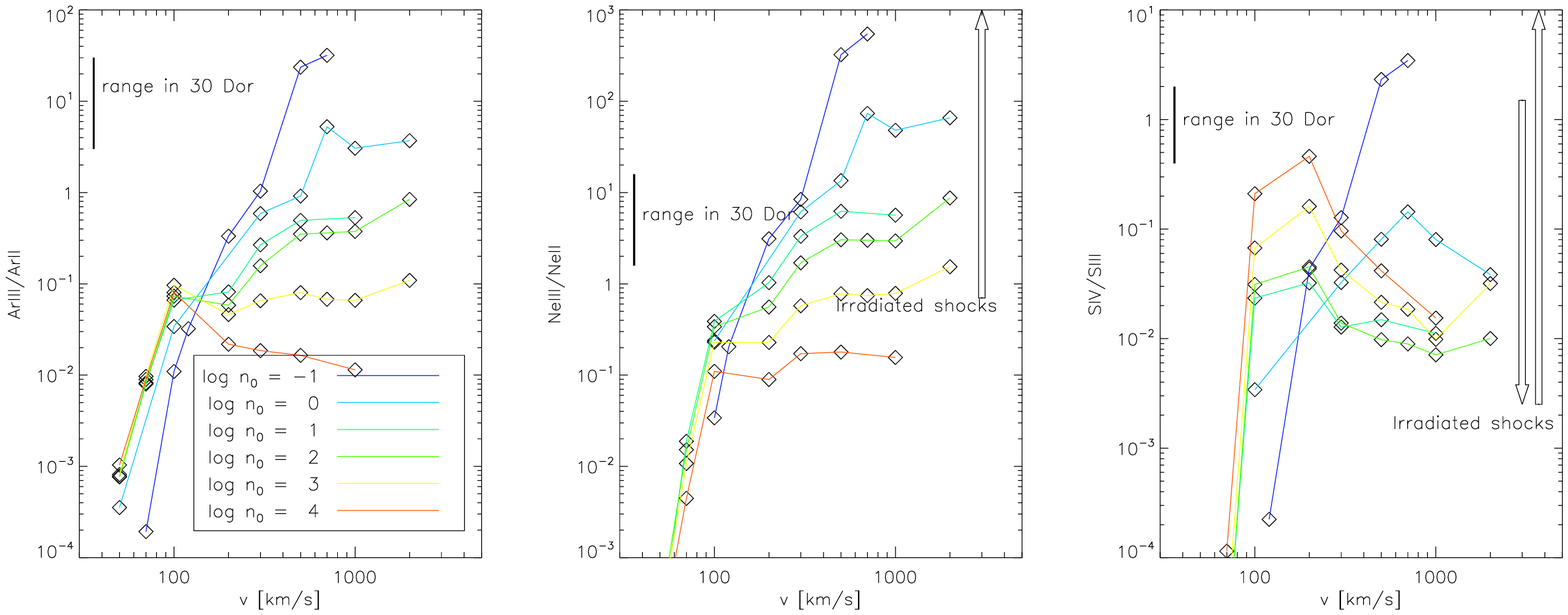}}}
\centerline{\resizebox{!}{3.5in}{\includegraphics{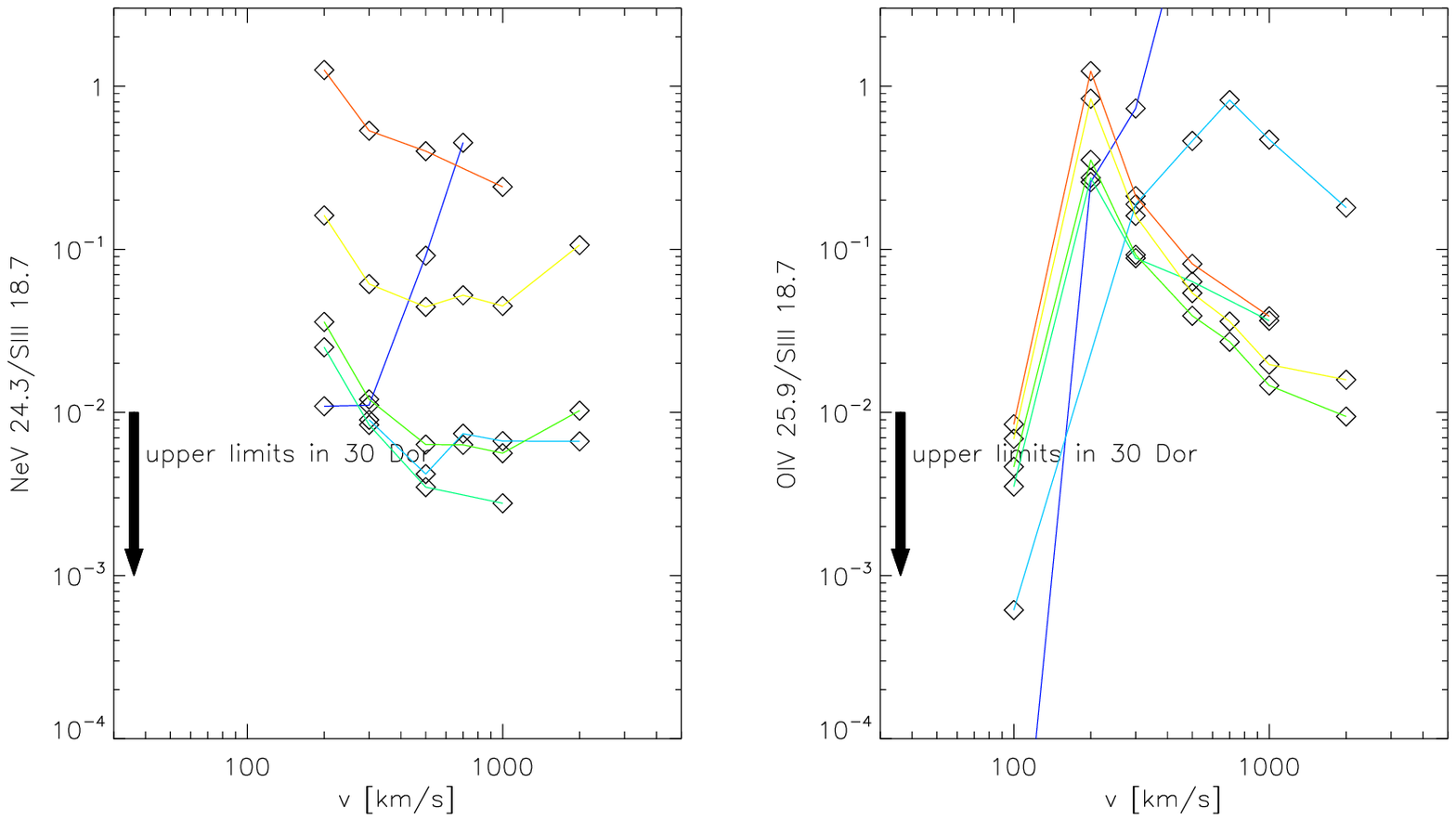}}}
\caption{\label{shockfig} Ratios of line strengths calculated from
  plane-parallel shock models (the actual models that were run are
  diamonds, connected for clarity).  Although the Ne, Ar, and S ratios
  can be produced by shocks of a few 100\kms\ in diffuse gas, those
  shocks should also produce [\ion{O}{4}] emission, which we do not
  detect in \dor.}
\end{figure}

We ran a set of shock models using John Raymond's code
\citep{raymond1,raymond2,hrh} with a range of velocities and pre-shock
densities.  The ratios of relevant fine-structure lines are shown in
Figure~\ref{shockfig}, along with the observed ranges in \dor.  To
first order, all three ratios can be matched with shocks on the order
of a few hundred \kms\ in quite low-density gas ($n<1$cm$^{-3}$).
However, we have significant nondetections of emission from the higher 
ionization species [\ion{O}{4}] and [\ion{Ne}{5}], with 3-$\sigma$ upper limits
less than 10$^{-2}$ relative to \siiil\ across the map.  The predicted
ratios for those two species relative to \siii\ are also calculated,
and shocks of this order should produce easily detectable
[\ion{O}{4}].  Shocks can also produce [\ion{Ne}{5}] \citep[e.g. in
  supernova remnants][]{rakowski}, but it is more usually used as an
indicator of very hard radiation from AGN or WR stars \citep[][ and
  Table~\ref{ip}]{abel}.  Thus we conclude that in \dor\ as elsewhere,
although shocks certainly exist, the ionization structure of the gas
is dominated by photoionization and not shock activity.  Furthermore,
while the effects of hard radiation from WR stars affects the
ionization balance in some parts of the nebula, the radiation field on
$\sim$0.5pc scales is not apparently hard enough for very high
ionization species to be very important.

It is particularly interesting to consider the ``hot'' region on the
eastern side of the bubble where the photoionization models are driven
to high $T_{eff}$ and moderate $U$ by low \siv/\siii\ and moderate
\neiii/\neii.  That low \siv/\siii\ ratio makes this part of \dor\ more
consistent with shock models than other parts, although the
nondetections of [\ion{O}{4}] and [\ion{Ne}{5}] are still problematic.
We suggest that of anywhere in the nebula, the eastern end of the
bubble may have the highest likelihood of excitation by shocks, as
would be expected from winds of the WN6 star located there hitting the
denser sides of an evacuated bubble.

\onecolumn
\twocolumn


\section{Conclusions}
\label{conclusions}

We present an infrared data cube of \dor, observed with the InfraRed
Spectrograph and Multiband Imaging Spectrometer on the {\it Spitzer}
Space Telescope.  Aromatic dust emission features are of modest
strength in the \dor\ region, concentrated in the arc-like two-lobed
ridge coincident with CO emission and partially encircling the central
R136 cluster.  Detailed analysis of the dust content of the region
from these features will follow in a subsequent publication, although
the high average dust temperature is immediately obvious from the shape of
the MIPS/SED spectrum.  Of the pure rotational lines of molecular
hydrogen, only S(3)$\lambda$9.67\um\ is detected with any significance,
peaking on the bright ``ridge'' that dominates the morphology of the
region at many wavelengths (see Fig~\ref{atlinemaps3}).  Low-ionization 
atomic lines are present
but not particularly strong: [\ion{Si}{2}] is detected in the
outskirts of the mapped region, outside the region of strongest
centimeter continuum (highest emission measure ionized gas).
[\ion{Fe}{2}] $\lambda$26.0\um\ is only tentatively detected in two
locations.

Two hydrogen recombination lines Hu$\alpha$ (7-6) and Hu$\gamma$ (9-6)
can be mapped over a large fraction of the observed area, and from
these a relative measure of the extinction can be calculated.
Independent estimates of the extinction are calculated from fitting
the strength of the 10\um\ silicate feature, and from the ratio of
H$\alpha$ to centimeter continuum.  While none of the three extinction
measures is extremely high signal-to-noise, the locations where all
three show higher extinction should be quite secure - in particular
three distinct locations in the west, southwest, and northeast,
outside of the ``ridge'' (Fig~\ref{threecolorextinction}).

The strongest atomic or molecular features in the data cube are the
moderate-ionization ionic lines \ariil, \ariiil, \neiil, \neiiil,
\siiil, \siiill, and \sivl.  These lines are most sensitive to
physical conditions in the ionized gas: the ionization parameter $U$
and hardness of the ionizing field, parametrized by the radiation
temperature $T_{rad}$.  We fit the Neon and Sulphur ratios with
photoionization models to derive a 2-dimensional map of $U$ and
$T_{rad}$.  We find that the excitation generally follows the ``ridge
and bubble'' morphology, and that under the assumptions of pure
photoionization, there are ``hot spots'' of hardened ionizing field in
the east of the bubble and between the two lobes of the ridge
(corresponding to two molecular clouds).  We also compare the line
ratios to shock models and find poorer agreement with the data.  In
particular the nondetections of emission from more highly ionized species [\ion{O}{4}]
and [\ion{Ne}{5}] suggest that photoionization dominates over
collisional excitation by shocks.
Overall, the local effects of hot stars in \dor\ (such as the single WR
star on the eastern side of the bubble) appear to dominate over any
large-scale trend with distance from the central cluster R136.

\appendix
\twocolumn

\section{Flux calibration and adjustment}
\label{appendix1}

\subsubsection{Background subtraction}

Our off-target background observations bracket the data set in time.
We chose to take an average (with minima and maxima trimmed) of all
the background records and subtract that from all of the data records.
It was possible that the two sets of background records could be
appreciably different, reflecting a gradual change in the instrument
(e.g. electronic drift), and that a linear interpolation of the two
sets over time would be more appropriate as the background for each
data record.  Indeed, the mean flux level of our background records
tends to increase with time, varying by 8\% between the first and last
exposures.  Therefore, the average of the second set of background
records has a higher mean flux level than the average of the first
set.

We tested three different background subtractions on several apertures 
from the SL1 module: using only the first background set, using only 
the second, and using the mean of the two sets.  The difference between 
the resulting
spectra was very small.  Specifically, the difference in continuum
levels between typical spectra using the different background
subtractions was on the order of 2\%.  For very low signal-to-noise
spectra extracted from faint areas of the map, the continuum
difference could reach 10\%.  We emphasize that, even then, the shape
of the spectrum and the strength of the emission line features did not
appreciably change.  For the highest signal-to-noise spectra, the
continuum difference was reduced to about 0.1\%.  We conclude that
using the mean of all background records is adequate.

\subsubsection{Flux calibration}\label{fluxcal}

The IRS SL slit is approximately 3.6 arcseconds wide, and the LL is
approximately 10.6 arcseconds wide \citep[see][Table 7.5]{som}.  Some
fraction of the instrument point-spread function (PSF) falls outside
of the slit, depending on wavelength.  For example, the full width
half max (FWHM) of the PSF at 14.5\um, at the upper wavelength range
of the SL first order, is about 3.74 arcseconds \citep[see][\S
  8.1.2.1]{som}.  
Optimized for point sources, the Spitzer Space Center pipeline adjusts the 
data for this loss of flux.  However, in the case of a spatially uniform extended 
source, there is no net loss of flux.  CUBISM can apply a slit loss correction factor 
which re-corrected the data to the original levels.  In the case of \dor, an extended 
sources, we opted to do so.
However, it is important to
note that in some areas of the map, there are bright features which
are neither point sources nor uniformly extended emission, which have
some net loss of flux.  The magnitude of this loss is at most 36\%
\citep[see][Slide 13]{sheth}\footnote{Available at:
  ssc.spitzer.caltech.edu/sust/workshop/2006data2/talks/kartik.pdf} .
This loss may be responsible for some of the mismatches we experienced
between the flux densities in the four different modules at some
points in the map.

\subsubsection{Final adjustments to the spectra}\label{finaladjustments}

The wavelength range of the spectra needs to be trimmed.  The full
range of the IRS in each of the four modules includes bins where the
response is unreliable.  In the LL modules, where the spectrum of \dor\ 
is generally smooth, the drop in response at the long-wave edge of the
LL2 module and the increase in noise at the long-wave edge of the LL1
module are evident.
We trimmed the
LL modules based on these observations.  The spectrum of \dor\ is more
complex in the SL modules, however, with many emission lines and an
unclear continuum.  In order to determine the range of wavelengths
which can be trusted in each SL module, we examined the IRS \lr\ 
staring mode spectrum of $\alpha$ Lacertae, an A1 dwarf star whose
smooth spectrum is well known.  
In each of the SL modules, the spectrum departs
from its smooth curve at the margins.  We trimmed our SL spectra to
the wavelengths where the IRS spectrum of $\alpha$ Lac stayed faithful
to the smooth curve.  We also found, based on some spectra from \dor,
that the limit at the long-wave edge of the SL1 module based on the
spectrum of $\alpha$ Lac was not conservative enough; consequently, we
decreased the trusted upper limit on wavelength for the SL1.  We could
not use the spectrum of $\alpha$ Lac to determine the reliable
wavelengths for the LL modules because the stellar spectrum has low
signal-to-noise at such long wavelengths.


After extracting our spectra, we found that some of the SL spectra
were suspiciously low, even negative for some counts.  This may be a
problem with background subtraction, and thus requires an additive
correction.  We raised both SL modules by the additive quantity
necessary to make the floor of the SL2 spectrum (the median of the five 
lowest counts) non-negative.  Over 72\% of the map, this correction was 
unnecessary.  Over the remaining 28\%, mainly in areas of low signal, 
the SL spectrum was raised by a mean level of 5.5 MJy/sr, with standard
deviation of 6.5 MJy/sr.  For comparison, the mean level of the 
background spectrum in the SL is 3.2 MJy/sr
and the mean level of the background-subtracted SL spectrum averaged 
across a large region of 30 Doradus is about 100 MJy/sr.

The different modules of an IRS spectrum do not always match up well
in flux density.  The mismatches between the SL1 and SL2 modules, and
between the LL1 and LL2 modules, were rarely serious.  However, we
noticed more consistent mismatches between the SL1 and LL2 modules.
The primary cause is probably that the emission in the \dor\ region is
often best characterized as neither a point source nor as a uniform
extended source, but instead as something in between.  We have treated
it as a uniform extended source (see \S\ref{fluxcal}), likely losing
some flux.  This would call for a multiplicative correction.  We found
logarithmic fits to the continuum emission close to the junction in
each module and multiplied the SL spectra by the ratio of the fits in
the overlapping wavelengths.  The mean value of the multiplicative 
correction is 1.2 with a standard deviation of 1.2.


\begin{figure}
\epsscale{1.05}
\plotone{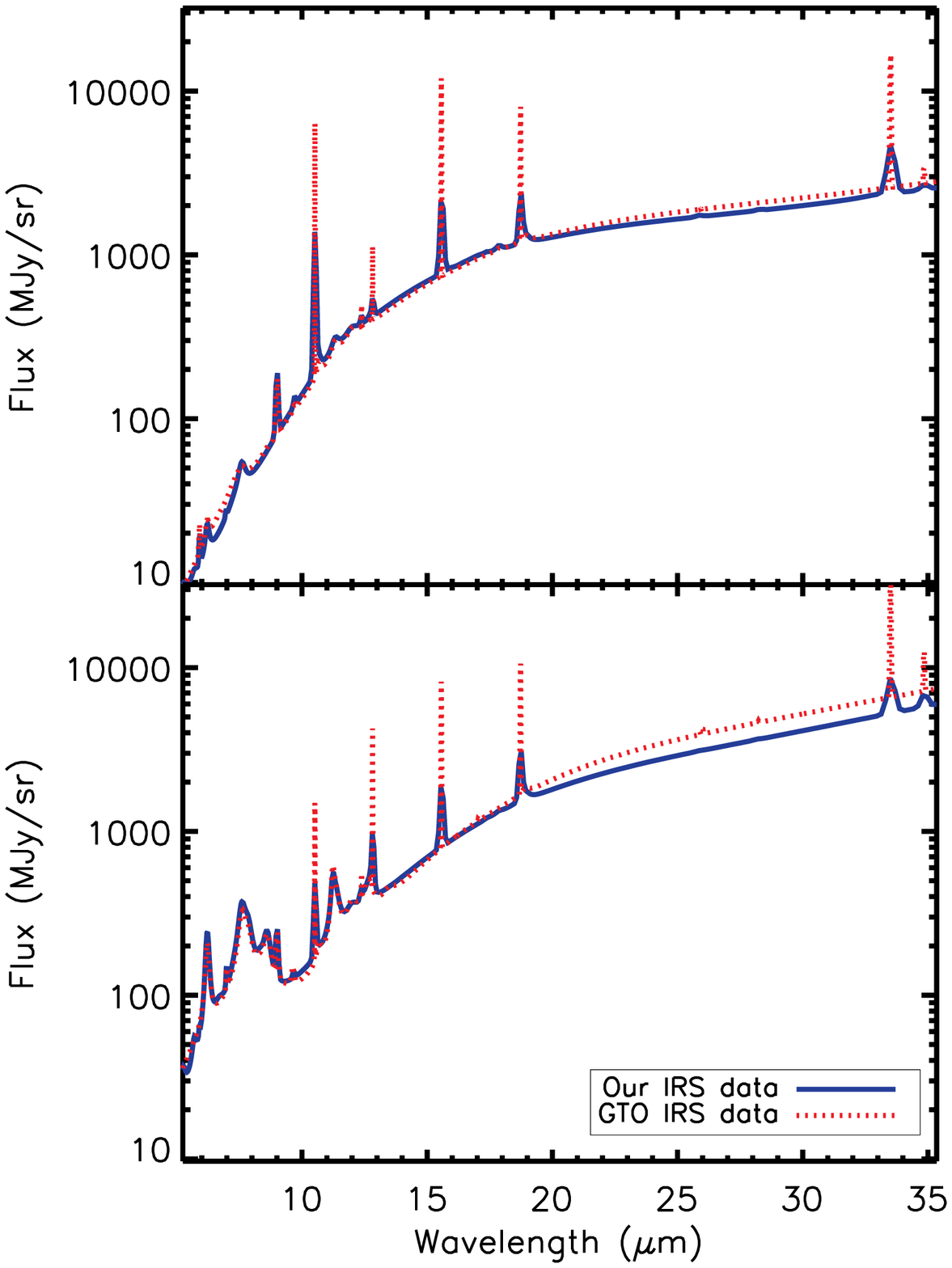}
\caption{\label{brandlcomp_total}In blue,
 the fit to our \lr\ spectra.  In red,
the same type of fit applied to the \hr\ GTO spectra, with \hr\ data at
long wavelengths combined with \lr\ data at short wavelengths (the
transition is at about 10\um).  Top: a relatively faint area in the
southern part of the \dor\ nebula.  Bottom: a brighter area, where the
\lr\ spectrum suffers from partial saturation and falling response in
the LL1 module.
}
\end{figure}

\begin{figure}
\epsscale{2.2}
\plottwo{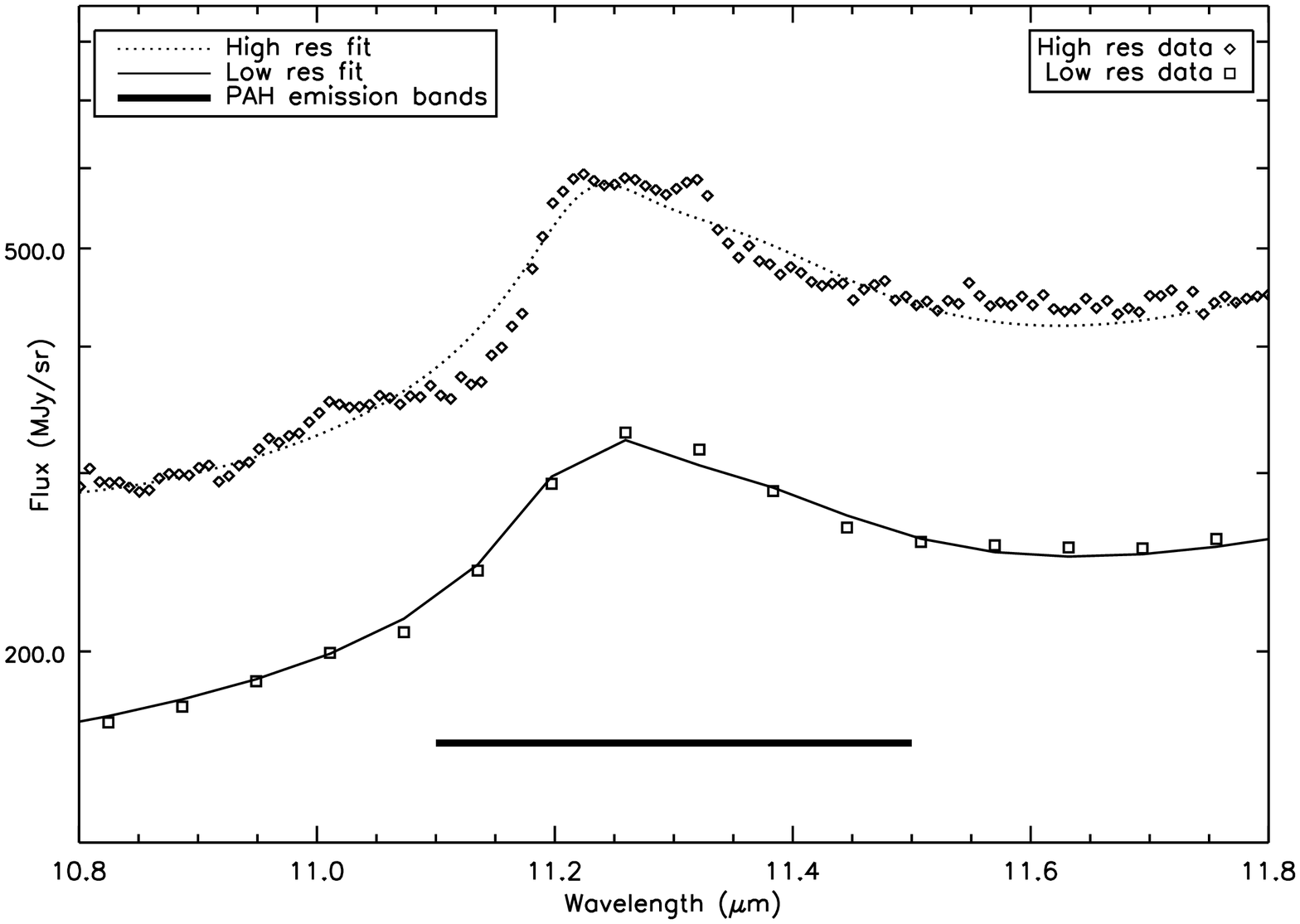}{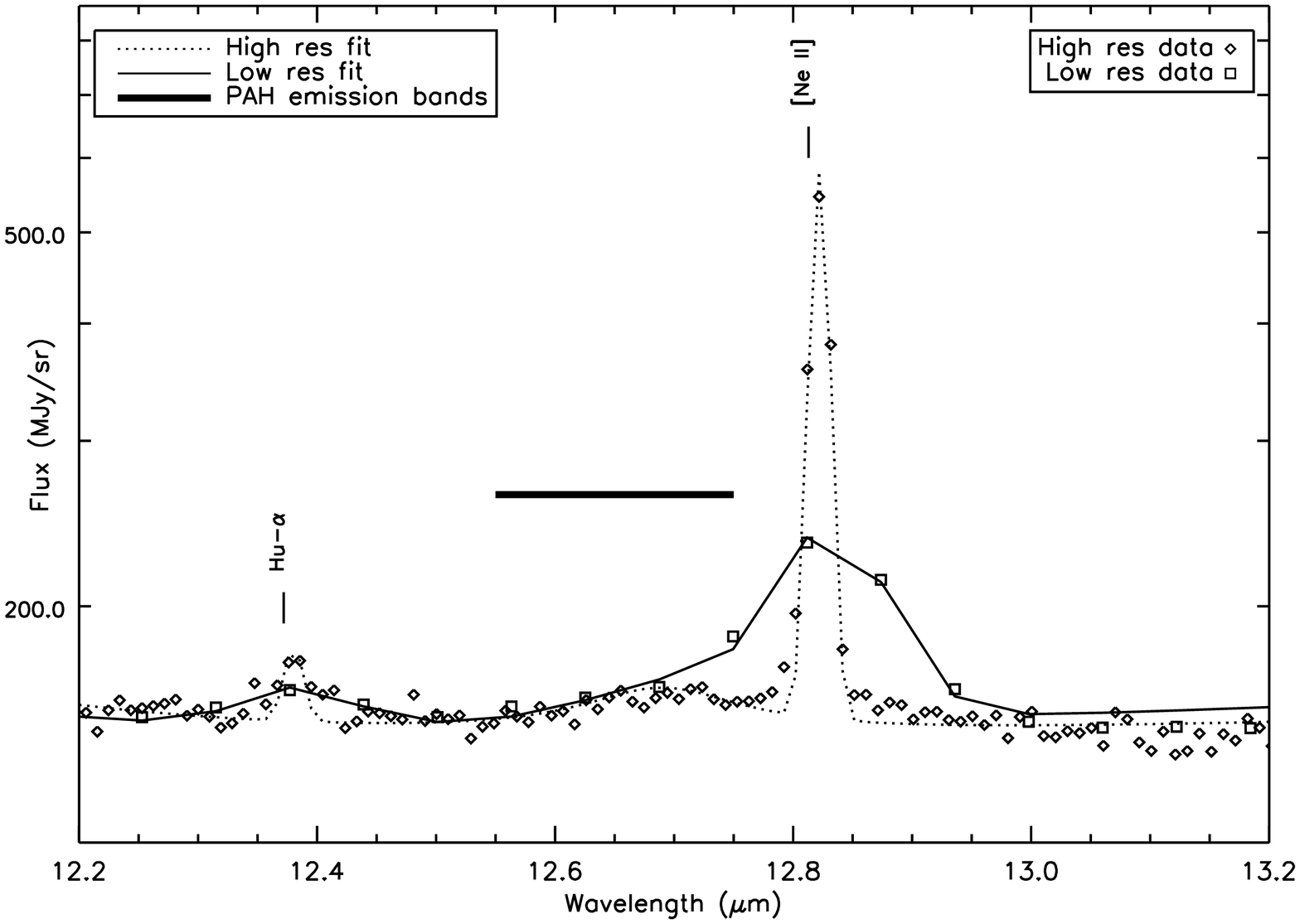}
\caption{\label{brandlcomp_singleline}Two comparisons of our
  low-resolution IRS data to high-resolution IRS data from
  \citet{lebouteiller08}.  (The two panels are drawn from different
  regions on the sky, but for each panel the high and low resolution
  spectra are both extracted from the same aperture on the sky.)  Both
  the measured spectra (symbols) and our fits made with PAHFIT (solid
  and dotted lines) are displayed.  Note that while the \hr\ spectrum
  has higher resolution for the lines, and the continuum strength may
  vary, the overall strength of the lines agree.  Also note that the
  PAH has a fairly smooth appearance in the low-resolution plot and a
  more complex morphology in the high-resolution spectrum, which can
  cause problems when attempting to fit PAH strength in
  high-resolution.  See Appendix~\ref{qasec}.}
\end{figure}

\begin{figure}
\epsscale{1.1}
\plotone{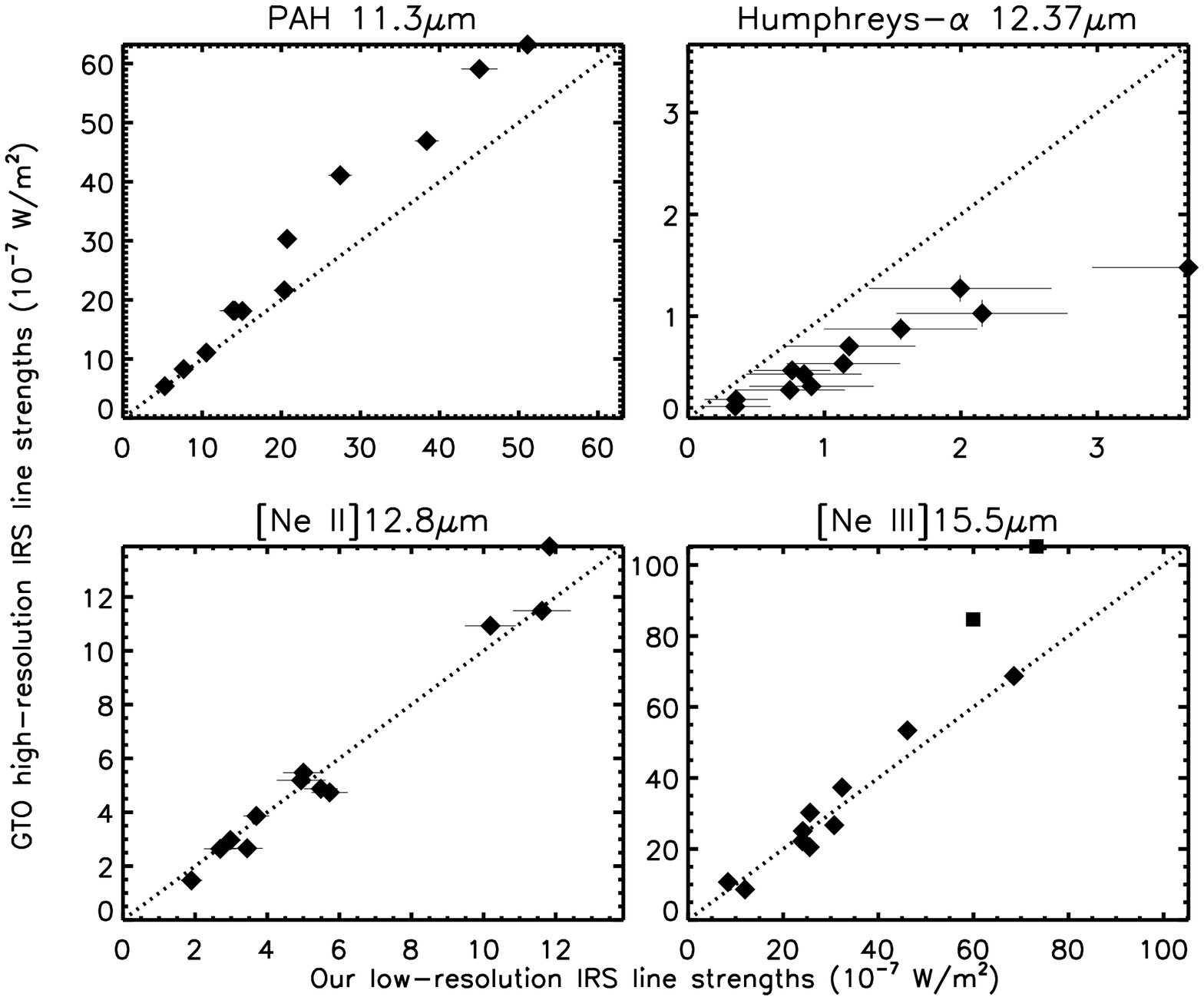}
\caption{\label{brandlcomp_linestrengths}Comparison of the fitted
  integrated strengths of four emission lines.  On the horizontal axis
  are the results from our \lr\ spectra.  On the vertical axis are the
  results of using the same type of fit on the GTO spectra, which are
  \hr\ longward of about 10 \um.  Each symbol represents one area of
  the \dor\ nebula, with horizontal and vertical error bars.  The
  discrepancies from the line of unity are discussed in
  Appendix~\ref{qasec}.  Among those discrepancies, the fitting
  algorithm we used is not designed to fit the more complex structure
  of PAHs that is resolved in the \hr\ spectrum.  Another reason for
  the discrepancy in the \hua\ is that our fits to the line also
  include the fainter \molhtwol\ emission line, which is not well resolved
  at \lr.  The square symbols in the lower right plot call attention
  to two outlying points discussed in Appendix ~\ref{qasec}.}
\end{figure}

\section{Quality Assurance}\label{qasec}
\label{appendix2}

In order to check our work on reducing the spectral map, we compared
our spectra, taken with the \lr\ modules of the IRS in mapping mode, to
spectra of several areas of the \dor\ nebula taken with both the low
and \hr\ modules of the IRS, in staring mode \citep [private
  correspondence and][]{lebouteiller08}\footnote{IRS GTO program,
  ID\#63, AOR keys 4382720, 12081152, and 12081408}.  These spectra
will henceforth be referred to as the GTO spectra.  The apertures they
span are displayed in green in Figure \ref{irscoverage}.  By selecting
the subsets of our map that coincide with the GTO apertures, we can
produce spectra which should be consistent with theirs.  We applied
PAHFIT, the same fitting algorithm that we use on our spectral map, to
the GTO spectra.  We made some adjustments to PAHFIT to allow for the
much higher spectral resolution in the GTO spectra redward of about
10\um, where there is coverage from the \hr\ modules.

In Figure \ref{brandlcomp_total}, the fitted spectra from two of these
apertures are shown.  The overall agreement in the shape of the
spectrum is clear, with a notable exception in the continuum from 20
to 32\um (likely the result of nonlinearity in the short-low detector response).
In the upper panel, taken from an aperture toward the south of the
\dor\ nebula, at around \radec{05}{38}{50}{-69}{06}{41}, where the 
continuum emission at long wavelengths is relatively faint, the 
agreement between the GTO spectrum and ours is
close, but the effect starts to become noticeable at higher flux densities.  In the lower
panel, taken from a brighter part of the \dor\ nebula at 
\radec{05}{38}{48}{-69}{04}{10} where the falling
response in the LL1 is more severe, the difference between the GTO
spectrum and ours is larger.

We are most concerned with whether we can obtain the same emission
line strengths by fitting the GTO spectra in the same manner as ours.
Figures \ref{brandlcomp_total} and \ref{brandlcomp_singleline} show
that the PAHs have similar profiles, while the unresolved atomic lines
are narrower with higher central intensities in the long-wave, \hr\ 
part of the GTO spectra.  The integrated strengths of several of the
long-wave emission lines of interest are plotted in Figure
\ref{brandlcomp_linestrengths}, for twelve GTO spectra and the
corresponding apertures in our map.  Perfect agreement falls along the
line of unity.

The fits to the PAH complex at 11.3\um\ agree well only at low
strength.  This may be because of detailed fine structure that appears
in the PAH complex at \hr.  In the \lr\ spectrum, only a broad feature
at 11.3\um, with a shoulder at 11.0\um, is apparent.  A Drude profile
is sufficient to fit that, and a more complex fitting profile would
not be appropriate because of the lack of resolution.  In \hr\ such as
in the GTO spectra, the shoulder at 11.0\um\ resolves into a
completely separate feature (see Figure \ref{brandlcomp_singleline}),
possibly another PAH feature.  This might also be the Paschen-$\gamma$
HI recombination line, although so significantly shifted from its rest
wavelength at 10.95$\mu$m as to make such an identification questionable.
The main PAH feature at 11.3\um\ displays a different profile that
would require a more complex fitting algorithm.
Thus, we are not concerned about this disparity.  The conclusion is
that PAHFIT is adequate for fitting PAHs at \lr, but may not
sufficiently describe the same features at \hr\ -- not surprising, as
the list of features used in PAHFIT was empirically tuned using
\lr\ spectra.

A systematic disparity is evident in the comparison of the
\hual\ fitted line strength.  Upon closer examination, two factors
surfaced.  Both support the trend that the fitted line strength at
\lr\ is higher than the line strength at \hr.  First, the fainter
molecular hydrogen line, \molhtwol, is resolved in \hr.  The fit to
the \hr\ \hua\ line therefore does not include any of the flux from
the molecular hydrogen line. Meanwhile, in the \lr\ spectrum, the flux
from both lines contributes to the same broad feature.  We are
confident that \hua\ contributes most of the flux we see in \lr, in
part because the line center more closely coincides with 12.37 than
with 12.28\um, and in part because, as we can see in the \hr\ spectra,
the \hua\ line is equal to or stronger than \molhtwo\ in all of the
areas of our map covered by the \hr\ spectra.  However, the flux of
both emission lines is contributing to the low-resolution fit, and so
to some extent, our fits to the line strength of the \hua\ line are
overestimated.  Quantitative comparison between high- and
low-resolution spectra and between fits with and without the molecular
hydrogen line show that the contamination of the \hua\ line strength
is at most $\sim$20\% (see \S\ref{extinction}).  This effect is
weakest in the parts of \dor\ dominated by ionized gas that are the
focus of this paper, but should be kept in mind for analysis of the
outer parts of the region and PDR physics in future papers.
The second factor that contributes to the disparity seen in the second
panel of Figure \ref{brandlcomp_linestrengths} is the PAH feature at
12.6\um.  As discussed above in the context of the PAH complex at
11.3\um, the fitting package PAHFIT is intended for the relatively
smooth features seen in \lr, and does not handle the complex
morphology of PAHs seen at \hr\ as well.  In the case of the PAH at
12.6\um, its line strength in the \hr\ fit is generally overestimated.
Its broad wings thus lead the strength of the \hua\ line to be
underestimated in the \hr\ fit.  The slight overestimation of the line
in the \lr\ spectrum and the slight underestimation of the line in the
\hr\ spectrum both contribute to the disparity.

The results for the forbidden neon lines, as for most of the stronger
lines, show a tighter correlation.  The close agreement in the fits to
the low and \hr\ spectra for the weaker \neiil\ is particularly
encouraging because one of our concerns in interpreting the neon ratio
was the possible entanglement of the \neii\ atomic emission line with
the above-mentioned PAH at 12.6\um.  In the \hr\ spectrum, the PAH
(with its often-unusual shape) and the atomic line are resolved.  As
in the case of the \hua\ line, the strength of the \neii\ line may be
slightly underestimated because of the poor, overestimated fit to the
PAH at 12.6\um.  The close agreement between the \lr\ and \hr\ fit,
across almost all of the twelve apertures, regardless of PAH strength
in each aperture, indicates that there are no systematic errors to the
\lr\ fit that do not also appear in the \hr\ fit.  Thus, our results may
share the slight bias toward underestimating the \neii\ strength, but
we are generally able to decompose \neii\ and the PAH at 12.6\um\ 
despite the lack of spectral resolution.

Finally, the correlation for the fitted line strength of \neiiil\ is
similarly good, with the exception of two apertures where the \hr\ 
fitted line strength is significantly higher than the \lr\ fitted line
strength.  These two apertures (displayed with squares rather than
diamonds in Figure \ref{brandlcomp_linestrengths}) are located in a
small region of the map that exemplifies the extreme of the trend
shown in Figure \ref{brandlcomp_total}, where the response in the LL1
module falls until the module finally saturates.
In these cases, our fit to the continuum
with thermal dust components is not very successful.  The fitted
continuum is dragged down in the area from 20 to 32\um, which forces
it up in the region of 15\um.  This causes our estimation of the
strength of \neiii\ to be underestimated.

For these two anomalous apertures, we conducted a test by removing the
faulty LL1 spectrum and fitting only the remaining spectrum, to get a
better estimation of continuum in the region of \neiii.  This
increased the fitted strength of \neiii, though not enough to get it
to agree with the value we obtained by fitting the high-resolution
spectrum.

The effect of long-wave line strengths being underestimated in our \lr\ 
spectra compared to the \hr\ GTO spectra is seen for \neiiil\ and
\siiill\ in some areas of the map.  Empirically, the line strengths of
\neiil\ and \siiil\ appear to be unaffected, as shown by their close
agreement with the GTO spectra.  It is crucial to note that this
effect is restricted to the small areas of the map where we can see
falling response in the LL1 module, as drawn in
Figure~\ref{saturationmap}.  In that localized area, we can expect our
values of \neiii\ and \siiill\ to be underestimated relative to the GTO
results by perhaps 30\%, as shown in Figure
\ref{brandlcomp_linestrengths} (the effect is very similar for
\siiill).  Outside these small areas of the map, the agreement between
our spectra and the high-resolution GTO spectra is excellent for these
two emission lines.

\ \\

\centerline{\bf{ACKNOWLEDGMENTS}}

This work benefited greatly from the generous contributions of several
individuals: Gary Ferland and his students for creating the Cloudy
photoionization code, John Raymond for giving us a copy of his shock
code, John Dickel and Jasmina Lazendic for giving us their calibrated
centimeter synthesis images, and Sean Points for giving us calibrated
MCELS H$\alpha$ data.
This publication would not have been possible without extensive use of
NASA's Astrophysics Data System Bibliographic Services, and the SIMBAD
database, operated at CDS, Strasbourg, France.
This publication makes use of data products from the Two Micron All
Sky Survey, which is a joint project of the University of
Massachusetts and the Infrared Processing and Analysis
Center/California Institute of Technology, funded by NASA and NSF.
Data from the {\it Spitzer} Space Telescope (operated by the Jet
Propulsion Lab for NASA) are taken from PID 30653 ``Stellar Feedback
on Circumcluster Gas and Dust in 30 Doradus, the Nearest Super-Star
Cluster'' and PID 20203 ``Spitzer Survey of the Large Magellanic
Cloud: Surveying the Agents of a Galaxy's Evolution (SAGE)''
GdM and RI were supported in part by Spitzer/NASA/JPL grant \#1288328,
and RI in part by a Spitzer fellowship to UVa. Meixner and Sewilo were
supported in part by {\it Spitzer}/NASA/JPL grants \#1288277 and
\#1275598.


{\it Facilities:} \facility{Spitzer (IRS)}



\begin{thebibliography}{}\setlength{\itemsep}{0cm}\setlength{\parsep}{0cm}

\bibitem[Abel \& Satyapal(2008)]{abel} Abel, N.~P., \& Satyapal, S.\ 2008, ArXiv e-prints, 801, arXiv:0801.2766 
\bibitem[Asplund et al.(2005)]{asplund05} Asplund, M., Grevesse, 
N., \& Sauval, A.~J.\ 2005, Cosmic Abundances as Records of Stellar 
Evolution and Nucleosynthesis, 336, 25 
\bibitem[Bernard et al(2008)]{sageism} Bernard, J.P.\ 2008, \aj\ submitted.
\bibitem[Brandner et al.(2001)]{brandner01} Brandner, W., Grebel, 
E.~K., Barb{\'a}, R.~H., Walborn, N.~R., \& Moneti, A.\ 2001, \aj, 122, 858 
\bibitem[Breysacher et al.(1999)]{breysacher} Breysacher, J., 
Azzopardi, M., \& Testor, G.\ 1999, \aaps, 137, 117 
\bibitem[Castelli \& Kurucz(2004)]{atlasref} Castelli, F., \& 
Kurucz, R.~L.\ 2004, ArXiv Astrophysics e-prints, arXiv:astro-ph/0405087 
\bibitem[Chiar \& Tielens(2006)]{chiar06} Chiar, J.~E., \& Tielens, A.~G.~G.~M.\ 2006, \apj, 637, 774
\bibitem[Chu et al.(2004)]{chu04snr} Chu, Y.-H., Gruendl, R.~A., Chen, C.-H.~R., Lazendic, J.~S., \& Dickel, J.~R.\ 2004, \apj, 615, 727 
\bibitem[Cox \& Raymond(1985)]{raymond2} Cox, D.~P., \& Raymond, J.~C.\ 1985, \apj, 298, 651 
\bibitem[Dickel et al.(1994)]{dickel94} Dickel, J.~R., Milne, D.~K., Kennicutt, R.~C., Chu, Y.-H., \& Schommer, R.~A.\ 1994, \aj, 107, 1067 
\bibitem[Dopita et al.(2006)]{dopita06} Dopita, M.~A., et al.\ 2006, \apj, 639, 788 
\bibitem[Dudik et al.(2007)]{dudik} Dudik, R.~P., 
Weingartner, J.~C., Satyapal, S., Fischer, J., Dudley, C.~C., \& 
O'Halloran, B.\ 2007, \apj, 664, 71 
\bibitem[Fazio et al.(2004)]{fazio_irac} Fazio, G.~G., et al.\ 
2004, \apjs, 154, 10 
\bibitem[Ferland et al.(1998)]{ferland} Ferland, G.~J., 
Korista, K.~T., Verner, D.~A., Ferguson, J.~W., Kingdon, J.~B., \& Verner, 
E.~M.\ 1998, \pasp, 110, 761 
\bibitem[France et al.(2007)]{france07} France, K., McCandliss, 
S.~R., \& Lupu, R.~E.\ 2007, \apj, 655, 920 
\bibitem[Galliano et al.(2008)]{galliano} Galliano, F., Dwek, 
E., \& Chanial, P.\ 2008, \apj, 672, 214 
\bibitem[Gordon et al.(2007)]{gordon_ge} Gordon, K.~D., et al.\ 
2007, \pasp, 119, 1019 
\bibitem[Hartigan et al.(1987)]{hrh} Hartigan, P., Raymond, 
J., \& Hartmann, L.\ 1987, \apj, 316, 323 
\bibitem[Hollenbach \& McKee(1989)]{hollenbach89} Hollenbach, D., \& 
McKee, C.~F.\ 1989, \apj, 342, 306 
\bibitem[Houck et al.(2004)]{houck_irs} Houck, J.~R., et al.\ 
2004, \procspie, 5487, 62 
\bibitem[Hubeny \& Lanz(1995)]{tlusty} Hubeny, I., \& Lanz, T.\ 1995, \apj, 439, 875 
\bibitem[Hunter (1999)]{hunter} Hunter, D.A.  1999, IAUS, 190, 217
\bibitem[Hyland et al.(1992)]{hyland92} Hyland, A.~R., Straw, 
S., Jones, T.~J., \& Gatley, I.\ 1992, \mnras, 257, 391 
\bibitem[Indebetouw et al.(2005)]{remyextinct} Indebetouw, R., et 
al.\ 2005, \apj, 619, 931 
\bibitem[Johansson et al.(1998)]{johansson} Johansson, L.~E.~B., 
et al.\ 1998, \aap, 331, 857 
\bibitem[Johnson(2004)]{johnson} Johnson, K.~E.  2004, \nar, 48, 1337
\bibitem[Kim(2007)]{kim07} Kim, S.\ 2007, The Seventh Pacific 
Rim Conference on Stellar Astrophysics, 362, 297 
\bibitem[Lacy et al.(2007)]{lacy07} Lacy, J.~H., et al.\ 2007, \apjl, 658, L45 
\bibitem[Lazendic et al.(2003)]{lazendic} Lazendic, J.~S., Dickel, J.~R., \& Jones, P.~A.\ 2003, \apj, 596, 287 
\bibitem[Lebouteiller et al.(2007)]{lebouteiller08} Lebouteiller, V., 
Bernard-Salas, J., Brandl, B., Whelan, D., Wu, Y., Charmandaris, V., \& 
Devost, D.\ 2007, ArXiv e-prints, 710, arXiv:0710.4549 
\bibitem[Lefloch et al.(2003)]{lefloch03} Lefloch, B., 
Cernicharo, J., Cabrit, S., Noriega-Crespo, A., Moro-Mart{\'{\i}}n, A., \& 
Cesarsky, D.\ 2003, \apjl, 590, L41 
\bibitem[Lodders(2003)]{lodders03} Lodders, K.\ 2003, \apj, 591, 
1220 
\bibitem[Lodders(2007)]{lodders07} Lodders, K.\ 2007, ArXiv 
e-prints, 710, arXiv:0710.4523 
\bibitem[Lu et al.(2008)]{lu_sed} Lu, N., et al.\ 2008, ArXiv 
e-prints, 802, arXiv:0802.3723 
\bibitem[Maercker \& Burton(2005)]{maercker05} Maercker, M., \& 
Burton, M.~G.\ 2005, \aap, 438, 663 
\bibitem[Madden et al.(2006)]{madden06} Madden, S.~C., Galliano, 
F., Jones, A.~P., \& Sauvage, M.\ 2006, \aap, 446, 877 
\bibitem[Mart{\'{\i}}n-Hern{\'a}ndez et al.(2002)]{martinh02} 
Mart{\'{\i}}n-Hern{\'a}ndez, N.~L., et al.\ 2002, \aap, 381, 606 
\bibitem[Meixner et al.(2006)]{meixner} Meixner, M., et al.  2006, \aj, 132, 2268
\bibitem[Mendoza \& Zeippen(1982)]{mendoza} Mendoza, C., \& 
Zeippen, C.~J.\ 1982, \mnras, 199, 1025 
\bibitem[Morisset et al.(2004)]{morisset04} Morisset, C., 
Schaerer, D., Bouret, J.-C., \& Martins, F.\ 2004, \aap, 415, 577 
\bibitem[Molinari \& Noriega-Crespo(2002)]{molinari02} Molinari, 
S., \& Noriega-Crespo, A.\ 2002, \aj, 123, 2010 
\bibitem[Neufeld et al.(2006)]{neufeld06} Neufeld, D.~A., et al.\ 
2006, \apj, 649, 816 
\bibitem[Neufeld et al.(2007)]{neufeld07} Neufeld, D.~A., 
Hollenbach, D.~J., Kaufman, M.~J., Snell, R.~L., Melnick, G.~J., Bergin, 
E.~A., \& Sonnentrucker, P.\ 2007, \apj, 664, 890 
\bibitem[Noriega-Crespo et al.(2004)]{alberto04} Noriega-Crespo, 
A., et al.\ 2004, \apjs, 154, 352 
\bibitem[Noriega-Crespo et al.(2004)]{alberto04b} Noriega-Crespo, 
A., Moro-Martin, A., Carey, S., Morris, P.~W., Padgett, D.~L., Latter, 
W.~B., \& Muzerolle, J.\ 2004, \apjs, 154, 402 
\bibitem[Okamoto et al.(2001)]{okamoto01} Okamoto, Y.~K., Kataza, H., Yamashita, T., Miyata, T., \& Onaka, T.\ 2001, \apj, 553, 254 
\bibitem[Parker(1993)]{parker} Parker, J.~W.\ 1993, \aj, 106, 560 
\bibitem[Peck et al.(1997)]{peck} Peck, A.~B., Goss, W.~M., Dickel, H.~R., Roelfsema, P.~R., Kesteven, M.~J., Dickel, J.~R., Milne, D.~K., \& Points, S.~D.\ 1997, \apj, 486, 329 
\bibitem[Peeters et al.(2002)]{peeters} Peeters, E., et al.\ 
2002, \aap, 381, 571 
\bibitem[Poglitsch et al.(1995)]{poglitsch} Poglitsch, A., 
Krabbe, A., Madden, S.~C., Nikola, T., Geis, N., Johansson, L.~E.~B., 
Stacey, G.~J., \& Sternberg, A.\ 1995, \apj, 454, 293 
\bibitem[Rakowski et al.(2007)]{rakowski} Rakowski, C.~E., 
Raymond, J.~C., \& Szentgyorgyi, A.~H.\ 2007, \apj, 655, 885 
\bibitem[Raymond(1979)]{raymond1} Raymond, J.~C.\ 1979, \apjs, 39, 1 
\bibitem[Rieke et al.(2004)]{rieke_mips} Rieke, G.~H., et al.\ 
2004, \apjs, 154, 25 
\bibitem[Rubio et al.(1998)]{rubio98} Rubio, M., Barb{\'a}, 
R.~H., Walborn, N.~R., Probst, R.~G., Garc{\'{\i}}a, J., \& Roth, M.~R.\ 
1998, \aj, 116, 1708 
\bibitem[Rosa \& Mathis(1987)]{rosa} Rosa, M., \& Mathis, J.~S.\ 1987, \apj, 317, 163 
\bibitem[Shaver et al.(1983)]{shaver83} Shaver, P.~A., McGee, R.~X., Newton, L.~M., Danks, A.~C., \& Pottasch, S.~R.\ 1983, \mnras, 204, 53 
\bibitem[Schaefer(2008)]{schaefer} Schaefer, B.~E.\ 2008, \aj, 
135, 112 
\bibitem[Schaerer \& de Koter(1997)]{costarref} Schaerer, D., \& 
de Koter, A.\ 1997, \aap, 322, 598 
\bibitem[Sheth (2006)]{sheth} Sheth, K.  2006, in Spitzer Data Analysis Workshop \#4.
\bibitem[Simpson et al.(2007)]{simpson07} Simpson, J.~P., Colgan, 
S.~W.~J., Cotera, A.~S., Erickson, E.~F., Hollenbach, D.~J., Kaufman, 
M.~J., \& Rubin, R.~H.\ 2007, ArXiv e-prints, 708, arXiv:0708.2103 
\bibitem[Smith (2007a)]{smith07pahfit} Smith, J.D.T., et al.  2007, \apj, 656, 770
\bibitem[Smith (2007b)]{smith07cubism} Smith, J.D.T., et al.  2007, \pasp, 119, 1133
\bibitem[Smith et al.(2002)]{lindasmith} Smith, L.~J., Norris, 
R.~P.~F., \& Crowther, P.~A.\ 2002, \mnras, 337, 1309 
\bibitem[Sofia \& Jenkins(1998)]{sofia98} Sofia, U.~J., \& 
Jenkins, E.~B.\ 1998, \apj, 499, 951 
\bibitem[Spitzer Observer's Manual 7.1 (2006)]{som} Spitzer Science Center.  2006, Spitzer Space Telescope Observer's Manual, Version 7.1, Spitzer Science Center
\bibitem[Stasi{\'n}ska \& Schaerer(1997)]{stasinska} 
Stasi{\'n}ska, G., \& Schaerer, D.\ 1997, \aap, 322, 615 
\bibitem[Sturm et al.(2000)]{sturm00} Sturm, E., Lutz, D., 
Tran, D., Feuchtgruber, H., Genzel, R., Kunze, D., Moorwood, A.~F.~M., \& 
Thornley, M.~D.\ 2000, \aap, 358, 481 
\bibitem[Tayal \& Gupta(1999)]{tayal} Tayal, S.~S., \& Gupta, 
G.~P.\ 1999, \apj, 526, 544 
\bibitem[Townsley et al.(2006)]{townsley06} Townsley, L.~K., 
Broos, P.~S., Feigelson, E.~D., Brandl, B.~R., Chu, Y.-H., Garmire, G.~P., 
\& Pavlov, G.~G.\ 2006, \aj, 131, 2140 
\bibitem[Tsamis \& P{\'e}quignot(2005)]{tsamis05} Tsamis, Y.~G., 
\& P{\'e}quignot, D.\ 2005, \mnras, 364, 687 
\bibitem[Vermeij et al.(2002)]{vermeij02} Vermeij, R., Peeters, E., Tielens, A.~G.~G.~M., \& van der Hulst, J.~M.\ 2002, \aap, 382, 1042  
\bibitem[Walborn (1991)]{walborn} Walborn, N.R.  1991, IAUS, 148, 145
\bibitem[Weingartner \& Draine(2001)]{weingartner01} Weingartner, J.~C., \& Draine, B.~T.\ 2001, \apj, 548, 296.
\bibitem[Werner et al.(2004)]{werner04} Werner, M.~W., et al.\ 
2004, \apjs, 154, 1 

\end{thebibliography}
\end{document}